\documentclass[longauth]{aa}

\usepackage{graphicx}
\usepackage{txfonts}
\usepackage{hyperref}
\hypersetup{colorlinks=true,linkcolor=blue,citecolor=blue}
\usepackage{natbib}  
\bibpunct{(}{)}{;}{a}{}{,}  
\usepackage{enumitem}
\usepackage[dvipsnames]{xcolor}
\usepackage{comment}
\usepackage{afterpage}
\usepackage{siunitx}
\usepackage{array}
\usepackage[normalem]{ulem} 
\usepackage[varvw]{newtxmath}
\usepackage{stfloats}
\DeclareUnicodeCharacter{2212}{-}
\usepackage{xspace}

\newcommand{\LePhare}{\texttt{LePHARE}\xspace}
\newcommand{\SEPP}{\texttt{SourceXtractor++}\xspace}
\newcommand{\treecorr}{\texttt{TreeCorr}\xspace}
\DeclareSIUnit\angstrom{\text {Å}}

\defcitealias{Jose16}{J16}
\defcitealias{Reed09}{R09}

\newcommand{\citeinprep}[1]{#1 et al. (in prep)}

\newcommand{\citepinprep}[1]{(#1 et al., in prep)}

\usepackage{cancel} 
\usepackage{bm} 
\newcommand{\revisionadded}[1]{
    \ifmmode
        \textcolor{Green}{\bm{#1}}
    \else
        \textbf{\textcolor{Green}{#1}}
    \fi
}
\newcommand{\revisionremoved}[1]{
    \xspace
    \ifmmode
        \textcolor{BrickRed}{\cancel{#1}} 
    \else
        \textcolor{BrickRed}{\sout{#1}} 
    \fi
    \xspace
}

\begin{document}
    \title{Tracing the galaxy-halo connection with galaxy clustering in COSMOS-Web from $z = 0.1$ to $z \sim 12$}

   \author{
        L.~Paquereau\inst{\ref{IAP}}\fnmsep\thanks{\email{louise.paquereau@iap.fr} \\ ORCID: 0000-0003-2397-0360}
        \and C.~Laigle\inst{\ref{IAP}}
        \and H.~J.~McCracken\inst{\ref{IAP}}
        \and M.~Shuntov\inst{\ref{DAWN},\ref{NBI}}
        \and O.~Ilbert \inst{\ref{LAM}}
        \and H.~B.~Akins\inst{\ref{UAT}}
        \and N. Allen \inst{\ref{UAT}, \ref{DAWN}}
        \and R.~Arango-~Togo \inst{\ref{LAM}}
        \and E.~M.~Berman\inst{\ref{NortheasternU}}
        \and M.~Béthermin\inst{\ref{Strasbourg}}
        \and C.~M.~Casey\inst{\ref{UAT}, \ref{DAWN}}
        \and J.~McCleary\inst{\ref{NortheasternU}}
        \and Y.~Dubois\inst{\ref{IAP}}
        \and N.~E.~Drakos\inst{\ref{HawaiiHilo}}
        \and A.~L.~Faisst\inst{\ref{Caltech}}
        \and M.~Franco\inst{\ref{UAT}}
        \and S.~Harish\inst{\ref{Rochester}}
        \and C.~K.~Jespersen\inst{\ref{Princeton}}
        \and J.~S.~Kartaltepe\inst{\ref{Rochester}}
        \and A.~M.~Koekemoer\inst{\ref{STScI}}
        \and V.~Kokorev\inst{\ref{UAT}}
        \and E.~Lambrides\inst{\ref{NASAGoddard}}
        \and R.~Larson\inst{\ref{UAT}}
        \and D.~Liu\inst{\ref{Nanjing}}
        \and D.~Le Borgne\inst{\ref{IAP}}
        \and J.~S.~W.~Lewis\inst{\ref{IAP}}
        \and J.~McKinney\inst{\ref{UAT}}
        \and W.~Mercier \inst{\ref{LAM}}
        \and J.~D.~Rhodes\inst{\ref{NASA}}
        \and B.~E.~Robertson\inst{\ref{UnivCalifornia}}
        \and S.~Toft\inst{\ref{DAWN}, \ref{NBI}}
        \and M.~Trebitsch\inst{\ref{MeudonLUX}}
        \and L.~Tresse\inst{\ref{LAM}}
        \and J.~R.~Weaver\inst{\ref{UMASS}}
    }

   \institute{
        Institut d’Astrophysique de Paris, UMR 7095, CNRS, Sorbonne Université, 98 bis boulevard Arago, F-75014 Paris, France\label{IAP}
        \and Cosmic Dawn Center (DAWN), Denmark\label{DAWN}
        \and Niels Bohr Institute, University of Copenhagen, Jagtvej 128, 2200 Copenhagen, Denmark \label{NBI}
        \and Aix Marseille University, CNRS, CNES, LAM, Marseille, France\label{LAM}
        \and Department of Astronomy, The University of Texas at Austin, 2515 Speedway Blvd Stop C1400, Austin, TX 78712, USA \label{UAT}
        \and Purple Mountain Observatory, Chinese Academy of Sciences, 10 Yuanhua Road, Nanjing 210023, China \label{Nanjing}
        \and Department of Astronomy and Astrophysics, University of California, Santa Cruz, 1156 High Street, Santa Cruz, CA 95064 USA\label{UnivCalifornia}
        \and Department of Physics and Astronomy, University of Hawaii, Hilo, 200 W Kawili St, Hilo, HI 96720, USA\label{HawaiiHilo}
        \and Jet Propulsion Laboratory, California Institute of Technology, 4800 Oak Grove Drive, Pasadena, CA 91109\label{NASA}
        \and Laboratory for Multiwavelength Astrophysics, School of Physics and Astronomy, Rochester Institute of Technology, 84 Lomb Memorial Drive, Rochester, NY 14623, USA\label{Rochester}
        \and Space Telescope Science Institute, 3700 San Martin Drive, Baltimore, MD 21218, USA\label{STScI}
        \and Kapteyn Astronomical Institute, University of Groningen, P.O. Box 800, 9700AV Groningen, The Netherlands\label{Groningen}
        \and Caltech/IPAC, 1200 E. California Blvd. Pasadena, CA 91125, USA\label{Caltech}
        \and Northeastern University, 100 Forsyth St. Boston, MA 02115, USA \label{NortheasternU}
        \and NASA Goddard Space Flight Center, Code 662, Greenbelt, MD, 20771, USA \label{NASAGoddard}
        \and Université de Strasbourg, CNRS, Observatoire astronomique de Strasbourg, UMR 7550, 67000 Strasbourg, France \label{Strasbourg}
        \and Department of Astrophysical Sciences, Princeton University, Princeton, NJ 08544, USA \label{Princeton}
        \and LUX, Observatoire de Paris, Université PSL, Sorbonne Université, CNRS, 75014 Paris, France \label{MeudonLUX}
        \and Department of Astronomy, University of Massachusetts, Amherst, MA 01003, USA\label{UMASS}
   }

   \date{Received; accepted}

 
\abstract{
We explore the evolving relationship between galaxies and their dark matter halos from $z \sim 0.1$ to $z \sim 12$ using mass-limited angular clustering measurements in the 0.54 deg$^2$ of the COSMOS-Web survey, the largest contiguous JWST extragalactic survey. This study provides the first measurements of the mass-limited two-point correlation function at $z \ge 10$ and a consistent analysis spanning 13.4 Gyr of cosmic history, setting new benchmarks for future simulations and models. Using a halo occupation distribution (HOD) framework, we derive characteristic halo masses and the stellar-to-halo mass relationship (SHMR) across redshifts and stellar mass bins. Our results first indicate that HOD models fit data at $z \ge 2.5$ best when incorporating a non-linear scale-dependent halo bias, boosting clustering at non-linear scales ($r = 10-100$\,kpc). 
We find that galaxies at $z \ge 10.5$ with $\log(M_\star / M_\odot) \ge 8.85$ are predominantly centrals in halos with $M_{\rm h} \sim 10^{10.5}\,M_\odot$, achieving a star formation efficiency (SFE) $\varepsilon_{\rm SF} = M_\star / (f_b M_{\rm h}) $ up to 1 dex higher than at $z \le 1$. 
The high galaxy bias at $z \ge 8$ suggests that these galaxies reside in massive halos with intrinsic high SFE, challenging stochastic SHMR scenarios. 
Our SHMR evolves significantly with redshift, starting very high at $z \ge 10.5$, decreasing until $z \sim 2 - 3$, then increasing again until the present. Current hydrodynamical simulations fail to reproduce both massive high-$z$ galaxies and this evolution, while semi-empirical models linking SFE to halo mass, accretion rates, and redshift align with our findings. We propose that early galaxies ($z > 8$) experience bursty star formation without significant feedback altering their growth, driving the rapid growth of massive galaxies observed by JWST. Over time, increasing feedback efficiency and exponential halo growth suppress star formation. At $z \sim 2 - 3$ and after, halo growth slows down while star formation continues, supported by gas reservoirs in halos.}


\maketitle



\section{Introduction} \label{sec:intro}

The importance that dark matter plays in the formation and evolution of galaxies is no longer in doubt. In the classic \cite{White&Rees1978} paradigm, galaxies form at the center of haloes of dark matter. Dark matter shapes the distribution of galaxies on large scales \citep{Davis1985} and this large-scale environment plays a key role in the galaxy assembly process. Thus, to improve our understanding of galaxy formation and evolution over cosmic time, it is crucial to explore the connection between galaxies and their dark matter halos (see \citealt{Wechsler&Tinker2018} for a review). The halo mass influences the efficiency of gas cooling, which is essential for the formation of galaxies \citep{White&Frenk1991}. Additionally, processes regulating star formation, such as mergers \citep{Lacey&Cole1993}, feedback mechanisms (e.g., supernovae feedback in low-mass halos, and feedback from active galactic nuclei in high-mass halos, see \citealt{Bower2006}), gas accretion rates \citep[e.g.,][]{Conroy&Wechsler2009}, or galaxy dynamics \citep[e.g.,][]{Rodriguez-Gomez2017} are also likely influenced by halo mass.

Estimating the mass of dark matter halos hosting galaxies is therefore crucial. While direct methods like weak gravitational lensing can measure halo masses, they are typically limited to low redshifts due to the need for well-resolved sources. Indirect methods, such as abundance matching and clustering, can be applied over a larger range in redshift. Clustering measures the spatial distribution of galaxies using the two-point auto-correlation function, which quantifies the excess of galaxy pairs relative to a random distribution. By comparing how galaxies are clustered to the clustering of halos of different masses, we can infer the typical halo mass hosting them. Galaxy clustering studies began in the 1970s when simple power-law models were fitted to correlation functions measured using catalogs derived by photographic plates \citep{Tosuji&Kihara1969, Groth&Peebles1977, Peebles1980}. These works laid the foundation for modern approaches, which now use the halo occupation distribution (HOD) framework \citep[][]{Peacock2000, CooraySheth02, Smith2003}. HOD models, derived for stellar mass-selected samples, provide a powerful tool for estimating halo masses and other properties by assuming a halo model that populates halos of a certain mass with central and satellite galaxies. \cite{Asgari2023} provide a recent review of the halo model formalism.

With these models, clustering analyses have provided significant insights into the galaxy-halo connection, mostly constraining its evolution at $z \le 3$. The James Webb Space Telescope (JWST) has now opened a new observational window reaching back to $z \ge 10$, revealing the first galaxies formed in the universe \citep[e.g.,][]{Finkelstein2022, Naidu2022b, CurtisLake2023, Carniani2024, Kokorev2024d}. Early observations indicated an overabundance of massive, bright galaxies at $z \ge 10$ compared to predictions (e.g., in COSMOS, \citealt{Casey2024, Shuntov2025, Franco2024}); massive quiescent galaxies at $z \ge 4$ \citep[e.g.,][]{Carnall2024, Weibel2024b}; or red, compact sources with active galactic nuclei (AGN) emission \citep[e.g.,][]{Labbe2023, Greene2024, Matthee24_LRD}. Such discoveries challenge predictions of stellar mass assembly in hierarchical models of structure formation and raise new questions about the relationship between galaxies and their halos during reionization, as theoretical models struggle to reproduce these observations. These challenges extend to hydrodynamical simulations, which are calibrated on low-redshift data and face inherent limitations due to resolution and volume: small-box, high-resolution simulations such as \textsc{Sphinx} \citep{Katz2023} and {\sc Obelisk} \citep{Trebitsch2021} provide insights into the formation of low-mass galaxies at high-$z$ but lack the large-scale coverage needed to study global trends, while large-volume simulations like {\sc TNG} \citep{Pillepich2018, Nelson2018, Marinacci2018, Naiman2018, Springel2018} and {\sc Horizon-AGN} \citep{Dubois2014} are less reliable in capturing the detailed baryonic processes at play during the reionization era. As a result, key aspects of the galaxy-halo connection, including the evolution of the stellar-to-halo mass relationship and star formation efficiency, are still not fully settled at $z > 3$. 

Recent JWST studies have focused on 1-point statistics, such as the ultraviolet (UV) luminosity function \citep[UVLF; e.g.,][]{Finkelstein2023, Harikane2023, Finkelstein2024, Chemerynska2024} and the stellar mass function \citep[SMF; e.g.,][]{NavarroCarrera2024, Weibel2024, Shuntov2025}, which are important quantities for understanding galaxy properties. However, these measurements can suffer from degeneracies, depending on the assumptions made about galaxy evolution such as star formation histories, dust attenuation, or its connection with large-scale structures (e.g., \citealt{Munoz23} and \citealt{Gelli24} discuss the importance of galaxy bias in this context). While 1-point statistics provide useful constraints, 2-point statistics, like galaxy clustering, offer more detailed information, allowing for the mapping of high-redshift galaxies to their halos and large-scale environments. High-$z$ clustering studies have used UV magnitudes (e.g., \citealt{Dalmasso2024a} for samples at $z \sim 8$; \citealt{Dalmasso24} for $z \sim 11$), but these samples are often incomplete, as well as Lyman-break galaxy samples at $z \sim 5-8$ \citep{Ouchi2004, Lee2006, Overzier2006, Hatfield18, BaroneNugent14}. 
While stellar masses become less reliable at high-$z$ due to the increasing uncertainties on the assumptions mentioned above, they offer complementary insights. In contrast to UV magnitudes, they provide a cumulative view of galaxy growth and are less sensitive to bursty star formation activity. Furthermore, the small fields used in these JWST studies are prone to biases due to cosmic variance, which can affect the results and limit our understanding of the broader galaxy population \citep{Ucci2021, Steinhardt2021, Jespersen2024}.

In this study, we use the COSMOS-Web survey \citep{Casey23_CWeb}, the largest contiguous extragalactic survey observed with JWST to date. We investigate the galaxy-halo connection from the local universe to $z \sim 12$ using photometric redshifts in COSMOS-Web. Spanning an area of $0.54 \, \rm{deg}^2$, this survey reduces cosmic variance, enabling us to probe a wide range of densities and large-scale environments. We perform a consistent angular galaxy clustering analysis from $z = 0.1$ to $z > 10$, using stellar mass-limited, complete samples. By applying HOD models, we extract characteristic halo masses and study the stellar-to-halo mass relationship (SHMR) across this broad redshift range, providing insights into the star formation efficiency in the early universe and its evolution through cosmic time.

This paper is organized as follows. We begin in Sect.~\ref{sec:obs} by describing the COSMOS-Web observations and catalogs, and discussing how we construct a mass-complete galaxy sample. Section \ref{sec:measurements} presents the theory used to model galaxy clustering measurements using the HOD formalism. In Sect.~\ref{sec:results}, we present the galaxy clustering measurements and their corresponding fits, comparing our findings with the literature and examining the evolution of best-fit parameters such as the characteristic halo masses and the SHMR. We place these results into a broader context and discuss their implications for the galaxy-halo connection from cosmic dawn to the local universe in Sect. \ref{sec:discussion}. Finally, Sect.~\ref{sec:conclusion} summarizes and concludes our analysis.

Throughout this work, we adopt AB magnitudes \citep{OkeGunn1983_ABmag} and Planck18 cosmology \citep{Planck18}. Stellar masses are computed assuming a \cite{Chabrier03} initial mass function (IMF).

\section{Observations} \label{sec:obs}

\subsection{The COSMOS-Web survey and catalogs} \label{subsec:CWebsurvey}

COSMOS-Web \citep[][GO$\#$1727, PI: Casey \& Kartaltepe]{Casey23_CWeb} is a JWST survey that covers a contiguous area of $0.54 \, \deg^2$ in the COSMOS field in four NIRCam filters (F115W, F150W, F277W, F444W). Imaging from one MIRI filter, F770W, is also provided over a non-contiguous area of $0.19 \, \deg^2$. These data reach a $5\sigma$ depth of $\sim 28.1$ AB magnitudes in the F444W band and $\sim 25.7$ in F770W, measured in "empty" apertures of 0.15 arcsec radius (in this context, apertures are chosen which contain no objects in order to provide a realistic estimation of the background noise). Complete details of the image processing will be presented in a forthcoming paper \citepinprep{Franco}. Here, we use the complete COSMOS-Web NIRCam survey area, observed at three dates (January 2023, April 2023, and January 2024), along with observations from April 2024 that completed visits missed in earlier epochs due to issues such as guide star failures.
This work uses a new COSMOS catalog created by combining the COSMOS-Web JWST data with the rich multi-wavelength coverage already present in COSMOS \citepinprep{Shuntov}. A brief overview of these 30 imaging bands can be found below:
\begin{itemize}
    \item Near-UV imaging with the $u$-band from the CFHT Large Area $u$-band Deep Survey \citep[CLAUDS,][]{Sawicki19};
    \item Ground-based optical imaging from the Hyper Suprime-Cam Subaru Strategic Program (HSC-SSP) third data release, with broad bands $g,r,i,z,y$ and three narrow bands \citep{Aihara18b, Aihara22};
    \item Images from the Subaru Suprime-Cam with 12 medium bands in the optical between $4266 \, \si{\angstrom}$ and $8243 \, \si{\angstrom}$ \citep{Taniguchi07, Taniguchi15};
    \item Near-infrared imaging, including ground-based data from final "legacy" data release DR6 of the UltraVista survey \citep{McCracken12, MilvangJensen2013}, comprising four broad bands $Y, J, H, K_s$ and one narrow band at $1.18 \, \si{\micro\meter}$;
    \item Space-based optical imaging with the HST ACS F814W band \citep{Koekemoer07}.
\end{itemize}
Object detection follows a "hot/cold" strategy using the python implementation of \texttt{SExtractor} \citep{Bertin_SE}, called \texttt{SEP} \citep{Barbary2016}, applied to a combined $\sqrt{\chi^2}$ image of all NIRCam bands \citep[e.g.,][]{Szalay1999}. First, the "cold" mode detects and deblends bright and extended sources with a shallow threshold. Next, a second "hot" mode run is performed on the same image, with the bright sources masked, to detect faint, isolated sources and push the detection to fainter magnitudes. We note that this detection image covers only the $0.5 \, \deg^2$ COSMOS-Web area, unlike the UltraVISTA-selected COSMOS2020 catalog which covers  $\sim 1.5 \, \deg^2$ of the COSMOS area \citep{weaver_cosmos2020_2022}. However, NIRCam data are much deeper than UltraVISTA, allowing us to reach considerably lower stellar mass and higher redshift limits than COSMOS2020. 

The 30-band COSMOS-Web photometry is extracted using a new model fitting version of \texttt{SExtractor}, \SEPP \citep{Kummel20_SEPP, Bertin20_SEPP}. The main advantage of this new package is that there is no longer any need to resample input images to equivalent pixel scales (which would certainly be problematic for combining $0.030 \, \arcsec$ NIRCam images with $0.15 \, \arcsec$ dataset used in COSMOS2020). Instead, profiles can be fitted across many images with different resolutions and sensitivities. For each object, a Sérsic model \citep{Sersic1963} convolved with the point spread function (PSF) is fitted on all NIRCam bands simultaneously, while fluxes and magnitudes in each band are extracted using these fitted morphological parameters. To minimize photometric errors caused by blending, overlapping sources are grouped together and fitted simultaneously. Full details about COSMOS-Web catalogs can be found in \citeinprep{Shuntov}.

Photometric redshifts and physical parameters are determined using \LePhare \citep{Arnouts02_LP, Ilbert06_LP}. This tool fits galaxy spectral energy distribution (SED) templates generated from stellar population models from the \cite{BruzualCharlot03} library and a diverse set of star formation histories, as detailed in \cite{Ilbert15}. It includes star and quasi-stellar object (QSO) templates. \LePhare additionally incorporates the effects of dust attenuation by adjusting the color excess $E(B-V)$ from 0 to 1.2 considering three attenuation curves \citep{Calzetti00, Arnouts13, Salim2018}; accounts for emission lines \citep[following a method similar to][]{Schaerer&deBarros09}; and adds the intergalactic medium absorption through the analytical correction of \cite{Madau95}. After marginalization over templates and dust laws, a redshift likelihood distribution is returned, and physical parameters are subsequently derived at the fixed redshift. Photo-$z$ and stellar mass estimates for each object are then provided as the medians of the resulting probability density functions, hereafter named PDF($z$), with $1\sigma$ confidence interval given by the $16^{\rm th}$ and $84^{\rm th}$ percentiles. The photo-$z$ accuracy is then assessed by comparing to a compilation of about $12,000$ spectroscopic redshifts up to $z = 8$ from various programs in COSMOS, that will be presented in \citeinprep{Khostovan}. It results in a photo-$z$ precision, quantified by $\sigma_{\textrm{NMAD}} = 1.48 \times \textrm{median} \left(\frac{| \Delta z - \textrm{median}(\Delta z) |}{1 + z_{\textrm{spec}}}\right)$ with $\Delta z = z_{\textrm{phot}} - z_{\textrm{spec}}$, that achieves values better than 0.015 for galaxies in the range $23 < m_{\textrm{F444W}} < 25$. For fainter galaxies at $25 < m_{\textrm{F444W}} < 28$, the precision remains below 0.030, with outlier fraction (outliers defined by $|\Delta z| > 0.15(1 + z_{\textrm{spec}})$) lower than 10\%.

\subsection{Sample selection} \label{subsec:sample}

\subsubsection{Completeness}  \label{subsubsec:completeness}

To reliably measure galaxy clustering and to be able to interpret these results, we must be certain that our catalogs contain representative samples of the parent galaxy populations. We begin by carefully cleaning the sample. We remove objects classified as stars in \texttt{LePHARE}, based on their better $\chi^2$ fit to star or brown dwarf SED templates, along with a compactness criterion. We exclude as well objects with a radius smaller than twice the FWHM in the F115W image, and those flagged as hot pixels or other image artifacts. Moreover, we mask the region around bright stars in the NIRCam images: objects contained inside it are excluded from the analysis ($\sim$ 2\% of the total population). The same masking technique is applied to the optical images, specifically based on HSC stars that are very bright. This last procedure results in the removal of $\sim$15\% of the objects from the initial catalog.

\begin{figure}[t]
    \centering
    \includegraphics[width=0.85\columnwidth]{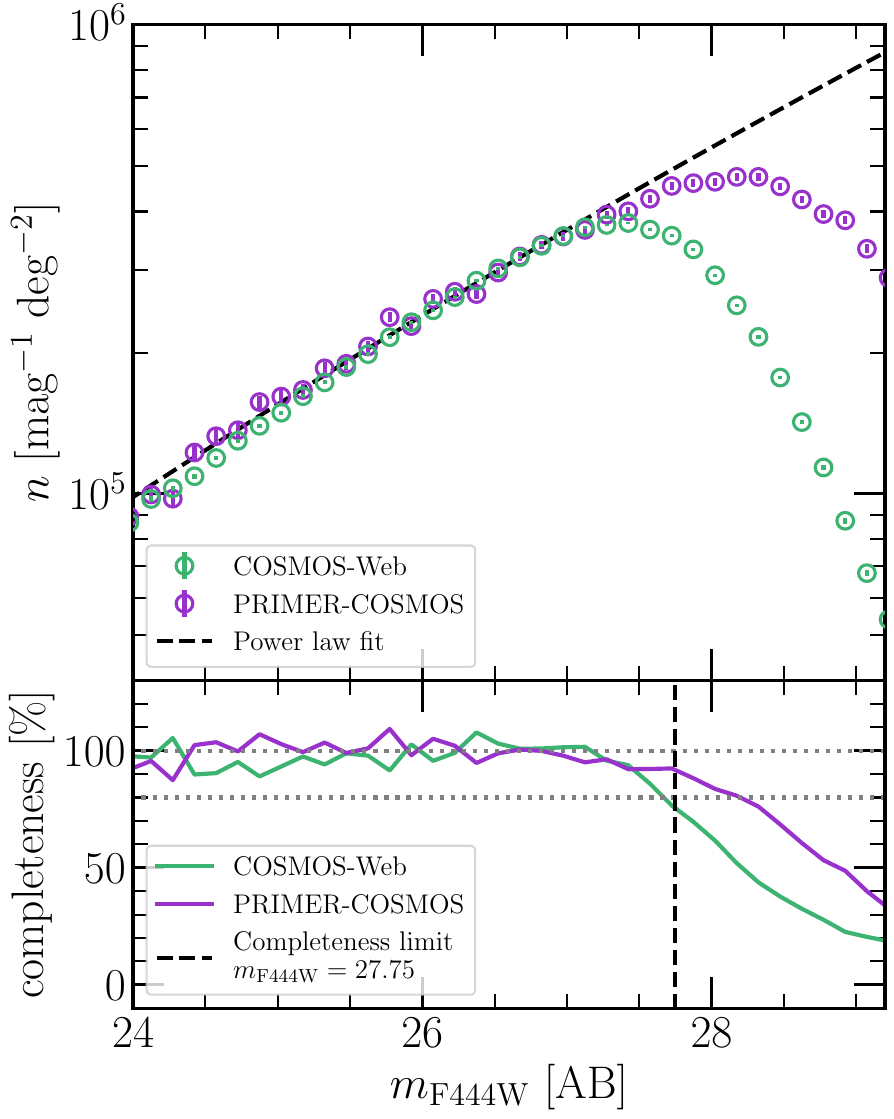}
    \caption{\textit{Top:} Galaxy number counts for COSMOS-Web and PRIMER-COSMOS, and the fitted power law in the range $23.0 \le \textrm{m}_{\textrm{F444W}} \le 27.0$ (dashed line). \textit{Bottom:} Magnitude completeness for COSMOS-Web and PRIMER-COSMOS, defined respectively as the ratio of number counts in COSMOS-Web over PRIMER-COSMOS and as the number counts in PRIMER-COSMOS over the fitted power law. The dotted black line shows the magnitude cut for COSMOS-Web where completeness falls to 80\%, used for this work.
    }
    \label{fig:magcut}
\end{figure}

We consider our sample incomplete above magnitude $m_{\textrm{F444W}}^\textrm{lim} = 27.75$. This value is derived from the number counts completeness of the survey compared to a deeper JWST survey in the same field, PRIMER-COSMOS, covering a common area of $\sim 200\,\textrm{arcmin}^2$ (following the same method as \citealt{Shuntov2025} where more details about PRIMER-COSMOS can be found). Figure \ref{fig:magcut} shows galaxy counts from both surveys with a fitted power law in the range $23.0 \le m_{\textrm{F444W}} \le 27.0$. The bottom panel shows the magnitude completeness for both surveys, defined respectively as the ratio of number counts in COSMOS-Web over PRIMER-COSMOS, and as the number counts in PRIMER-COSMOS over the power law fit. We choose the magnitude cut when the COSMOS-Web completeness drops below 80\% ($m_{\textrm{F444W}}^\textrm{lim} = 27.75$).

\begin{figure}[h]
    \centering
    \includegraphics[scale=0.53]{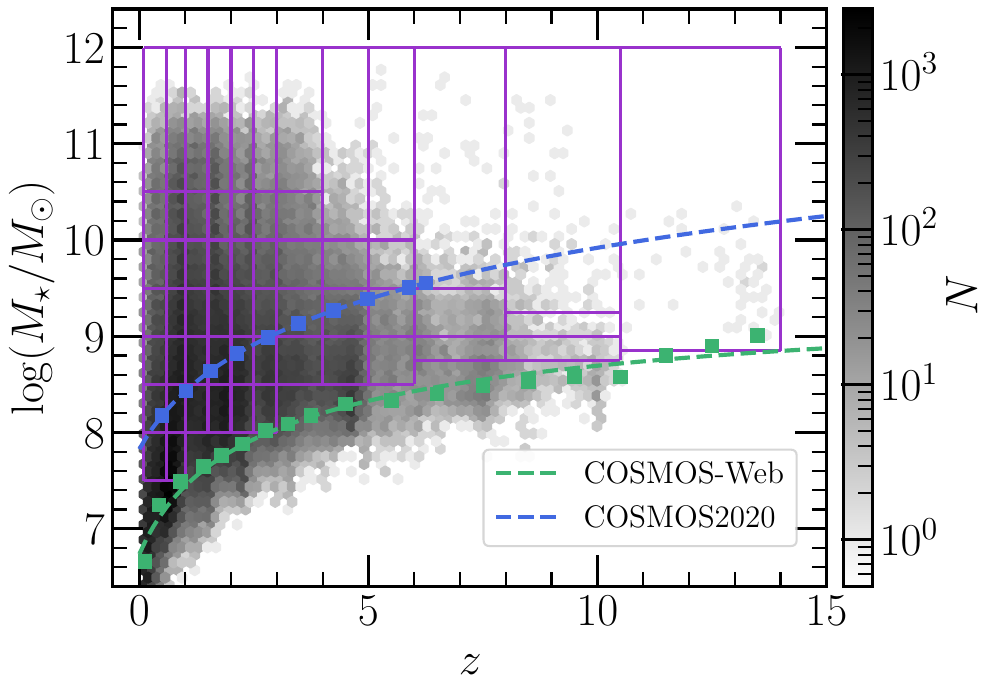}
    \caption{Stellar mass distribution as a function of redshift of COSMOS-Web sources, with the mass completeness limits for COSMOS-Web in green and COSMOS2020 in blue as given in \cite{weaver_cosmos2020_2022}. All galaxies after magnitude filtering only are shown. Redshift bins and stellar mass thresholds, used for clustering measurements, are also represented in solid purple lines.}
    \label{fig:SM_completeness}
\end{figure}

Finally, to select stellar mass thresholds for our clustering measurements, we need to compute the completeness of the galaxies in stellar mass. We use the empirical method from \cite{Pozzetti10}: for each galaxy in a given redshift bin, we determine the mass that can be observed at the magnitude limit in a chosen band. Here, we choose the reddest NIRCam filter F444W, which probes the post-Balmer break region where the emission from older stars dominates at $z > 3$, along with the limit magnitude computed above.
\begin{equation} \label{eq:SM_completeness}
    \log M_\star^{\textrm{lim}} = \log M_\star - 0.4 \times (\textrm{m}_\textrm{F444W}^\textrm{lim} - \textrm{m}_{\textrm{F444W}})
\end{equation}
The stellar mass completeness is then defined as the $M_\star^{\textrm{lim}}$ within which 90\% of galaxies lie. Figure \ref{fig:SM_completeness} shows the number density of galaxies as a function of estimated photometric redshift, with COSMOS-Web and COSMOS2020 (from \citealt{weaver_cosmos2020_2022}) stellar mass completeness curves. Thanks to the much deeper COSMOS-Web NIRCam observations, mass limits are approximately one order of magnitude lower than COSMOS2020, allowing us to probe lower-mass and higher-redshift galaxies.

\subsubsection{High-redshift selection}  \label{subsubsec:highzselection}

Since the earliest JWST observations, several high-redshift candidates identified using photometric redshifts have, in fact, turned out to be heavily dust-obscured lower-redshift galaxies \citep[e.g.][]{Naidu22a, Zavala23, Fujimoto2023}. To remove these interlopers from our clustering sample at high-$z$, we perform a more advanced cleaning for the redshift bins $z \ge 4$. 

Firstly, we identified that a significant fraction of sources in the high-z sample were actually hot pixels (image artifacts). These objects are flagged in the catalog using a method based on size and compactness, as described in \citeinprep{Shuntov}, and are excluded from the sample across all redshift bins.

Secondly, some sources are better fitted by a QSO template at lower redshifts, indicating contamination, likely due to AGN-dominated emission. Such sources can mimic high-redshift galaxies, falsifying photo-$z$ and stellar mass estimates because \texttt{LePHARE} cannot distinguish between stellar and AGN components in the SED. This is also particularly true for AGN-dominated objects with significant dust obscuration (e.g., Little Red Dots; \citealt{Matthee24_LRD}). To exclude AGN-dominated contaminants, we apply criteria based on their spectral and morphological properties, following the methodology of \citet{Shuntov2025} and \citet{Akins2024}. Specifically, we remove sources at $z \ge 4$ that are (1) compact, with an effective radius $R_{\rm eff} < 0.1"$ (FWHM of F277W's PSF) and a flux ratio in F444W $ 0.5 < F(0.2")/F(0.5") < 0.7$; and (2) either with a better AGN SED fit $\chi^2_{\rm AGN} < \chi_{\rm GAL}^2$; or red with $m_{\rm F277W} - m_{\rm F444W} > 1.5$ indicating AGN emission in the rest-frame optical. To summarize, the criteria is $\rm( (AGN \cup Red) \cap Compact)$.

At higher redshifts, we found that assumptions in the SED fitting code, such as the amount of dust attenuation or emission lines allowed, can significantly impact photo-$z$ estimates at high redshift. For instance, some high-$z$ candidates shift to low-$z$ if we increase the covered range in $E(B-V)$. Examining specific objects, we observed that using the redshift at maximum likelihood ($z_{\rm chi2}$) might be more reliable than the median of the probability density function ($z_{\rm PDF}$) employed in this work. Discrepancies between these two estimates arise because $z_{\rm chi2}$ corresponds to the redshift of the template that minimizes the $\chi^2$, which may be the only model at that redshift, while $z_{\rm PDF}$ represents the median of the summed probabilities from all templates at a given redshift, making it generally more robust. However, we consider that the limited variety of templates at high-$z$, due to the relatively recent observation of early galaxies and the current lack of well-established SED models for these populations, can excessively bias the PDF toward low-$z$ solutions, even when the best-fit template corresponds to a higher redshift. A comparison between $z_{\rm PDF}$ and $z_{\rm chi2}$ is shown in App.~\ref{app:zPDFvszchi2}.

\begin{figure}[t]
    \centering
    \includegraphics[scale=0.5]{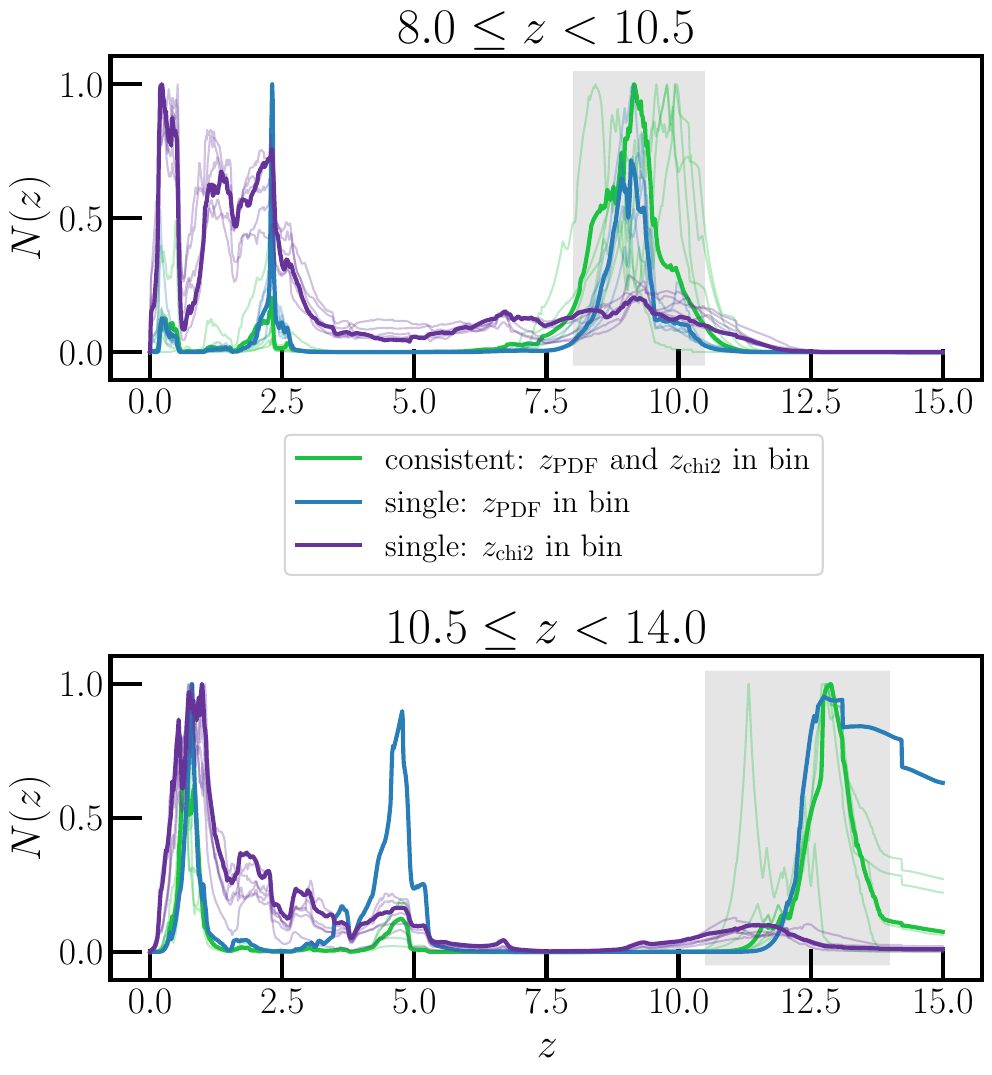}
    \caption{Stack of PDF(z) in our two highest redshift bins for galaxies selected either by $z_{\rm PDF}$, $z_{\rm chi2}$, or both. Thin lines are stacks of 15 galaxies and solid lines of 50 galaxies.}
    \label{fig:highz_stackPDF}
\end{figure}

For the $8 \le z < 10.5$ and $10.5 \le z < 14$ bins, we then include galaxies with either $z_{\rm PDF}$ or $z_{\rm chi2}$ within the bin, followed by cleaning to remove obvious contaminants. To illustrate the differences, we stack PDF($z$) distributions for galaxies where both photo-$z$ are in the bin versus only one: see Fig.~\ref{fig:highz_stackPDF}. When both are in the bin, the stacked PDF($z$) shows a strong peak in the high-$z$ bin with a smaller counterpart at lower $z$. For galaxies with only $z_{\rm PDF}$ in the bin, we see a strong high-$z$ peak and a narrower low-$z$ peak where $z_{\rm chi2}$ lies. Conversely, for those with only $z_{\rm chi2}$ in the bin, the probability density spans $z=0$ to 5, with a small bump in the high-$z$ bin. At $z \ge 10.5$, we see the contribution from the two main low-$z$ interloper populations: one at $z \sim 3-5$, corresponding to strong emission-line galaxies or dusty star-forming galaxies that boost NIRCam photometry and mimic a blue continuum slope \citep{Naidu22a, Zavala23}, and another at $z \sim 1$, caused by massive dusty galaxies with Balmer breaks falling into the near-infrared bands, mimicking the Lyman break combined with dust reddening.

\begin{figure}[t] 
    \centering
    \includegraphics[scale=0.5]{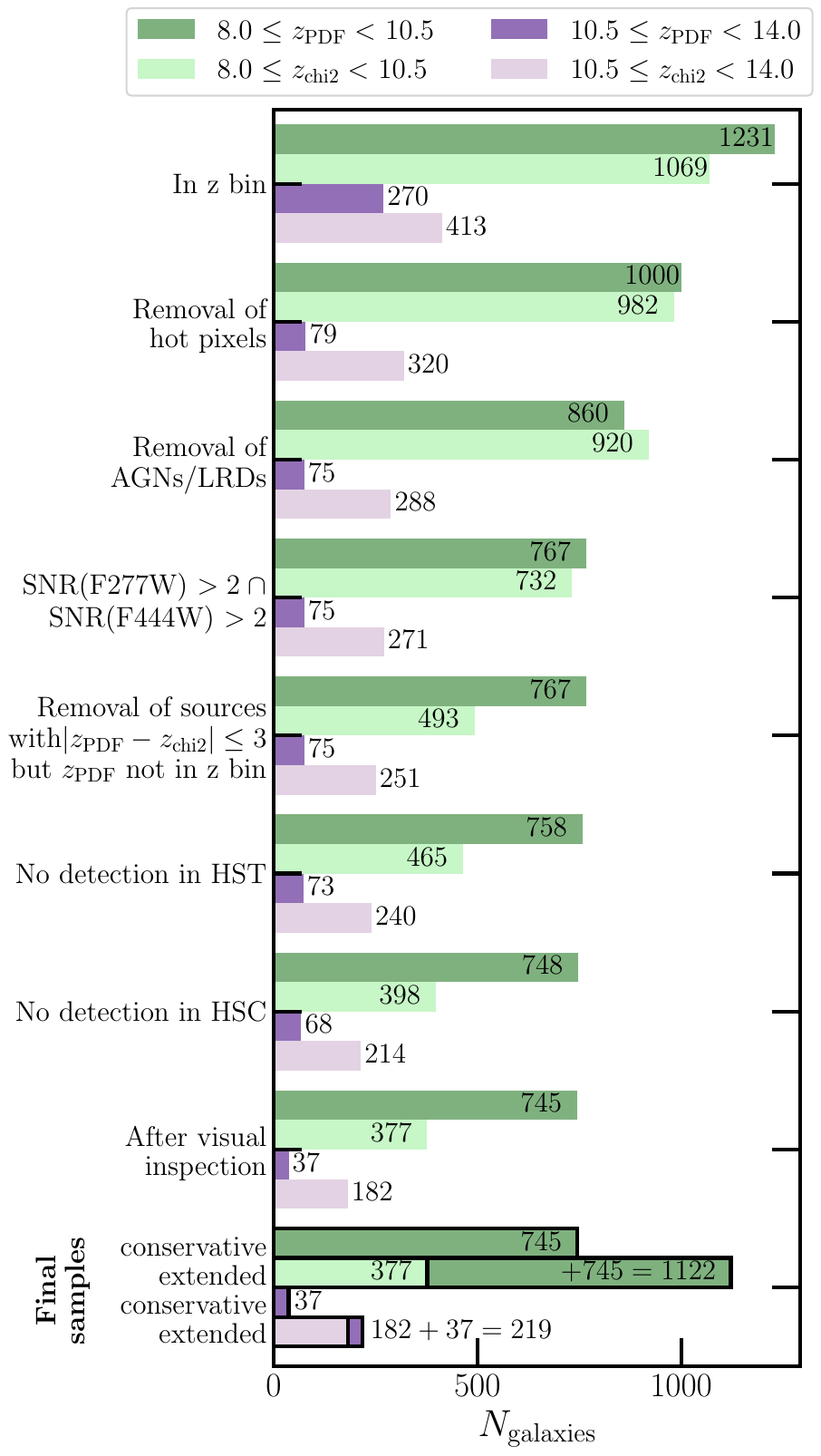}
    \caption{Number counts of galaxies in our two highest redshift bins after the different cleaning steps applied one after the other. The top bar (dark colors) represents galaxies selected by $z_{\rm PDF}$, and the bottom bar (light colors) $z_{\rm chi2}$-selected galaxies.}
    \label{fig:highz_cleaningsteps}
\end{figure}

To refine the sample and remove obvious low-$z$ contaminants, we apply the following steps:
\begin{enumerate}
    \item \textbf{S/N criteria.} Sources must have S/N(F277W) $\ge 2$ and S/N(F444W) $\ge 2$ as the Lyman break for $z < 14$ is expected in bluer wavelengths.
    \item \textbf{Neighboring redshift bins.} For galaxies selected based only on $z_{\rm chi2}$, we exclude those where $z_{\rm PDF}$ and $z_{\rm chi2}$ are close ($| z_{\rm PDF} - z_{\rm chi2} | \leq 3$) but $z_{\rm PDF}$ falls outside the redshift bin of interest, typically in an adjacent $z$ bin. We consider these objects to be more likely true intermediate-$z$ galaxies identified by $z_{\rm PDF}$ rather than low-$z$ interlopers incorrectly assigned to high redshift. They are then included in the appropriate previous redshift bins.
    \item \textbf{Detection in optical bands.} We run the package \texttt{photutils} on optical cutouts from HST, HSC, and a stack of HSC bands to perform source detection. We also add F115W for galaxies at $z \ge 10.5$ as their Lyman-break falls after this band. Galaxies with a $\ge 2 \sigma$ detection are removed.
    \item \textbf{Visual inspection.} We manually inspect all cutouts and SED fits left of $z \ge 8$ sources to remove image artifacts, incorrectly deblended sources, or false detections.
\end{enumerate}
Figure \ref{fig:highz_cleaningsteps} shows the evolution of galaxy number counts in these high-z bins after each cleaning step. These steps remove from 5 to 20\% of sources selected by $z_{\rm chi2}$, while those selected by $z_{\rm PDF}$ appear more robust. After applying these criteria, we obtain 715 and 26 galaxies with both $z_{\rm PDF}$ and $z_{\rm chi2}$, 30 and 11 galaxies with only $z_{\rm PDF}$, and 377 and 182 galaxies with only $z_{\rm chi2}$ in the $8 \leq z < 10.5$ and $10.5 \leq z < 14$ bins, respectively. We define two final samples for our work: the "conservative sample", including only $z_{\rm PDF}$-selected galaxies consistent with lower-z bins, and the "extended sample", which adds $z_{\rm chi2}$-selected galaxies. Since $z_{\rm PDF}$ is more conservative as it accounts for all templates over redshift, we adopt it as the correct redshift estimate for the other low redshift bins.

Appendix \ref{app:crosscorr} presents the cross-correlations between all redshift bins. The ratio of the auto-correlation to the redshift bin cross-correlations becomes particularly significant at $z \ge 3$. At lower redshifts, the auto-correlation is, on average (across all scales), 1 to 2 dex higher than the cross-correlation. However, at $3 \le z < 8$, this ratio drops below 1 dex for 3 to 4 cross-correlations with low-$z$ bins. The trend becomes even more pronounced when considering more massive galaxies. At the highest redshifts ($z \ge 8$), the ratio continues to decrease, with the auto-correlation being, on average, only about three times higher than the cross-correlation. While we do not expect zero cross-correlations due to the magnification bias, this effect cannot account for the signal we observe as it is expected to be smaller, with values below $w(\theta) < 10^{-2}$ \citep[e.g.,][]{Hildebrandt2009, Xu2023}. This trend persists regardless of the selection or cleaning criteria applied to this version of the catalog.

At $z > 10$, the high-redshift sample remains uncertain, as these sources are only constrained by two or three photometric bands. While the absence of detections in bluer bands supports their high-redshift nature, the lack of data leads to poorly constrained SED fits and, consequently, more uncertain stellar mass estimates. We therefore advise interpreting the $z \ge 10$ results with caution, as they remain preliminary measurements subject to sample contamination. A more detailed inspection of $z \ge 10$ sources in COSMOS-Web will be presented in \citeinprep{Franco}.

\subsubsection{Redshift - stellar mass binning and distributions}  \label{subsubsec:Nz}

We make clustering and number density measurements in mass-selected and volume-limited samples. Mass thresholds are determined starting from the stellar mass completeness limit, and then arbitrarily set to achieve a sufficient number of galaxies for clustering analysis. Additionally, redshift width bins are defined by hand to ensure a roughly similar number of galaxies in each bin up to $z \sim 4$, facilitating comparisons. For $z \ge 4$, we chose bins so that there are at least 100 galaxies in each bin at the end of the cleaning process, a limit that has been decided empirically to ensure enough statistics for measuring clustering. These bins are illustrated in Fig.~\ref{fig:SM_completeness}, and the corresponding number of galaxies in each redshift-mass bin after cleaning are detailed in App.~\ref{app:tableHODparams}. We exclude objects in bins $z < 8$ whose PDF($z$) has more than 70\% of its distribution outside $z \pm \Delta z_{\rm bin}$. This criterion removes objects with poorly constrained PDF($z$), often contaminants or unlikely to belong to the redshift bin. Approximately 20\% of objects are removed per bin, but we confirmed that changes in clustering measurements remain within the error bars after this exclusion. Ultimately, the final dataset used for this work comprised a total of $\sim 240,000$ galaxies above the completeness limits. 

Modeling the angular correlation function requires accurate redshift distributions $N(z)$. We use the method derived in \cite{Ilbert21_Nz}, which takes advantage of photo-$z$ likelihoods returned by SED fitting codes such as \LePhare. For each redshift and stellar mass selected sample, $N(z)$ can be estimated by stacking individual posterior probabilities $\mathcal{P}(z|\textbf{o})$ for a galaxy inside the bin to have a redshift $z$ considering colors (in the full covered wavelength range) and magnitude (in a reference band) vector $\textbf{o}=(\textbf{c}, m_0)$. This posterior, by marginalizing over $N_{\rm SED}$ SED templates, can be written as the product of the photo-$z$ likelihood $\mathcal{L}(\textbf{o}|z)$ and a prior probability of $z$ given a reference magnitude $\textrm{Pr}(z|m_{0,i})$ (see Eq.~\ref{eq:Nz}).
\begin{equation} \label{eq:Nz}
    N(z) = \sum_{i}^{N_{\rm SED}} \mathcal{P}_i(z|\textbf{o}) = \sum_{i}^{N_{\rm SED}} \mathcal{L}_i(\textbf{o}|z) \textrm{Pr}(z|m_{0,i})
\end{equation}
\cite{Ilbert21_Nz} shows that an appropriate prior would be as below, calculated within magnitude bins centered at $m_0$:
\begin{equation} \label{eq:Nz_prior}
    \textrm{Pr}(z|m_0) = \sum_{i}^{N_{\rm SED}} \mathcal{L}_i(\textbf{o}|z) \Theta(m_{0,i}|m_0)  \, ,
\end{equation}
where $\Theta(m_{0,i}|m_0) = 1$ if the object magnitude $m_{0,i}$ is within the magnitude bin, and 0 otherwise. The authors show that this prior can have a significant impact on the median redshift estimate, quantified as $\mu (\delta z) \sim 0.01 (1+z)$ for an Euclid configuration. This bias could be due to the SED fitting directly (not adapted templates, degeneracies, etc.), a non-adequate photo-$z$ prior, or more. They propose a correction, however, that is not applicable to COSMOS-Web currently because this technique assumes overestimated photo-$z$ uncertainties and a training set of spec-$z$ representing all galaxy populations in the sample. Moreover, \cite{Shuntov22} showed that for the COSMOS2020 field, this bias could lead to differences in the clustering of the order of 3\%, which is below the error bars we have for clustering measurements. 

\begin{figure}[t]
    \centering
    \includegraphics[scale=0.5]{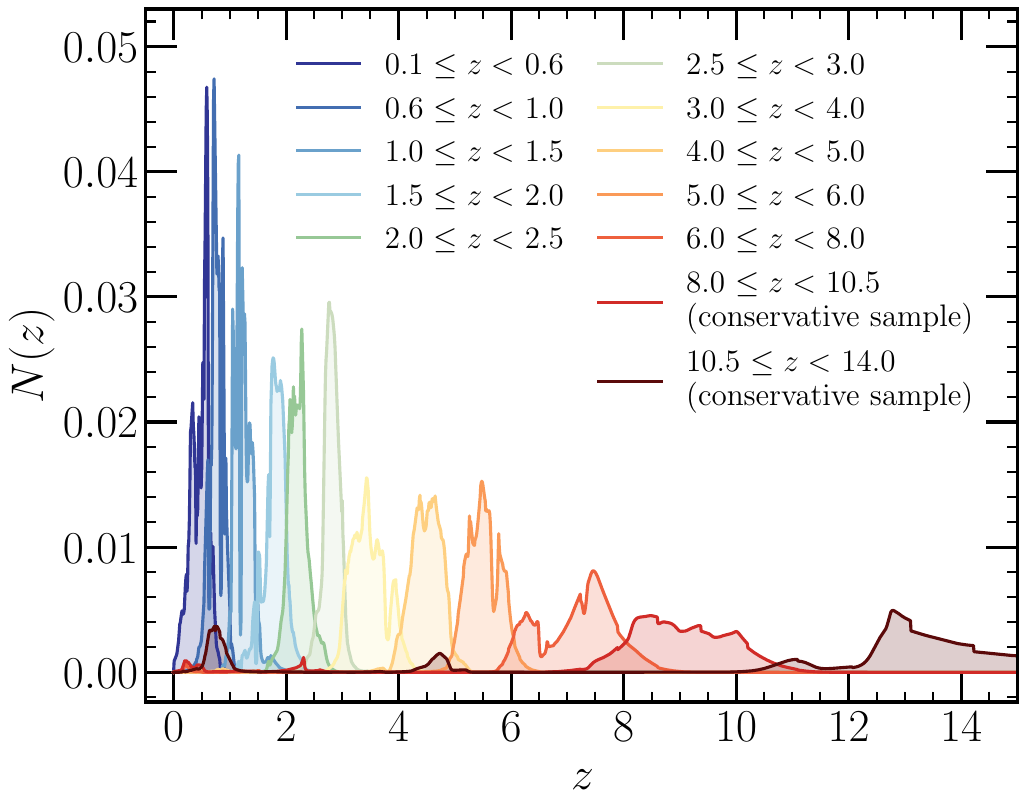}
    \caption{Redshift distributions for galaxies above the stellar mass completeness limit, in the redshift bins chosen for this work.}
    \label{fig:Nz_allgals}
\end{figure}

An example of resulting redshift distributions for all galaxies above the stellar mass completeness limit for each redshift bin in our study is shown in Fig.~\ref{fig:Nz_allgals}. The sharp peaks in some distributions likely result from large-scale structures in the survey area; for example some of these peaks at $z < 3$ are confirmed by the COSMOS spec-$z$ compilation \citepinprep{Khostovan}. Alternatively, they may arise from \LePhare fitting technique, where the code can converge on a narrow range of values when it gets stuck in local minima, particularly when handling low S/N sources. We can see in the $10.5 < z < 14$ redshift distribution the contributions from SED solutions corresponding to galaxies at $z \sim 1$ and $z \sim 3 - 5$, as discussed in Sect.~\ref{subsubsec:highzselection}.

\section{Measurements and modeling} \label{sec:measurements}

\subsection{Galaxy clustering measurements} \label{subsec:clust_mes}

In this work, clustering measurements are performed using the package \treecorr\footnote{\url{https://github.com/rmjarvis/TreeCorr}} \citep{Jarvis_treecorr}. To compute the angular correlation function $w(\theta)$ between galaxy positions separated by an angle $\theta$, we employ the \cite{LandySzalay93} estimator, which involves comparing the observed galaxy catalog with a randomly distributed one: 
\begin{equation} \label{eq:wtheta}
w(\theta) = \frac{DD(\theta) -2DR(\theta) +RR(\theta)}{RR(\theta)} \, .
\end{equation}
Here, for each angular bin $[\theta, \theta + \delta \theta]$, $DD$ is the number of galaxy pairs in the observed catalog, $RR$ the number of pairs in a random catalog and $DR$ between both catalogs. This random catalog is created using the code \texttt{venice}\footnote{\url{https://github.com/jcoupon/venice}}. It is based on the exact same area of the survey (after applying HSC and JWST masked regions), and composed of $\sim 50$ times the total number of objects in the survey. We measure this function across a range of angular scales determined by the resolution and covered area of the survey, specifically $10^{-5} \le \theta \le 0.1$ degrees, with a number of theta bins $N_\theta \in [10, 15]$.

Statistical errors associated with the correlation function are determined using the jackknife resampling method implemented in \texttt{TreeCorr}. The entire area is divided in $N_{\rm patches} = 20$ sub-samples of $\sim 90\, \text{arcmin}^2$ each (except for the two highest-$z$ bins where we reduced $N_{\rm patches}$ to 9 and 7 respectively), and the angular correlation function is computed by excluding one patch at a time. This yields a covariance matrix expressed as:
\begin{equation} \label{eq:cij}
C_{ij} = \frac{N_{\rm patches}-1}{N_{\rm patches}} \sum_{k=1}^{N_{\rm patches}} (w_k(\theta_i) - \overline{w}(\theta_i))^T (w_k(\theta_j) - \overline{w}(\theta_j)) \, ,
\end{equation}
where $\overline{w}$ is the mean correlation function and $w_k$ the estimate of $w(\theta)$ when the $k$-th patch is excluded. This allows us to address Poisson noise in the count of galaxy pairs and cosmic variance at large angular scales arising from the finite survey area. However, it is important to note that uncertainties due to cosmic variance may still be underestimated as the patches cannot be fully independent, as some correlated large-scale structures could contaminate the covariance of the data.

Although COSMOS-Web is one of the largest surveys observed by JWST, its spatial coverage is still limited, which can bias clustering measurements -- particularly at large scales, where the amplitude may be underestimated. This effect is called the "integral constraint" bias \citep{Groth&Peebles1977} and a constant correction factor can be expressed in terms of survey area $\Omega$ as
\begin{equation} \label{eq:wIC}
w_{\rm IC} = \frac{1}{\Omega^2} \int \int w(\theta) d\Omega_1  d\Omega_2 \, .
\end{equation}
\cite{RocheEales99} proposed to use the random catalog to compute this term as a function of the true correlation function $w_{\rm true}$:
\begin{equation} \label{eq:wIC_obs}
w_{\rm IC}(w_{\rm true}, RR) = \frac{ \sum w_{\rm true} RR(\theta) }{\sum RR(\theta)} \, ,
\end{equation}
\begin{equation} \label{wIC_wtrue}
\textrm{where} \; w_{\rm true}(\theta) = w_{\rm mes}(\theta) + w_{\rm IC} \, .
\end{equation}
In our case, we apply this factor only to the model angular correlation function before fitting and not directly to the measurements, according to $w_{\rm mod}(\theta) = w_{\rm mod, true}(\theta) - w_{\rm IC}(w_{\rm mod, true}, RR)$. Its impact becomes significant at scales above $0.02$\,deg, and reaches up to one order of magnitude at the largest scales.

Finally, we also compute the number density of galaxies for each sample, in each redshift-mass bin. This is defined as the number of galaxies $N_{\rm g}$ in the sample limited by the stellar mass threshold $M_{\star, \rm th}$, divided by the comoving volume probed by the redshift bin $[z_{\rm min}, z_{\rm max}]$:
\begin{equation} \label{eq:ng_obs}
n_{\rm g}^{\rm obs}(M_\star \ge M_{\star, \rm th}) = N_{\rm g} /  \left( \Omega \int_{z_{\rm min}}^{z_{\rm max}} \frac{{\rm d}V}{{\rm d}z} {\rm d}z \right) \, .
\end{equation}
Errors on galaxy number densities $\sigma_n$ are computed by combining Poisson count noise $\sigma_{\rm Pois}$ and an estimation of the cosmic variance $\sigma_{\rm cv}$ uncertainty: $\sigma_{\rm n}^2 = \sigma_{\rm Pois}^2 + \sigma_{\rm cv}^2$. Since galaxies cluster, field-to-field variance (related to cosmic variance), is greatly in excess of Poisson noise. It is higher for smaller volumes due to the lack of representative sampling of different density fluctuations \citep{Vujeva2024}, which makes COSMOS-Web less impacted by cosmic variance than other deep JWST surveys. Cosmic variance also scales as a function of mass and redshift, since both dark matter fluctuations and galaxy bias change rapidly as a function of these two quantities \citep{Moster_CV, Steinhardt2021}. Although \cite{Moster_CV} provided a popular cosmic variance calculator, it is only calibrated to low-$z$ clustering, and does not generalize well to higher redshifts, where it provides excessively high estimates of cosmic variance/clustering \citep{Weibel2024}. Here we instead follow the recommendations of \cite{Jespersen2024} and calibrate the cosmic variance to the \textsc{UniverseMachine} simulation suite \citep{Behroozi19}, which has been calibrated to clustering measurements at much higher redshifts than those considered by \cite{Moster_CV}. Our approach also directly incorporates any possible scatter in the stellar-to-halo mass relation, which is of the order of 0.3 dex, as well as possible effects from assembly bias \citep{Jespersen2022, Chuang2024}. To fit a smooth estimate of the cosmic variance, we sample number counts on the same field sizes, redshift bins, and mass limits as used in this work, and then fit a power law in mass with a redshift-dependent normalization and slope, as described by \cite{Jespersen2024}.

\subsection{Modeling the correlation function} \label{subsec:clust_model}

\subsubsection{The HOD model} \label{subsubsec:HOD}

The halo occupation distribution (HOD) analytical model is a fundamental tool used to populate dark matter halos of a certain mass with galaxies, distinguishing between central and satellite galaxies. This approach is extensively utilized in the literature, assuming that galaxy properties are primarily determined by the mass of their host halo. In this work, we adopt the HOD model described by \cite{Zheng05}, defined by five parameters ($M_{\rm h, min}, \, M_{\rm h, 1}, \, \sigma_{\log M}, \, \alpha, \, M_{\rm h,0}$) that characterize the occupation of central galaxies (defined as the most massive galaxy in a halo, typically residing in its center) and satellite galaxies (less massive objects, orbiting within the same halo as the central one) within halos of a given mass. While HOD models simplify the galaxy-halo connection by focusing solely on mass, it is acknowledged that other factors, such as halo formation history, can also influence the galaxy population within halos. More complex HOD models, such as the one by \cite{Leauthaud11}, exist; however, since we aim to apply these models to high-z clustering with a low number of data points, we chose one with fewer free parameters.

In the \cite{Zheng05} model, the occupation of central galaxies at a certain halo mass $M_{\rm h}$ is modeled by a smoothed step function:
\begin{equation} \label{eq:HOD_Nc}
    N_{\rm c}(M_{\rm h}) = \frac{1}{2} \left[ 1 + {\rm erf} \left( \frac{\log M_{\rm h} - \log M_{\rm h, min}}{\sigma_{\log M}} \right) \right] \, ,
\end{equation}
where $\sigma_{\log M}$ represents the scatter in the SHMR and $M_{\rm h, min}$ is the characteristic halo mass for which 50\% of halos host at least one central galaxy. The satellite occupation follows a power-law of slope $\alpha > 0$, with the characteristic halo mass $M_{\rm h,1}$ to host a satellite, for halos of mass greater than the $M_{\rm h,0}$ satellite cutoff mass:
\begin{equation} \label{eq:HOD_Ns}
    N_{\rm s}(M_{\rm h} \ge M_{\rm h,0}) = \left( \frac{M_{\rm h} - M_{\rm h,0}}{M_{\rm h,1}} \right)^\alpha \quad ; \quad N_{\rm s}(M_{\rm h} < M_{\rm h,0}) = 0 \, .
\end{equation}
The total number of galaxies in a halo of mass $M_{\rm h}$ is then given by :
\begin{equation} \label{eq:HOD_Ntot}
    N_{\rm tot}(M_{\rm h}) = N_{\rm c}(M_{\rm h}) \times (1 + N_{\rm s}(M_{\rm h}))\, .
\end{equation}
Some galaxy properties can be directly derived from the HOD model, such as the mean number density of galaxies $n_{\rm g}$, the fraction of satellites $f_{\rm sat}$, or the mean galaxy bias $b_{\rm g}$:
\begin{equation} \label{eq:HOD_ng}
    n_{\rm g}(z) = \int_{0}^{\infty} {\rm d}M_{\rm h} \frac{{\rm d}n}{{\rm d}M_{\rm h}}(M_{\rm h}, z) \: N_{\rm tot}(M_{\rm h})\, ,
\end{equation}
\begin{align} \label{eq:HOD_satfrac}
    f_{\rm sat}(z) & = 1 - f_{\rm cen}(z) \, , \\
    & = \frac{1}{n_{\rm g}(z)} \: \int_{0}^{\infty} {\rm d}M_{\rm h} \frac{{\rm d}n}{{\rm d}M_{\rm h}}(M_{\rm h}, z) \: N_{\rm s}(M_{\rm h})\, ,
\end{align}
\begin{equation} \label{eq:HOD_galbias}
    b_{\rm g}(z) = \int_{0}^{\infty} {\rm d}M_{\rm h} \frac{{\rm d}n}{{\rm d}M_{\rm h}}(M_{\rm h}, z) \: \frac{N_{\rm tot}(M_{\rm h})}{n_{\rm g}(z)} \: b_{\rm h}(M_{\rm h}, z)\, .
\end{equation}

We use the package \texttt{halomod} \citep{Murray20_halomod}\footnote{\url{https://github.com/halomod/halomod}} for computing the HOD model. Key elements of this model include the halo mass function $\frac{{\rm d}n}{{\rm d}M_{\rm h}}(M_{\rm h}, z)$ from \cite{Behroozi13_HMF}, the large-scale halo bias $b_{\rm h}(M_{\rm h}, z)$ from \cite{Tinker10}, a halo concentration-mass relationship from \cite{Duffy08}, a halo exclusion model based on double ellipsoids \citep{Tinker2005}, and a \cite{NFW1997} halo profile. The galaxy power spectrum, incorporating both 1-halo and 2-halo terms, is derived from these occupation distributions (see \citealt{Asgari2023} for a detailed derivation). The angular correlation function is finally computed using the \cite{Limber1954} approximation, based on the power spectrum and the observed redshift distribution of the galaxy sample.

\subsubsection{Non-linearities in the halo bias} \label{subsubsec:halobias}

One key component of the halo model is the halo bias, which quantifies how halos are biased tracers of the underlying dark matter distribution. The most widely used is the linear bias from \cite{Tinker10}, calibrated on large scales from N-body simulations. While effective on large scales, this linear bias encounters significant discrepancies in the transition region between the 1-halo and 2-halo terms. Studies such as \cite{Mead15} and \cite{Jose13} show that it under-predicts galaxy clustering in these "quasi-linear" regions, at scales of $r \sim 10-100\,\mathrm{kpc}$, with discrepancies reaching up to 30\%. This deviation is attributed to the breakdown of linear perturbation theory at these scales, since halos form through the non-linear collapse of over-densities in the dark matter fluctuation field, requiring a non-linear approach. This is coupled to scale-dependent variations, that primarily emerge from the shape of the halo profile. On large scales, these variations are negligible, allowing the halo center power spectrum to be simply expressed in terms of the linear matter power spectrum. However, at scales comparable to the size of individual halos, the halo profile becomes non-uniform, resulting in deviations from linear theory. This effect is expected to increase with redshift and halo mass, as the formation of rarer and then more highly biased halos in over-densities amplifies these non-linearities.

However, incorporating non-linearities into the halo model formalism for galaxy studies is challenging, and only a few models of non-linear halo bias exist to date (e.g., \citealt{Reed09, Jose16, MeadVerde21}). Non-linear bias is typically constructed by measuring the ratio of the halo power spectrum to the matter power spectrum, $b_{\rm h} = (\xi_{\rm hh}^{\rm sim}/\xi_{\rm mm})^{1/2}$, in N-body simulations across different redshifts and halo mass ranges, and comparing this to the large-scale ratio. A model for the halo bias $b_{\rm h}$ is provided by \cite{Jose16} (hereafter, J16), expressed as:
\begin{equation}
    b_{\rm h}(r, M_{\rm h}, z) = b_{\rm LS}(M_{\rm h}, z) \times \zeta(r, M_{\rm h}, z) \, ,
\end{equation}
where $b_{\rm LS}$ is the large-scale halo bias from \cite{Tinker10}, and $\zeta(r, M_{\rm h}, z)$ is a non-linear, scale-dependent correction. Using the MS-W7 simulations \citep{Guo2013, Pike2014} for $3 \le z < 5$, \citetalias{Jose16} derived a fitting function for $\zeta$ and showed that $b_{\rm h}$ is strongly scale-dependent and non-linear at scales of $r \sim 0.5$–$10,h^{-1}\,\rm Mpc$, with stronger effects at higher redshifts and with increasing halo mass as halos become rarer. Further details of this function are provided in App.~\ref{app:clust_SDNL}. The halo correlation function $\xi_{\rm hh}$ is then expressed as:
\begin{align}
    1 + \xi_{\rm hh}(r, M_{\rm h}', M_{\rm h}'', z) = \; & [1 + b_{\rm h}(r, M_{\rm h}', z) \, b_{\rm h}(r, M_{\rm h}'', z) \, \\
    & \xi_{\rm mm}(r, z)] \times \Theta[r - r_{\rm min}(M_{\rm h}', M_{\rm h}'')] \notag \, ,
\end{align}
where $\Theta(r)$ is the Heaviside function accounting for halo exclusion, suppressing the correlation at $r < r_{\rm min}$, with $r_{\rm min} = \min[r_{200}(M_{\rm h}'), r_{200}(M_{\rm h}'')]$ and $r_{200}(M_{\rm h})$ as the halo virial radius. This model of halo exclusion (from \citealt{vandenBosch2013}), used by \citetalias{Jose16} in its calibrations, is adapted to the halo finder used in our mass function calculations and prevents the exponential increase of the non-linear halo bias at small scales. Incorporating this correction into the halo model boosts the 2-halo term of the galaxy correlation function at intermediate scales for $z \ge 2$–3, as demonstrated by \cite{Jose17, Harikane18, MeadVerde21}, while leaving the 1-halo term unchanged since halo bias does not affect the internal galaxy distribution within a halo. Observational clustering studies at $z \sim 3$ \citep{Jose13, BaroneNugent14, Durkalec15, Dalmasso24} have found that the 1-to-2 halo transition break becomes less pronounced, with clustering measurements following a power-law across all scales rather than a two-component model. We suggest that this effect arises because non-linear processes cause early, massive halos to be more clustered and biased relative to dark matter, thereby boosting the 2-halo term of galaxy clustering at intermediate scales, as modeled here by the \citetalias{Jose16} NL-SD model.

In this work, we perform HOD modeling with and without \citetalias{Jose16} non-linear halo bias for galaxies at $z \ge 2.5$ and $z \ge 0.1$ respectively, after implementing the correction in \texttt{halomod}. However, our primary goal is not to achieve precise modeling of the correlation function with this correction, but to emphasize the importance of accounting for non-linear effects on the halo bias at these scales and high redshifts. There are several potential limitations to our implementation of their model. First, \citetalias{Jose16} fitting functions were calibrated using the MS-W7 simulation, which assumes a different cosmology and halo mass function than the one adopted here. This discrepancy affects the rarity of halos of a given mass, a key parameter in the computation of the correction term. Additionally, while the \citetalias{Jose16} model was originally calibrated for $3 \le z \le 5$, we extrapolate it to $z > 5$, though this approach inherently reduces its accuracy.

\subsubsection{MCMC fitting} \label{subsubsec:fitting}

We ran a Markov Chain Monte Carlo (MCMC) algorithm to fit HOD model parameters to our angular clustering measurements, using the \texttt{emcee} package \citep{emcee}. This method minimizes the $\chi^2$ value of clustering and number density measurements, within each redshift and stellar mass-limited bins:
\begin{align}
    \chi^2 = & \sum_{i,j}^{N_\theta} [w_{\rm obs}(\theta_i) - w_{\rm mod}(\theta_i)] \: C^{-1}_{ij} \: [w_{\rm obs}(\theta_j) - w_{\rm mod}(\theta_j)] \\
    & + \left( \frac{n_{\rm g}^{\rm obs}(> M_{\star, \rm th}) - n_{\rm g}^{\rm mod}(> M_{\star, \rm th})}{\sigma_{n}} \right)^2 \notag \, ,
\end{align}
where the model angular correlation functions have been corrected for the integral constraint as mentioned in Section \ref{subsec:clust_mes}. We excluded the smallest ($\theta < 2 \times 10^{-4} \,{\rm deg}$, up to $z = 4$) and largest ($\theta > 0.1 \,{\rm deg}$, all $z$) angular scales from the fits, as they are likely affected by the survey limits.

We chose to fit only the first three HOD parameters $M_{\rm h,min}$, $M_{\rm h,1}$, and $\alpha$ in our model. The parameter $\sigma_{\log M}$ has a negligible impact on the correlation function relative to our error bars and is thus more difficult to constrain with our data. It is also slightly degenerate with $M_{\rm h,min}$, as a higher scatter implies a lower $M_{\rm h,min}$ and vice versa. Therefore, we prioritize accurately constraining $M_{\rm h,min}$. Values for $\sigma_{\log M}$ range from 0.15 to 0.5\,dex in both observations and simulations, with a different evolution with redshift and halo mass (see \citealt{Porras-Valverde2024} for a compilation), but discrepancies can arise from the fact that it can contain both stellar mass measurement errors and intrinsic scatter in SHMR. We adopted a value of $\sigma_{\log M} = 0.20$\,dex, which balances various results \citep[e.g.,][]{Zheng05, Conroy2006, Harikane16, Cowley2019, Shuntov22}. Following other studies \citep{Hatfield18}, we fixed $M_{\rm h,0}$ using the following equation:
\begin{equation} \label{eq:M0}
    \log(M_{\rm h,0} / M_\odot) = 0.76 \log(M_{\rm h,1} / M_\odot) + 2.3 \; ,
\end{equation}
which has been derived in simulations up to $z \le 3$ by \cite{Conroy2006} and found in \cite{Contreras2023} trends. We still use it for $z > 3$, knowing that our tests with \texttt{halomod} showed that varying this parameter had a negligible effect on clustering results relative to our error bars in these regimes. The HOD parameters were re-fitted for each redshift bin as we opted not to model their intrinsic redshift evolution.

The MCMC used 30 walkers for a maximum of 6000 steps, stopping earlier if convergence criteria were met: $30\tau< N_{\rm iter}$ and $\Delta \tau < 15\%$ with $\tau$ the chain's auto-correlation time. The chain started with random initial positions based on flat priors, chosen from typical values fitted on other studies in the literature \citep[e.g.][]{Zheng07, Harikane16, Shuntov22}: $\log(M_{\rm h,min}/M_\odot) \in [7, 15]$, $\log(M_{\rm h,1}/M_\odot) \in [\log(M_{\rm h,min}/M_\odot), 16]$ and $\alpha \in [0.1, 2]$. 

We also investigated the impact of our covariance matrices on the fit. If the covariance matrix shows excessively high or low correlations in the data relative to the mean, it could misguide the fit. To reduce these effects, we optimized the number of patches used in the jackknife computation. However, for some bins with unreliable covariance matrices (for the bins at $z \ge 8$), we used only diagonal elements in the $\chi^2$ minimization.

Best fit parameters were derived using a Gaussian Kernel Density Estimation (KDE) across the 3 joint posterior distributions. The KDE was implemented using \texttt{scipy.stats.gaussian\_kde}, which automatically adjusts the bandwidth for each dimension based on the data's covariance matrix. Their lower and upper asymmetric uncertainties were computed in the 68\% confidence interval. Model-derived quantities and their uncertainties are computed from the 16$^{\rm th}$, 50$^{\rm th}$, and 84$^{\rm th}$ percentiles of the MCMC sample distribution. Parameters for all redshift and mass bins are listed in App.~\ref{app:tableHODparams}.

\section{Results} \label{sec:results}

\subsection{Angular clustering in COSMOS-Web} \label{subsec:result_clust}

\subsubsection{Clustering of mass-selected galaxies} \label{subsubsec:result_clust_mes}

\begin{figure*}[pt!]
    \centering
    \includegraphics[scale=0.5]{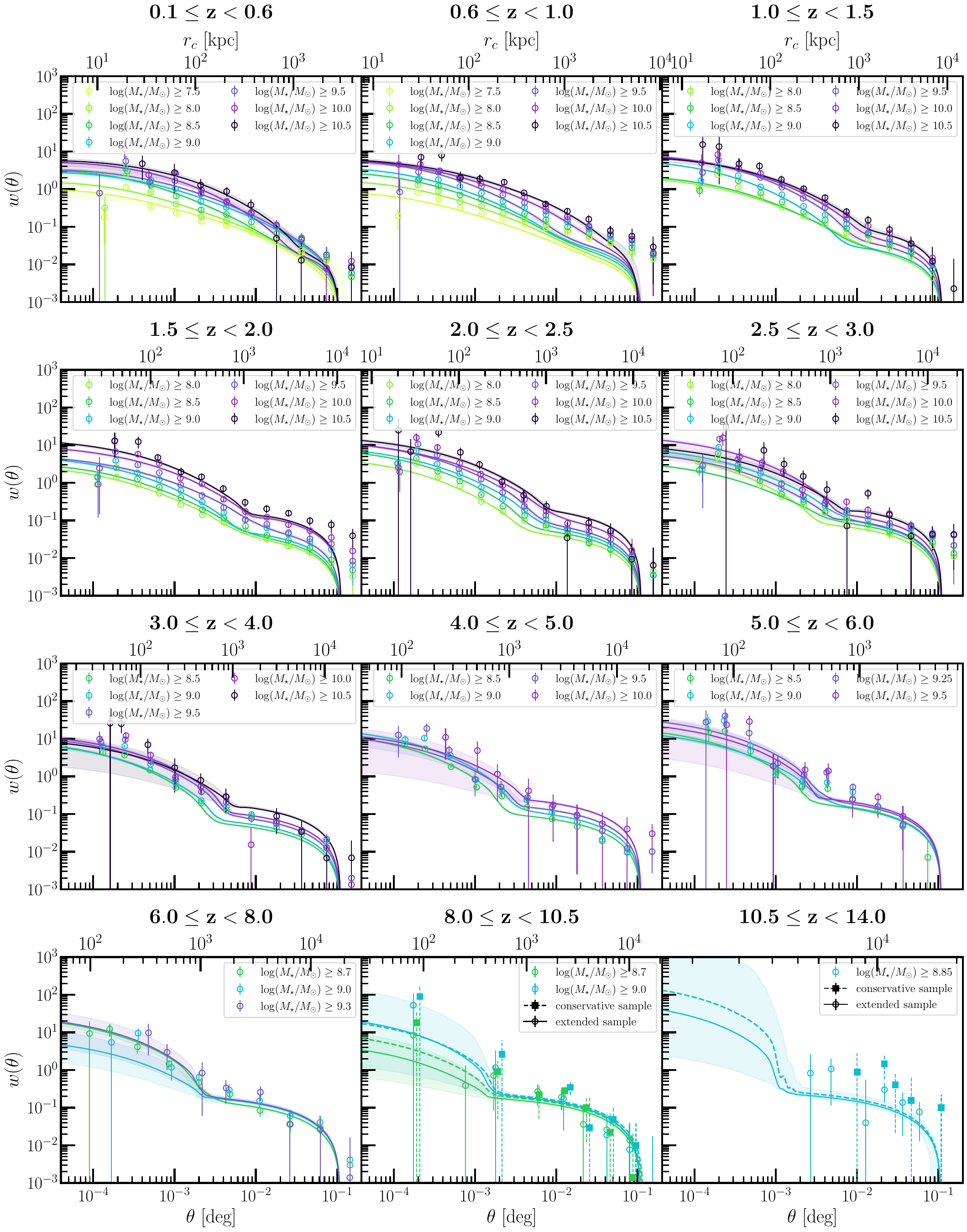}
    \caption{Angular auto-correlation function of galaxies in the COSMOS-Web survey, in redshift and mass limited bins. Solid lines show the HOD best-fit models, with their 1-$\sigma$ uncertainties in shaded regions. For the two highest redshift bins, both conservative and extended samples are represented, with HOD best-fit models in dashed and solid lines, respectively.}
    \label{fig:clust_allgals}
\end{figure*}

We present in Fig.~\ref{fig:clust_allgals} the auto-correlation function of COSMOS-Web galaxies from $z = 0.1$ to $14$, in redshift and mass-limited samples, as a function of the angular separation in degrees\footnote{We provide our clustering measurements at \url{https://github.com/LouisePaquereau/GalClustering_COSMOS-Web_Paquereau2025}}, along with their best-fit HOD models (discussed in Sect.~\ref{subsubsec:HODfits}). Across all redshifts, we observe the familiar trend of increasing clustering with higher galaxy stellar mass thresholds. This trend aligns with the expectation that massive galaxies typically trace denser and more clustered regions of the universe \citep[e.g.,][]{Peebles1980, Kaiser1984}.

The clustering deviates from a simple power-law at scales around $\theta \sim 10^{-2}$\,deg (equivalent to $\sim 0.3$\,Mpc in comoving coordinates at $z \sim 1$ for example). This break signifies the transition between the clustering of galaxies within the same halo (the "1-halo term") at small scales and the inter-halo clustering ("2-halo term") at larger scales. This behavior is more pronounced for massive galaxies, as their massive halos host more satellites, enhancing the 1-halo term. We note that clustering measurements drop at the lowest scales due to resolution and source deblending limitations in the survey, and at the highest scales (around $\theta \ge 0.02$ deg for the COSMOS-Web field) because of the limited size of the survey.

The unusually high clustering signal at large scales in the redshift bin $0.6 \le z < 1.0$ likely reflects the presence of the prominent filamentary structure known as the COSMOS Wall at $z = 0.73$ \citep{Iovino16_cluster07}. In the $5.0 \le z < 6.0$ bin, we observe clustering is elevated across all mass thresholds, potentially due to a field overdensity or SED fitting issues excluding sources near $z\sim 5$ (see the source overdensity just before $z = 5$ in Fig.~\ref{fig:SM_completeness}). Above $z = 6$, statistical uncertainties increase, complicating precise correlation function measurements. In the highest redshift bins, clustering follows a power law as the one-halo term becomes harder to probe. This agrees with other high-$z$ studies \citep[e.g.,][ for $3 \le z < 5$ passive galaxies]{Magliocchetti2023}; for instance, \citet{Jose17} propose that at $3 < z < 5$, non-linear scale-dependent halo bias amplifies power at intermediate scales ($\sim 0.5 - 1$\,Mpc), as discussed in Sect.~\ref{subsubsec:HODfits}. At $z \ge 8$, clustering amplitudes for fixed mass thresholds are higher than at lower redshifts ($+0.1$ to $0.5$\,dex at $r < 100$\,kpc), consistent with expectations. Observed high-z galaxies are rarer and represent the most massive and luminous systems, likely hosted by the most massive halos forming early in the universe \citep{Chiang17}.

\begin{figure*}[ht!]
    \centering
    \includegraphics[scale=0.475]{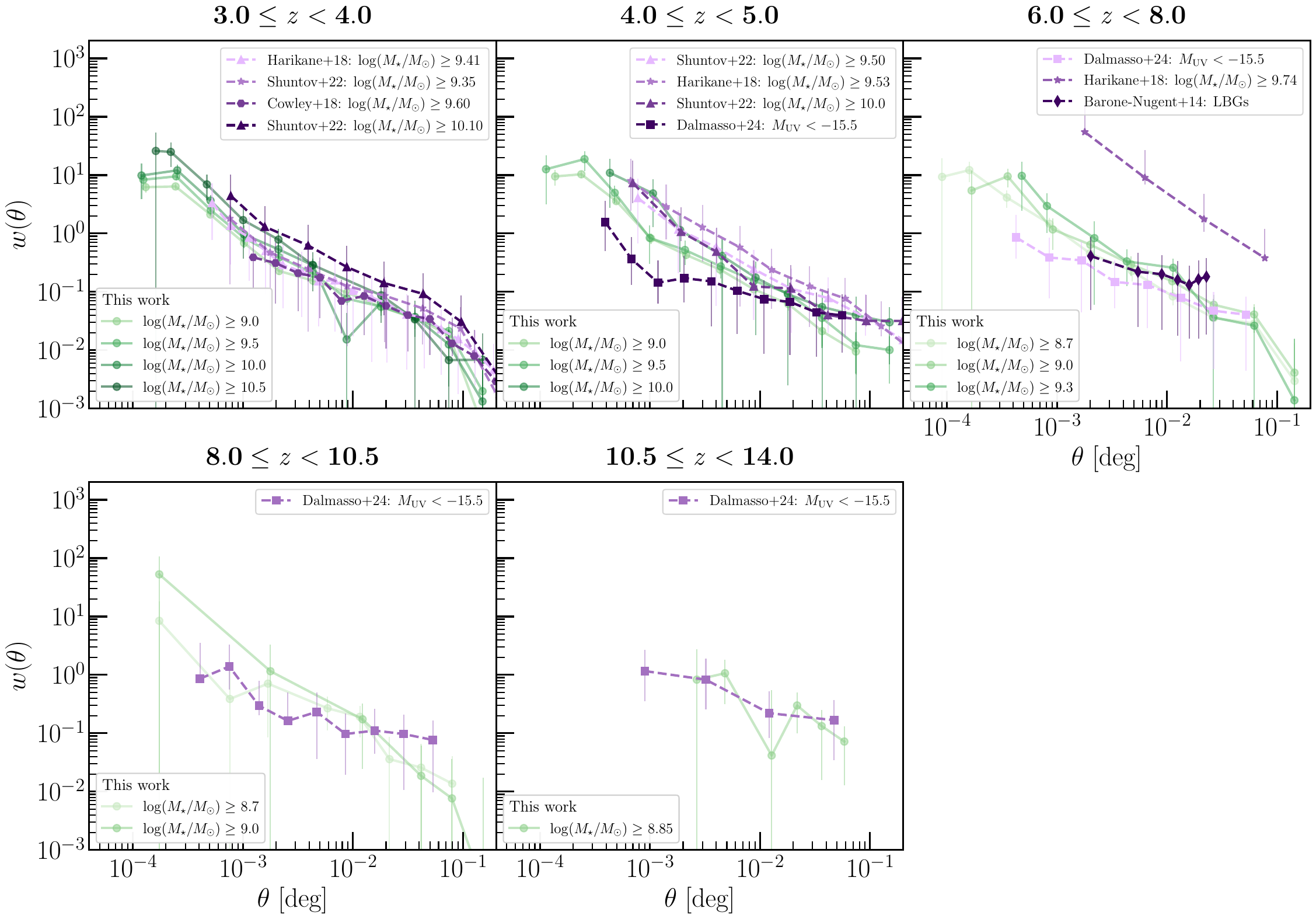}
    \caption{Comparison of COSMOS-Web's angular auto-correlation function with literature measurements \citep{BaroneNugent14, Harikane18, Cowley18, Shuntov22, Dalmasso24}. Some works used UV magnitude-limited samples or LBG samples, so direct comparison with mass-limited samples is not possible. However, we chose to represent them since they are the only existing clustering measurements at $z \ge 8$.}
    \label{fig:clust_literature}
\end{figure*}

\subsubsection{Comparison with the literature} \label{subsubsec:result_clust_lit}

While most of the surveys show very similar behaviors for clustering at low $z$, the high-$z$ regime has not been extensively explored yet, and some discrepancies are still visible between different data. In Fig.~\ref{fig:clust_literature}, we show a non-exhaustive comparison with literature measurements at high redshift, however with different sample selections: some are UV magnitude limited, others are Lyman-break selected galaxies (LBGs). 

For intermediate redshifts ($3 \leq z \leq 6$), our measurements align with those of \cite{Shuntov22} in the COSMOS field, \cite{Cowley18} in the SMUVS survey within COSMOS \citep{Ashby2018}, and \cite{Harikane18} for LBGs observed in the HSC Subaru Strategic Program over 100 deg$^2$ \citep{Aihara2018a}. In the $6 \le z < 8$ range, while the slope of our clustering measurements match those of \cite{Harikane18}, their amplitudes are more than 1\,dex higher. This discrepancy is most likely due to incompleteness or potential errors in stellar mass or redshift estimates in the HSC survey, which, despite pushing the limits of ground-based high-$z$ observations, probably detected only the brightest and most massive galaxies at $z \sim 6$, leading to this higher measured clustering amplitude. Finally, in the highest redshift bin, we compare our results with those of \cite{Dalmasso24} for UV-bright galaxies in JADES \citep{Eisenstein2023} at $z \sim 10.6$. Both the clustering amplitude and slope are similar. Given that JADES is deeper than COSMOS-Web and includes more JWST filters for sources at $z > 10$, this strengthens our confidence in our high-$z$ sample. However, because their sample is limited by UV magnitude, directly comparing their measurements with ours is not straightforward. Nonetheless, since their UV limit ($M_{\rm UV} < -15.5$) is broad, it is more likely to correspond to a low mass threshold (e.g., $10^{8} M_\odot$), with the resulting signal being primarily dominated by low-mass galaxies.

\subsubsection{HOD models} \label{subsubsec:HODfits}

The best-fit HOD models, along with their 1$\sigma$ uncertainties, are represented by solid curves in Fig.~\ref{fig:clust_allgals}, for the case where we did not correct for non-linear halo bias. Detailed best-fit parameters and uncertainties are provided in App.~\ref{app:tableHODparams}.

At low redshifts ($z < 3$), the HOD model closely matches the observational data across $4 \times 10^{-4} < \theta < 6 \times 10^{-2}$\,deg. However, the largest angular scales are not well-fitted due to the model's drop-off caused by the integral constraint. Similarly, at small scales ($\theta \sim 10^{-5}$\,deg) amplitudes are poorly constrained because of the finite survey resolution; however projection effects within a given redshift bin may create an apparent signal. Observed clustering at small scales is often higher than predicted, with a relative error $(w_{\rm obs} - w_{\rm mod})/w_{\rm mod} \ge 150$\% for scales $\theta < 10^{-3}$\, deg, suggesting that many galaxies are satellites and are more tightly clustered within halos. Consequently, the galaxy distribution profile likely differs from the one we assume in the model. For $z \ge 3$, discrepancies with the HOD model become more pronounced, especially at intermediate scales. Similar behavior has been reported in previous studies \citep[e.g.][]{Jose13, Harikane18} and may indicate the need to incorporate additional physical processes, such as non-linearities in the halo bias assumed in the HOD model \citep{Reed09, Jose16, Jose17}. In the bin $5 \le z < 6$, the shape of the modeled clustering does not align with the observations, which is likely due to issues with the measurements and galaxy sample at this specific redshift bin since it is resolved at $z > 6$. In some high-redshift bins, such as $z \ge 10.5$, the clustering amplitude is not as well constrained. This is most likely because there is also a constraint on number densities in the fit, which have smaller errors than clustering measurements since they are first-order statistics, less affected by field variations or sample size. As a result, number densities can have a stronger influence on the fit than clustering.

\begin{figure}[h]
    \centering
    \includegraphics[scale=0.47]{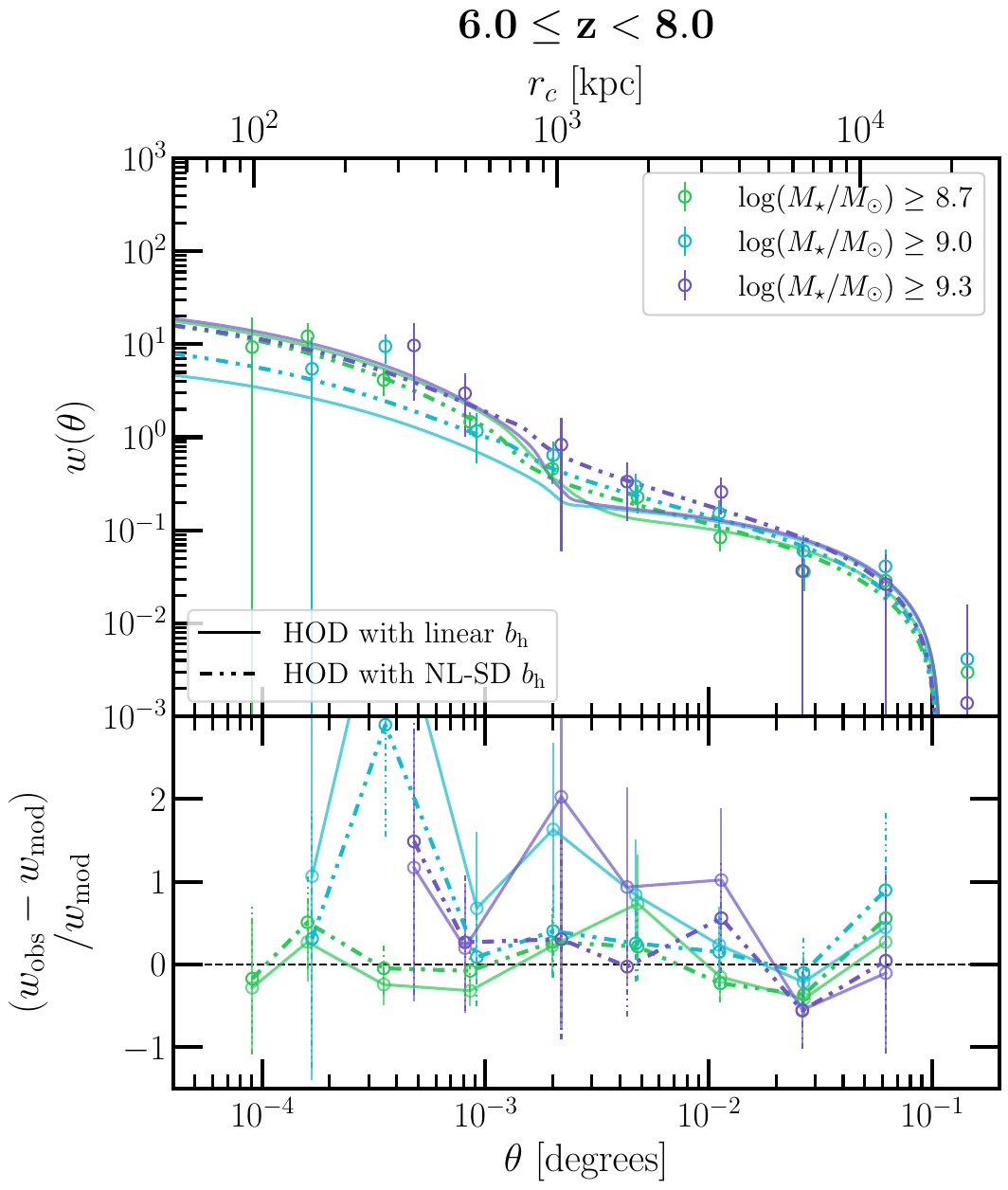}
    \caption{Angular auto-correlation function of galaxies in the bin $6 \le z < 8$. HOD best-fit models with and without a non-linear scale-dependent halo bias are shown in dash-dotted and solid lines, respectively. Relative errors are presented in the bottom panel.}
    \label{fig:clust6_SDNL}
\end{figure}

We explored HOD models incorporating the \citetalias{Jose16} non-linear scale-dependent halo bias (hereafter NL-SD halo bias) for galaxy clustering at $z \ge 2.5$, as described in Sect.\ref{subsubsec:halobias}. The best-fit models are presented in App.\ref{app:clust_SDNL}. For illustration, Fig.~\ref{fig:clust6_SDNL} shows results for the $6 \le z < 8$ bin, with and without the NL-SD halo bias. Adding this correction significantly reduces the relative error from 1 or 2 to nearly 0 across scales $10^{-3} < \theta < 2 \times 10^{-2}$\,deg. This improvement is consistent across all redshift bins at $z \ge 2.5$. We compared the $\chi^2$ values between the linear and NL-SD models, finding that $\chi^2_{\rm NL-SD} < \chi^2_{\rm linear}$ in 60\% of cases across the full $\theta$ range, with a mean $\Delta \chi^2 = \chi^2_{\rm linear} - \chi^2_{\rm NL-SD} = +4$. Some discrepancies arise for the most massive bins in certain redshift ranges (e.g., for samples with $\log(M_\star/M_\odot) \ge 10.5$ at $2.5 \le z < 3$ or $\log(M_\star/M_\odot) \ge 10$ at $4 \le z < 5$), where the non-linear correction overpowers the 1-halo term at intermediate scales, causing clustering to flatten at small scales. This behavior likely reflects limitations in the model calibration for very massive galaxies, where the non-linear correction is overestimated and the 1-halo term remains low, rather than a physical effect. A more robust implementation of the NL-SD bias within our model, considering, for example, the same assumptions in cosmology and the halo model when derivating the non-linear correction factor, will be required for accurate predictions of our high-redshift galaxy correlation functions. 

While both models yield similar halo mass estimates and HOD-derived quantities (e.g., Fig.~\ref{fig:M1Mmin_evol} shows a maximum difference of $\sim 0.25$ dex in $M_{\rm h,min}$), we proceed with the HOD best-fit model without NL-SD bias for simplicity and interpretability. Nevertheless, we argue that including a non-linear scale-dependent term in the halo bias and HOD formalism is likely essential to capture the power-law-like behavior of clustering observed at high redshift and to improve the modeling of halo and galaxy clustering.

\subsection{Characteristic halo masses} \label{subsec:halomass}

\begin{figure}[ht!]
    \centering
    \includegraphics[scale=0.5]{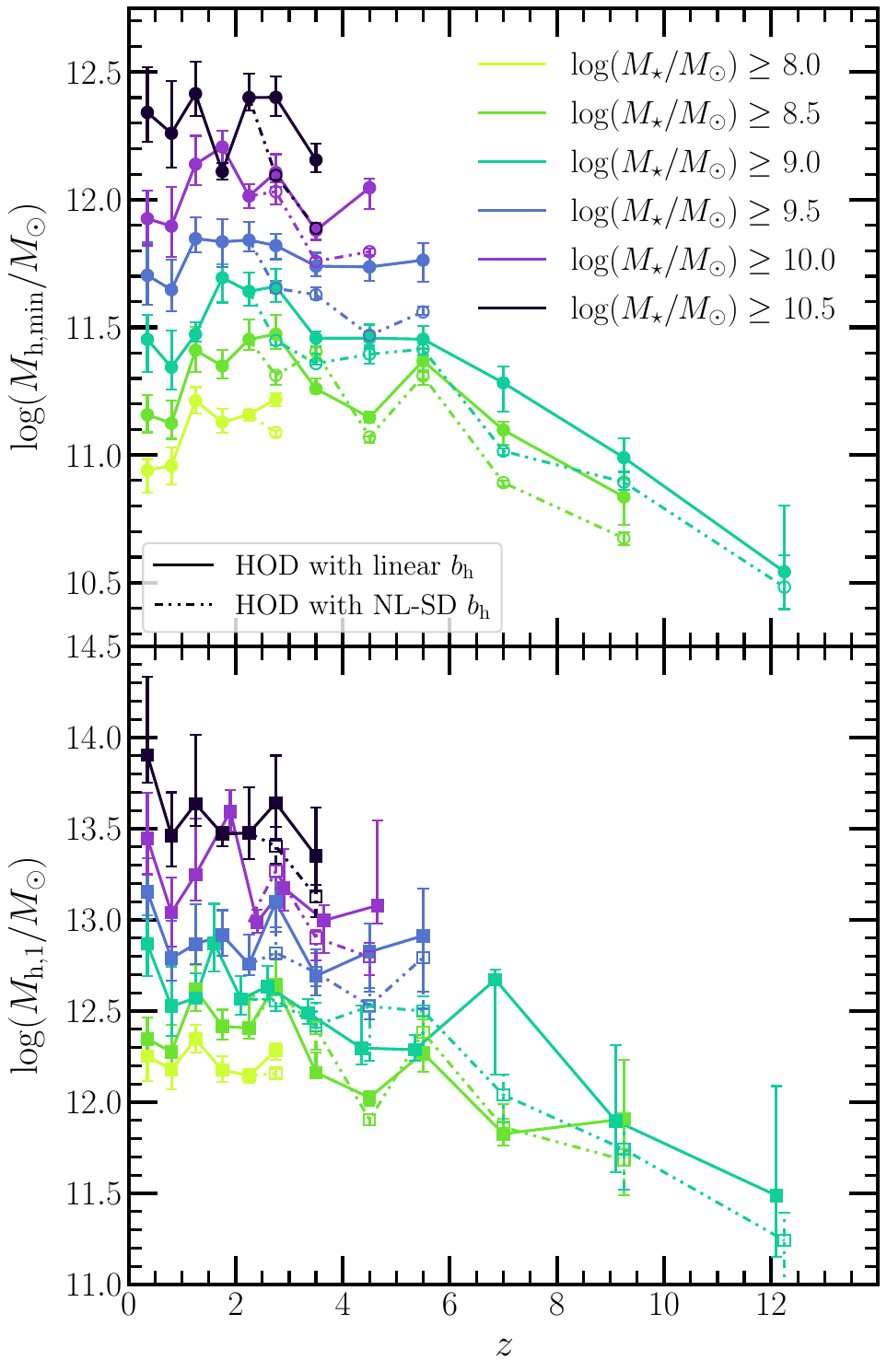}
    \caption{Redshift evolution of the characteristic halo masses fitted in the HOD model for stellar mass-limited galaxy samples. $M_{\rm h,min}$ is the characteristic halo mass to host a central galaxy, $M_{\rm h,1}$ to host a satellite. Errors bars are the 1$\sigma$ uncertainties from the MCMC sample distribution, and quantities are shown for HOD models with and without non-linear scale-dependent halo bias, respectively in dash-dotted and solid lines.
    }
    \label{fig:M1Mmin_evol}
\end{figure}

Characteristic halo masses of galaxy samples are estimated by fitting HOD-predicted clustering to the observations. Figure \ref{fig:M1Mmin_evol} illustrates the evolution with redshift of the two \cite{Zheng05} characteristic halo masses, $M_{\rm h,min}$ to host a central galaxy and $M_{\rm h,1}$ to host a satellite, with their error bars computed as the 16$^{\rm th}$ and 84$^{\rm th}$ percentiles of the MCMC sample distribution. Consistent with previous findings, there is a general trend indicating that more massive galaxies are hosted by more massive halos across all redshifts. Notably, $M_{\rm h,1}$ typically exceeds $M_{\rm h,min}$ by approximately 1 dex, a relationship also observed in prior studies \citep[e.g.][]{Hatfield19}. However, a new observation emerges: at a fixed stellar mass, $M_{\rm h,min}$ exhibits an increase with redshift, reaching a peak around redshift $z \sim 2-3$, before subsequently decreasing. For a more comprehensive analysis of this phenomenon, refer to Sect.~\ref{subsubsec:SHMR}. 

If we examine the characteristic halo masses at $z \ge 6$ (see App.~\ref{app:tableHODparams}), estimations are: $\log(M_{\rm h,min} / M_\odot) \simeq 10.99^{+ 0.08}_{- 0.13}$ for galaxies at $8.0 \le z < 10.5$ with $M_\star \ge 10^{9} M_\odot$; $\log(M_{\rm h,min} / M_\odot) \simeq 10.54^{+ 0.26}_{- 0.14}$ for galaxies at $10.5 \le z < 14$ with $M_\star \ge 10^{8.85} M_\odot$. We note that a similar halo mass estimate of $M_{\rm h} \simeq 10^{10.44} M_\odot$ for $10.5 \le z < 14$ galaxies is found when using only number densities through abundance matching.

We can estimate the integrated star formation efficiency (SFE) $\varepsilon_{\rm SF}$ needed to convert baryons to stellar mass, assuming a baryonic fraction from the cosmological model $\Lambda$CDM $f_{\rm b} \approx 0.18785$, as below:
\begin{align}
    \varepsilon_{\rm SF} & = \frac{M_\star}{f_{\rm b}  M_{\rm h}} \, ,\\
    & \simeq \frac{M_{\star, \rm th}}{f_{\rm b}  M_{\rm h,min}} \quad \textrm{or} \quad \simeq \frac{M_{\star, \rm med}}{f_{\rm b}  M_{\rm h,min}} \, ,
\end{align}
where $M_{\star, \rm th}$ is the stellar mass threshold and $M_{\star, \rm med}$ the median stellar mass of the galaxy sample. This describes how stellar mass grew integrated over the halo's lifetime. The obtained results agree with $\Lambda$CDM model predictions, however, a high $\varepsilon_{\rm SF} \sim 12 - 35\%$  (according to $\varepsilon_{\rm SF}$ definitions, see the next section for more details) is needed to explain the range of stellar mass that we observe at $z \ge 8$. 

\subsection{Stellar-to-halo mass relationship} \label{subsubsec:SHMR}

\begin{figure*}[ht!]
    \centering
    \includegraphics[scale=0.53]{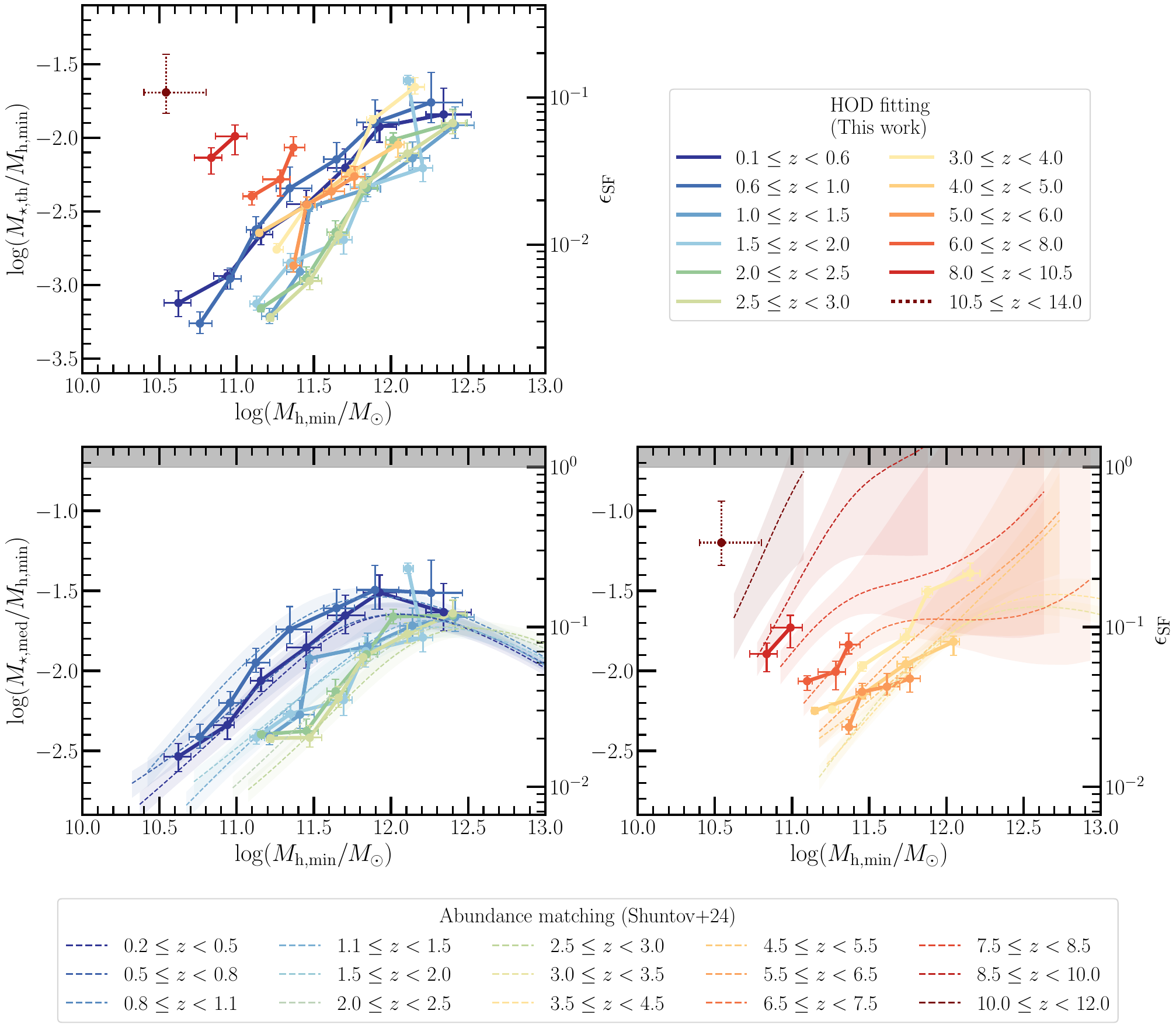}
    \caption{Stellar-to-halo mass relationship in COSMOS-Web determined by HOD fitting of our clustering measurements. The integrated star formation efficiency is also shown from $\varepsilon_{\rm SF} = M_\star / (M_{\rm h} \: f_{\rm b})$. The point at $10.5 \le z < 14$ is represented in dotted lines because it is considered less certain. \textit{Top panel:} SHMR defined as the ratio $M_{\star, \rm th}/M_{\rm h, min}$. \textit{Bottom panels:} SHMR defined $M_{\star, \rm med}/M_{\rm h, min}$ (solid lines). It is split into two panels, $z < 3$ in the bottom left, and $z \ge 3$ in the bottom right. The SHMR computed by abundance matching from \cite{Shuntov2025} is also shown in dashed lines.}
    \label{fig:SHMR}
\end{figure*}

We present in Fig.~\ref{fig:SHMR} the stellar-to-halo mass relationship (SHMR), computed firstly as the ratio of the threshold stellar mass of each sample $M_{\star, \rm th}$ to their associated halo mass $M_{\rm h,min}$, as a function of halo mass\footnote{We provide our SHMR measurements at \url{https://github.com/LouisePaquereau/GalClustering_COSMOS-Web_Paquereau2025}}. The SHMR as the ratio of the galaxy sample's median stellar mass $M_{\star, \rm med}$ over halo mass is also shown, as it gives different amplitudes and slightly different slopes. Both definitions are used in the literature \citep[e.g.][]{Harikane16, Zaidi2024}. While they both use the same halo mass, the first one gives a sense of the minimum star formation efficiency in producing central galaxies just above the threshold mass in halos of mass $M_{\rm h,min}$ and is then more influenced by the lower end of the stellar mass distribution in each sample. In contrast, the second definition connects the halo mass $M_{\rm h,min}$ to more massive central galaxies than the minimum mass required to be hosted by it, so it captures an average star formation efficiency in halos at this mass scale. Consequently, the latter SHMR tends to show a higher amplitude, as it reflects the contribution of more massive galaxies that dominate the median. 

We also display the SHMR derived from abundance matching in COSMOS-Web by \cite{Shuntov2025}, where the assumption that one halo hosts one galaxy is made. These two independent measurements of the SHMR show the same evolution with redshift, and their amplitudes are in agreement with each other in the case of the SHMR $M_{\star, \textrm{med}}/M_{\rm h, min}$. The discrepancy in amplitude with the $M_{\star,\textrm{th}}/M_{\rm h, min}$ relation can be explained by the fact that, while abundance matching counts galaxies above a stellar mass threshold, it associates one galaxy to one halo, focusing on centrals without explicitly including satellites or scatter, in contrast with our HOD model. As a result, the typical halo mass hosting a central galaxy is lower in abundance matching, which raises the SHMR and aligns with our measured one drawn using the median stellar mass, which better captures the typical stellar mass of centrals and/or accounts for scatter.

The integrated star formation efficiency $\varepsilon_{\rm SF}$ is also shown. We also note that the SHMR presented in this work is a rough estimate for central galaxies because we are using $M_{\rm h,min}$ which is the characteristic halo mass derived specifically for central galaxies. A more complex modeling of the SHMR (such as using \citealt{Leauthaud11} model) is beyond the scope of this work.

\subsection{Satellite fractions} \label{subsec:satfrac}

\begin{figure}[th!]
    \centering
    \includegraphics[scale=0.55]{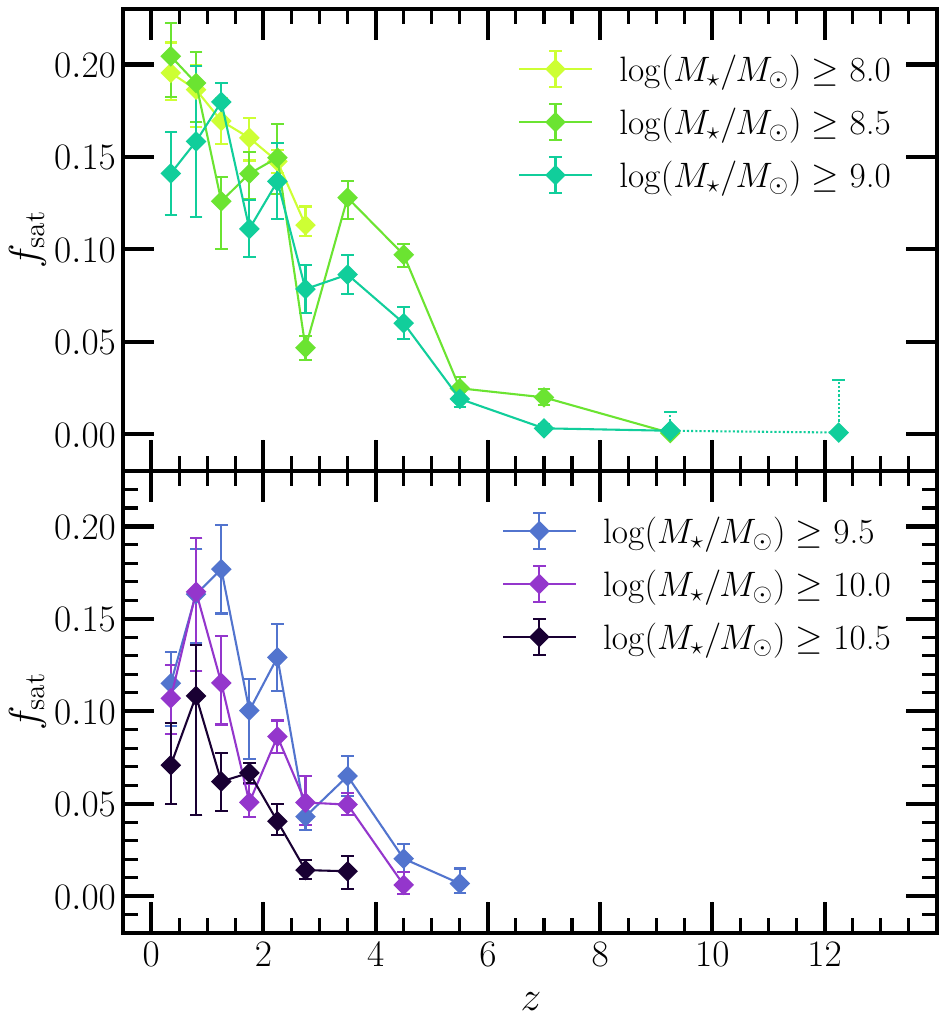}
    \caption{Redshift evolution of the HOD-derived satellite fraction for mass-limited samples of galaxies. The point at $10.5 \le z < 14$ is represented in dotted lines because it is considered less certain.}
    \label{fig:satfrac}
\end{figure}

Figure \ref{fig:satfrac} shows the evolution of the HOD-derived satellite fraction with redshift for fixed stellar mass thresholds, calculated using \texttt{halomod} based on Eq.~\ref{eq:HOD_satfrac}. In the HOD model, satellites are defined as galaxies that orbit around a central galaxy, which resides at the center of the halo's gravitational potential well. Generally, the fraction of satellite galaxies decreases with redshift, starting from 15-20\% at $z < 4$ and decreasing to 1-5\% at $z > 4$. As expected, the satellite fraction is higher for lower mass thresholds. The trends at $z < 1$ likely reflect a combination of limited survey volume at $z < 0.5$, which lowers the satellite fraction, and overdensities in the COSMOS field, such as the cluster at $z = 0.7$, which enhance it around $z \sim 1$. Another drop is seen at $2.5 \le z < 3.0$ for $\log(M_\star/M_\odot) \ge 8.5$ and $\ge 9.5$ bins (also seen in COSMOS2020 by \citealt{Shuntov22}), likely reflecting an overestimation of $M_{\rm h,1}$ in the HOD fit (see also Fig~\ref{fig:M1Mmin_evol}), rather than a physical effect. Then the fraction decreases sharply between $z = 1$ and $z = 2$, after which the decline becomes more gradual, eventually reaching a plateau at some mass thresholds and approaching nearly 0 at $z \ge 8$. The decrease is also seen in other clustering analyses \citep[e.g.,][]{Hatfield18, Harikane18, Ishikawa20}, and more recently in \cite{Zaidi2024} in UDS + COSMOS fields up to $z = 4.5$. For the high-$z$ regime, \cite{Bhowmick2018} for instance also found $f_{\rm sat} < 0.05$ for galaxies with mass $\log(M_\star/M_\odot) \ge 9.0$ at $z = 8$ and $z = 10$ using HOD modeling in the BlueTides simulation \citep{Feng2016}. As cosmic time progresses, the Universe's structure becomes more filamentary, small halos merge into larger and more massive ones, increasing the number of satellites per halo (see e.g. \citealt{White&Frenk1991, Bullock2001}). The almost negligible fraction of satellite galaxies at $z = 10$ could reflect the early universe's initial stages, where fewer structures have formed and merged. However, our sample is likely biased toward the most massive and brightest galaxies, which are typically centrals, since satellites with $M_\star > 10^9\,M_\odot$ at these redshifts are expected to be too rare to be detected by COSMOS-Web. \cite{Bhowmick2018} predicted a probability of less than 30\% of observing satellites of mass $\log(M_\star/M_\odot) \ge 9.0$ around centrals of mass $\log(M_\star/M_\odot) = 10.5$ at $z = 7.5$ according to JWST limitations, value that decreases with redshift. It is also important to note that our satellite fractions are lower (a difference of 0.01 to 0.05) than those reported in COSMOS2020 \citep{Shuntov22}, independently of redshift, likely because they used a different HOD model that better constrains satellite and central galaxies at low redshifts.

\subsection{Galaxy bias} \label{subsec:galbias}

Galaxy bias $b_{\rm g}$ is defined as the ratio between the clustering of galaxies $w_{\rm g}$ and the clustering of dark matter particles $w_{\rm DM}$ (see Eq.~\ref{eq:galbias}). It depends on various physical processes, such as feedback, gas cooling, star formation, and black hole accretion, which influence the distribution and evolution of galaxies differently than that of dark matter. 
\begin{equation} \label{eq:galbias}
    b_{\rm g} = \sqrt{\frac{w_{\rm g}}{w_{\rm DM}}}
\end{equation}
Estimations of this bias are derived using Eq.~\ref{eq:HOD_galbias} in the \texttt{halomod} package, based on the fitted halo occupation distribution for each galaxy sample.

Current halo and galaxy formation models predict that halos form in rare fluctuations of the primordial matter density field, leading to high bias and clustering in the most massive halos, where the first galaxies are expected to form \citep{Kaiser1984}. This suggests that bias should increase at earlier epochs. However, new studies of high-$z$ galaxies challenge this simple view. Two main scenarios are proposed to explain the abundance of UV-bright, massive galaxies observed at $z \ge 8$ with JWST. The first suggests higher scatter in the UV magnitude - halo mass relation \citep[or stochasticity, e.g.,][]{Mason2023, MirochaFurlanetto23, KravtsovBelokurov24}, allowing low-mass halos to host UV-bright galaxies due to starbursts that make them bright enough to be observed before their massive stars die \citep{SunG2023}. This scatter may be more pronounced in lower-mass halos, which dominate the early universe, as the galaxies inhabiting them could be more sensitive to feedback or environmental effects, triggering cycles of starbursts \citep{Gelli24}. Although our sample is mass-selected, at high redshift the SED-fitting estimation of the mass relies only on the rest-frame UV and optical. For example, at $z>10$, the reddest band probed by our set of filters is the rest-frame $u$. This could lead to overestimating the stellar mass of starbursting galaxies, hence artificially increasing the SHMR stochasticity as well. The second scenario suggests a tighter galaxy-halo connection with high star formation efficiency in massive halos \citep [e.g.,][]{Dekel23_FFB, Yung2024}. A study by \cite{Munoz23} showed that both scenarios are degenerate in the UV luminosity function but anticipated that bias measurements from JWST would help distinguish between them. A bias with little variation across stellar mass thresholds would support high stochasticity, with the apparent massive end being in fact dominated by lower-mass, less biased halos. In contrast, a high bias with strong mass variation would favor high star formation efficiency, suggesting that massive galaxies reside in massive halos.

\begin{figure}[h!]
    \centering
    \includegraphics[scale=0.5]{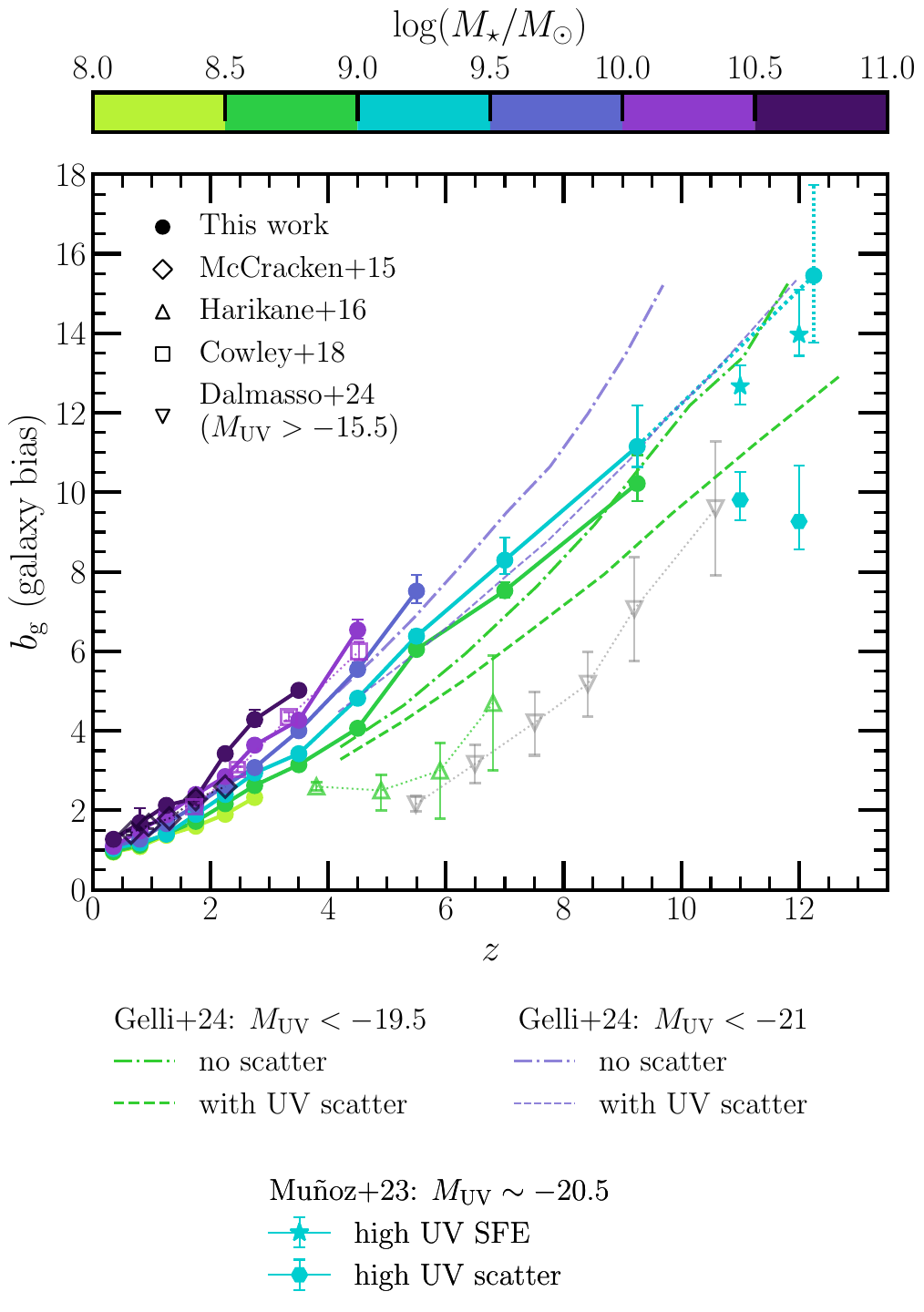}
    \caption{Evolution of the galaxy bias with redshift, for different stellar mass thresholds, derived from the HOD modeling of our clustering measurements. The point at $z \sim 12$ is represented in dotted lines because it is considered less certain. Predictions of the galaxy bias as a function of UV magnitude by \cite{Munoz23} and \cite{Gelli24} are superposed, with and without implementation of a large scatter in the $M_{\rm UV} - M_{\rm h}$ relationship to explain the overabundance of bright galaxies at high-$z$. Literature measurements from \cite{McCracken15, Harikane16, Cowley18} are shown in colors corresponding to their sample threshold mass, except for \cite{Dalmasso2024a}, which is in grey since it is based on an $M_{\rm UV}$ limited sample.}
    \label{fig:galbias}
\end{figure}

We represent in Fig.~\ref{fig:galbias} the evolution of our galaxy bias measurements with redshift for different stellar mass thresholds and model predictions for samples limited by UV magnitude from \cite{Munoz23} and \cite{Gelli24}. The first one has been calibrated on UV luminosity function measurements from HST \citep{Bouwens21} and early JWST observations \citep{PerezGonzales23, Finkelstein23, Harikane23b}, while \cite{Gelli24} calibrated its halo mass-dependent scatter based on the zoom-in simulation FIRE-2 following \cite{SunG2023}. To compare to our results, we computed simplified versions of the SMF and the UVLF. By matching the cumulative number of galaxies in our sample below a certain $M_{\rm UV}$ threshold, we determined the corresponding $M_\star$ threshold. We verified that using the median of the $M_{\rm UV} - M_{\star}$ distribution gives consistent results, with differences of less than 5\%. This procedure led to estimates presented in Tab.~\ref{tab:UVmasslink}. When comparing to the predictions, our bias measurements appear to support the second explanation: the hypothesis that high star formation efficiency in massive halos drives the observed abundance of galaxies at high redshifts. Specifically, our results for $\log(M_\star/M_\odot) \ge 8.5$ and $4 \le z < 10$ align more closely with the \cite{Gelli24} model without UV scatter for $M_{\rm UV} < -19.5$. The $z \simeq 12$ point was placed in the $\log(M_\star/M_\odot) \ge 9.0$ panel for visualization purposes, although it corresponds to galaxies with $\log(M_\star/M_\odot) \ge 8.85$, making it compatible with the no-scatter model for $M_{\rm UV} < -19.5$ galaxies. Similarly, for $\log(M_\star/M_\odot) \ge 9.5$ and $4 \le z < 6$, the model without scatter for $M_{\rm UV} < -21$ is within 1.5$\sigma$ of our measurements. Furthermore, the \cite{Munoz23} prediction without scatter falls within the 1$\sigma$ uncertainties at $z \sim 12$, while the one with scatter is $\sim 2\sigma$ lower.

However, strict conclusions require a mass-limited model of stochasticity at high redshift, ideally calibrated with COSMOS-Web data. In our analysis, we consider UV-bright galaxies detected at high-$z$ to be comparable to our massive galaxies. While these galaxies are likely star-forming and UV-emitting, selection effects may introduce biases, and a direct comparison with models based on $M_{\rm UV} - M_{\rm h}$ scatter is not straightforward. For instance, UV-limited samples could miss massive galaxies with low star formation rates (SFRs), whereas mass-limited samples might overlook low-mass galaxies that experience short-lived starbursts. Additionally, we did not fit the scatter $\sigma_{\log M}$ in the HOD, which may lead to bias if there is an intrinsic scatter present. Nonetheless, a comparison between dark matter clustering predictions from N-body simulations at $z \sim 12$ and our galaxy clustering measurements suggests an even higher ratio $w_{\rm g}/w_{\rm DM}$ than what we derive from the HOD model, indicating that our conclusions remain robust despite these potential biases.

\begin{table}[h!]
    \centering
    \setlength{\tabcolsep}{3pt}
    \renewcommand{\arraystretch}{1.5}
    \fontsize{9pt}{9pt}\selectfont
    \begin{tabular}{p{2.1cm}>{\centering\arraybackslash}m{2cm}>{\centering\arraybackslash}m{2cm}>{\centering\arraybackslash}m{2cm}}
        \hline
         & $M_{\rm UV} < -19.5$ & $M_{\rm UV} < -20.5$ & $M_{\rm UV} < -21.0$ \\
        \hline
        \hline
        $4 \le z < 6$ & 8.60 & 9.20 & 9.55 \\
        $6 \le z < 10.5$ & 8.70 & 9.0 &  9.40 \\
        $10.5 \le z < 14.0$ & --- & 8.85 & 9.10 \\
        \hline
    \end{tabular}
    \vspace{2mm}
    \caption{Stellar mass lower thresholds in units of $\log(M_\star/M_\odot)$ corresponding to $M_{\rm UV}$ upper thresholds, derived by matching cumulative numbers of galaxies in our sample as a function of UV-magnitude or stellar mass.} 
    \label{tab:UVmasslink}
\end{table}

\section{Discussion} \label{sec:discussion}

\subsection{What can clustering teach us about the galaxy - halo connection at $z \ge 6$ ?}  \label{subsec:SFEhighz} 

Recent observations, including our study, highlight the presence of highly efficient star formation and massive galaxies at $z \ge 6$. Our results, supported by clustering and HOD modeling, show that these high-$z$ galaxies are residing in more massive halos than those expected by pre-JWST galaxy formation models or simulations (see the discussion in Sect.~\ref{subsubsec:SHMR_comparison}). This is consistent with results from other JWST surveys. For instance, in the COSMOS-Web field, \cite{Casey2024} reported luminous and massive sources at $z > 10$ believed to be undergoing rapid starbursts, and \cite{Shuntov2025} reveals an overabundance of massive galaxies at high redshift from stellar mass function measurements in in the same field. 

To this day, several explanations have been proposed to account for these observations, which can be broadly divided into five categories: (I) an intrinsic high star formation efficiency at early times for example because of effective gas cooling, accretion, and fragmentation \citep[e.g.,][]{Fujimoto2024, Faisst2024}, feedback-free star formation \citep[hereafter FFB;][]{Dekel23_FFB, Li23_FFB}, high merger-induced star formation \citep{Duan2024}; (II) a stochastic and bursty star formation at high-z \citep[e.g.,][]{Mason2023}; (III) an overestimation of stellar masses e.g. because of a top-heavy IMF at high-$z$ \citep[see e.g.,][]{Trinca2024, Shengdong2025}; (IV) a different treatment of dust attenuation, eliminating the need for evolving SFE \citep{Ferrara2023}; (V) or a change in cosmology e.g. early dark energy \citep{Klypin21, Liu24}. Each of these scenarios translates in a different galaxy-halo connection, and some of them can lead to distinct clustering patterns due to differences in how galaxies populate halos of various masses, while other observables as the UV luminosity function are degenerate \citep{Munoz23}. 

Focusing on the FFB model, we find that our derived SFE and halo masses are compatible with its predictions. This model posits that at early times, star-forming clouds in halos with masses between $10^{10.8} \times [(1+z)/10]^{-6.2} \leq M_{\rm h} < 10^{12} M_\odot$ could achieve star formation efficiencies of up to $50$ to $100\%$, given a brief window of $\sim$1 Myr where feedback mechanisms are ineffective. Our estimated halo masses for galaxies with $M_\star > 10^{8.5-9}\,M_\odot$ at $z \ge 8$ fall within this predicted regime. Our derived SFE values at $z \ge 8$ agree with the FFB predictions assuming a FFB maximum efficiency around $\epsilon_{\rm FFB,max} \simeq 0.2$, and are lying about 2$\sigma$ above no-FFB expectations (calculated using the \texttt{ffb\_predict} code\footnote{\url{https://github.com/syrte/ffb_predict/}}, from \citealt{Li23_FFB}). These efficiencies naturally decrease with cosmic time as radiative feedback becomes more prominent and the star-forming clouds become less dense, aligning with the decline we see in the SHMR. Additionally, this model reproduces the COSMOS-Web SMF well when $\epsilon_{\rm FFB,max}$ increases with both redshift and stellar mass (see \citep{Shuntov2025} for the detailed analysis). Further investigations on this and other high-$z$ models are left for future work. We also note that stochasticity in the UV luminosity–halo mass relation, which may be connected to this bursty star formation regime, could influence the evolution of galaxy bias, which we explore in Sect.~\ref{subsec:galbias}.

Moreover, halo properties play a crucial role in regulating star formation. For instance, the halo accretion rate is one of the factors that controls the availability of gas for star formation \citep{White&Frenk1991}, while inflows and outflows are influenced by the depth of the halo's potential well, all of which can be related to halo mass. The high merger rates at early times \citep{Duan2024} further enhance star formation efficiency. Therefore, linking galaxies to their host halo mass is essential. Currently, this is often done using abundance matching, a method that relies on strong assumptions and is typically applied to incomplete samples, potentially introducing biases, and does not account for scatter in the UV magnitude - halo mass or stellar mass - halo mass relations. 

Thus, the information from galaxy clustering is currently the most robust way to connect galaxies at $z \ge 6$ to halos and estimate their host properties. Galaxies should not be viewed as isolated systems; rather, their star formation is influenced by the large-scale structures they live in. Additionally, we emphasize that any viable scenario for the early universe must simultaneously account for the high SFE observed at $z \ge 6$ and its gradual decline over time until approximately $z \sim 3$, as captured in the SHMR.

\subsection{The interplay between stellar and dark matter growth across time} \label{subsec:SHMR_explanations}

\subsubsection{Evolution of the SHMR with redshift} \label{subsubsec:SHMR_evolz}

The SHMR shows the interplay between dark matter accretion rate and conversion from gas to stars inside galaxies. Here, we propose a simple scenario to explain the evolution of SHMR from the early universe at $z > 10$ to the present.

At high redshift, the SHMR is greater than at lower redshifts for a given halo mass, indicating a higher star formation efficiency in the early universe. As discussed in Sect.~\ref{subsec:SFEhighz}, this high efficiency for $z > 8$ can be attributed to star formation occurring in bursty episodes with star-formation timescale shorter than SN explosion timescale leading to negligible (or none at all) impact of feedback from stellar or radiative processes \citep[FFB model][]{Dekel23_FFB}. \cite{renzini25} also suggested that the lack of angular momentum in the very early Universe could lead locally (at the globular cluster scale) to very high baryon concentrations and consequently to elevated SFE \citep[see also][]{loeb24}. In these very dense regions, gravitational acceleration could be also high enough to overcome momentum injection from stellar feedback, allowing efficient star formation \citep[e.g.][]{BoylanKolchin24}.

Over time, halos keeps accreting matter. The high accretion rate in the early universe leads to strong turbulence, which could prevent star formation due to the turbulent-pressure support \citep[e.g.][]{Andalman2024} and combined with outflows which drive part of the accreted gas out of the halos, the SHMR progressively decreases. As galaxies grow, feedback mechanisms such as stellar feedback or AGN activity become more efficient. In massive halos, AGNs will accumulate mass through accretion and expel or heat surrounding gas more forcefully \citep[e.g.,][]{Man&Belli2018}. In lower-mass halos, massive stars will die and generate strong stellar winds. Hence, star formation will become less and less efficient at a given stellar mass threshold, as shown in Fig.~\ref{fig:M1Mmin_evol}. As the ability to form stars from halo gas accretion diminishes, gas accumulates in halos faster than it can be converted into stars, further reducing the SHMR as time progresses.

However, at $z \sim 2-3$, this trend reverses, in particular for low-mass galaxies, and the observed SHMR begins to increase, indicating that the efficiency of central stellar mass growth becomes higher relative to dark matter growth. This corresponds to the redshift at which the halo growth rate slows and often reaches a plateau, so halo mass varies more gradually (see mass accretion rates from \citealp{Behroozi13_HMF} for example, or \citealp{Lilly13}). Meanwhile, delayed star formation can still occur due to the gas reservoir accumulated within halos, leading to an increase in the SHMR. This behavior is also in agreement with a type of ‘‘downsizing’’ scenario, where the peak of star formation shifts to lower-mass halos ($M < 10^{12} M_\odot$) as cosmic time progresses \citep[see e.g.][]{Cowie1996, Neistein2006, Conroy&Wechsler2009}. Although we cannot directly probe the evolution of the SHMR peak and its high-mass end, we observe that its amplitude at fixed halo mass (for $M_{\rm h} < 10^{12} M_\odot$) increases with time, indicating that in low-mass halos star formation becomes more efficient at later times. According to this scenario, these low-mass halos would then host younger galaxies than their massive counterparts, in opposition to the hierarchical scenario of structure formation \citep{DeLucia06}.

\begin{figure}[t!]
    \centering
    \includegraphics[scale=0.53]{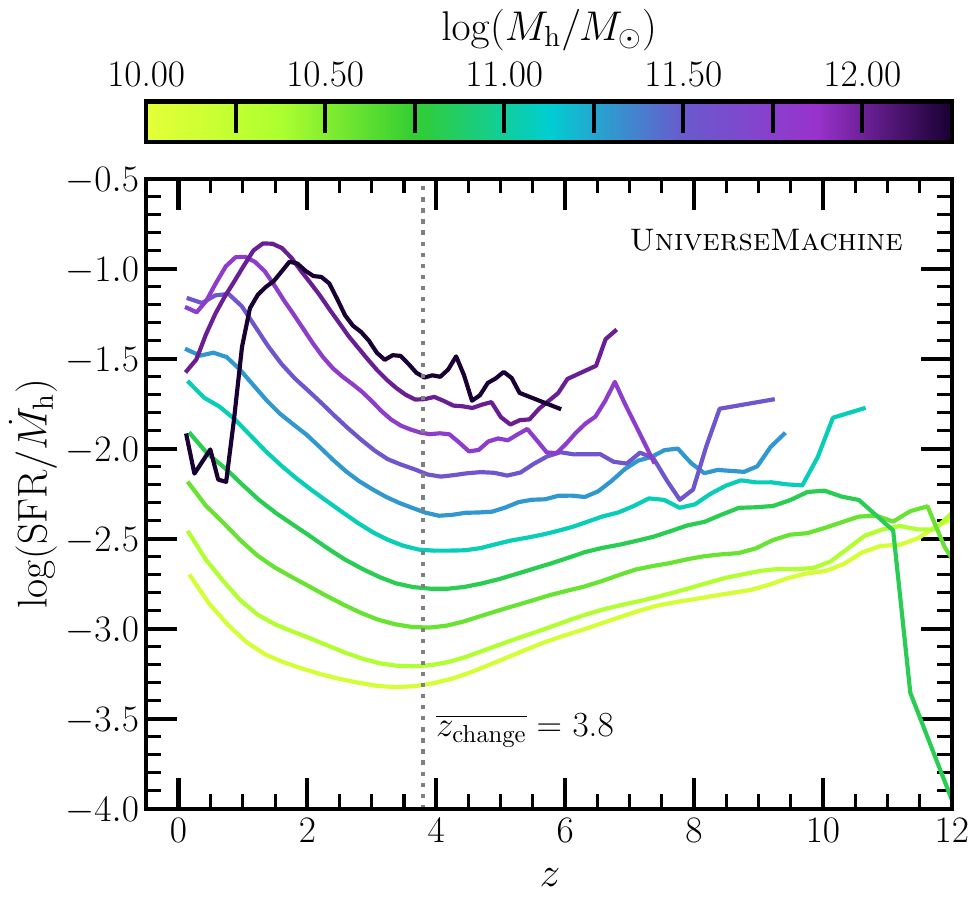}
    \caption{Evolution of the ratio between star formation rate and halo mass accretion rate as a function of redshift, in the semi-empirical model \textsc{UniverseMachine}. In halo mass bins of width 0.25 dex, the ratio is computed as the halo accretion rate from the analytical formula in \cite{Behroozi13_HMF} formula over the median SFR of galaxies hosted by halos in the mass bin. A transition in the ratio behavior is observed at a mean redshift $\overline{z_{\mathrm {change}}} = 3.8$ across all halo masses.}
    \label{fig:UM_SFRMAR}
\end{figure}

Using the semi-empirical model \textsc{UniverseMachine}\footnote{\url{https://halos.as.arizona.edu/UniverseMachine/DR1/JWST_Lightcones/}} \citep{Behroozi19} -- the only simulated catalog reproducing the same trend as our observations -- we show in Fig.~\ref{fig:UM_SFRMAR} the ratio of the star formation rate of galaxies to the halo mass accretion rate over time. In halo mass bins, halo mass accretion rates have been computed analytically with the formula of \cite{Behroozi13_HMF}, and we take the median SFR of galaxies hosted by these halos. Starting from high redshift, we observe a decreasing ratio, indicating that dark matter halos are growing faster than stellar mass. Around a mean redshift of $z = 3.8$ (with this transition redshift increasing for more massive halos), this trend reverses, showing an increase in the ratio as dark matter growth becomes less efficient than star formation. This trend holds up to a halo mass of $M_{\rm h} = 12.5$, beyond which the ratio only decreases from high to low redshift, but we do not explore this high-mass regime since it falls outside the scope of our observational data. This pattern further supports our proposed scenario, demonstrating the evolving relationship between galaxy and halo growth rates over time. We also compared this to measurements in COSMOS-Web, plotting the median SFR of galaxies at the threshold mass as a function of redshift and halo mass $M_{\rm h,min}$. The results are consistent with those in \textsc{UniverseMachine} in both amplitude and trend. However, due to the large scatter in median SFR and the fact that we do not have the results as a continuous function of $M_{\rm h,min}$ and $z$, we focus here on presenting the simulation results only for a clearer understanding of the scenario.

\cite{Shuntov2025} presents evolutionary tracks for halos (and their associated stellar mass) formed at $z = 10$, $z = 6$, and $z = 2$ with the same initial halo mass, constructed using the \cite{Dekel2013} halo mass growth function combined with a SFE derived from their SHMR. For instance, they show that a halo formed at $z = 6$ or $z = 10$ experiences an increase in SFE, which peaks and then declines once the halo reaches a mass of $\sim 10^{12.5}\,M_\odot$. As a result, such halos that formed before $z = 3.5$ would have built up most of their stellar mass by that time. Further investigations are needed to quantitatively describe this scenario, including, for example, analysis of halo and galaxy kinematics or gas properties that could explain this trend in the SHMR. However, this is beyond the scope of the current work.

\subsubsection{Comparison of our SHMR with observations, simulations, and models} \label{subsubsec:SHMR_comparison}

There is currently a significant discrepancy in the observations, models, and simulations regarding the evolution of SHMR with redshift. If we focus on the Universe at $z > 4$, some studies show no evolution \citep{SunFurlanetto2016, Stefanon2017, Stefanon2021}, while others see the SHMR decreasing with increasing redshift \citep{Tacchella2018, Zhu2020} or the opposite \citep{Finkelstein2015, Harikane16, Moster2018, Behroozi19}.\\

\textbf{Comparison with other observations.} Our findings are consistent with those in the COSMOS field, such as \cite{Shuntov22}, who report a similar trend in the SHMR for halos with $M_{\rm h} < 10^{12} M_\odot$ using the COSMOS2020 catalog. \cite{Leauthaud11}, using weak lensing measurements, also observed an increasing SHMR from $z = 2$ to $z = 0$, along with a shift toward lower masses for the pivot halo mass (the mass where star formation is most efficient). Recent results from \cite{Zaidi2024}, based on clustering and HOD modeling in the UDS + COSMOS fields up to $z = 4.5$, show a change in slope in the SHMR between $M_{\rm h} \sim 10^{10.5}$ and $10^{12.5} M_\odot$, with a 50\% increase in SHMR at $M_{\rm h} = 10^{11.5} M_\odot$ from $3.5 \le z <4.5$ to $0.2 \le z < 0.5$. These findings do not show significant discrepancies with ours, and differences in HOD models, SHMR definitions, and data completeness may explain minor variations.

At higher redshifts, to our knowledge, few studies probe the mass and redshift ranges we cover. \cite{Harikane16, Harikane18} found a similar trend, with a decrease in SHMR from $z = 7$ to $z = 4$ using clustering of Lyman-break galaxies in HSC data. Conversely, \cite{Stefanon2021} claims no redshift evolution in the SHMR from $z = 6$ to $z = 10$ using abundance matching. However, this study mixes different fields (HUDF, XDF, GOODS, and all five CANDELS fields) and relies on Spitzer/IRAC data, which has known limitations for galaxies at $z > 8$ due to factors like shallower depth (the program they use is $\sim 1.5$ magnitude shallower than COSMOS-Web's F444W depth) and challenges with deblending because of lower resolution. Their sample is likely incomplete, with only 800 Lyman-break galaxies, and their stellar mass function does not match ours in COSMOS-Web (as shown in \citealt{Shuntov2025}), being higher at $5.5 \le z < 7.5$ and lower at $7.5 \le z < 10$, which suggests a lower SHMR at high redshifts than what we find.\\

\begin{figure*}[ht!]
    \centering
    \includegraphics[scale=0.5]{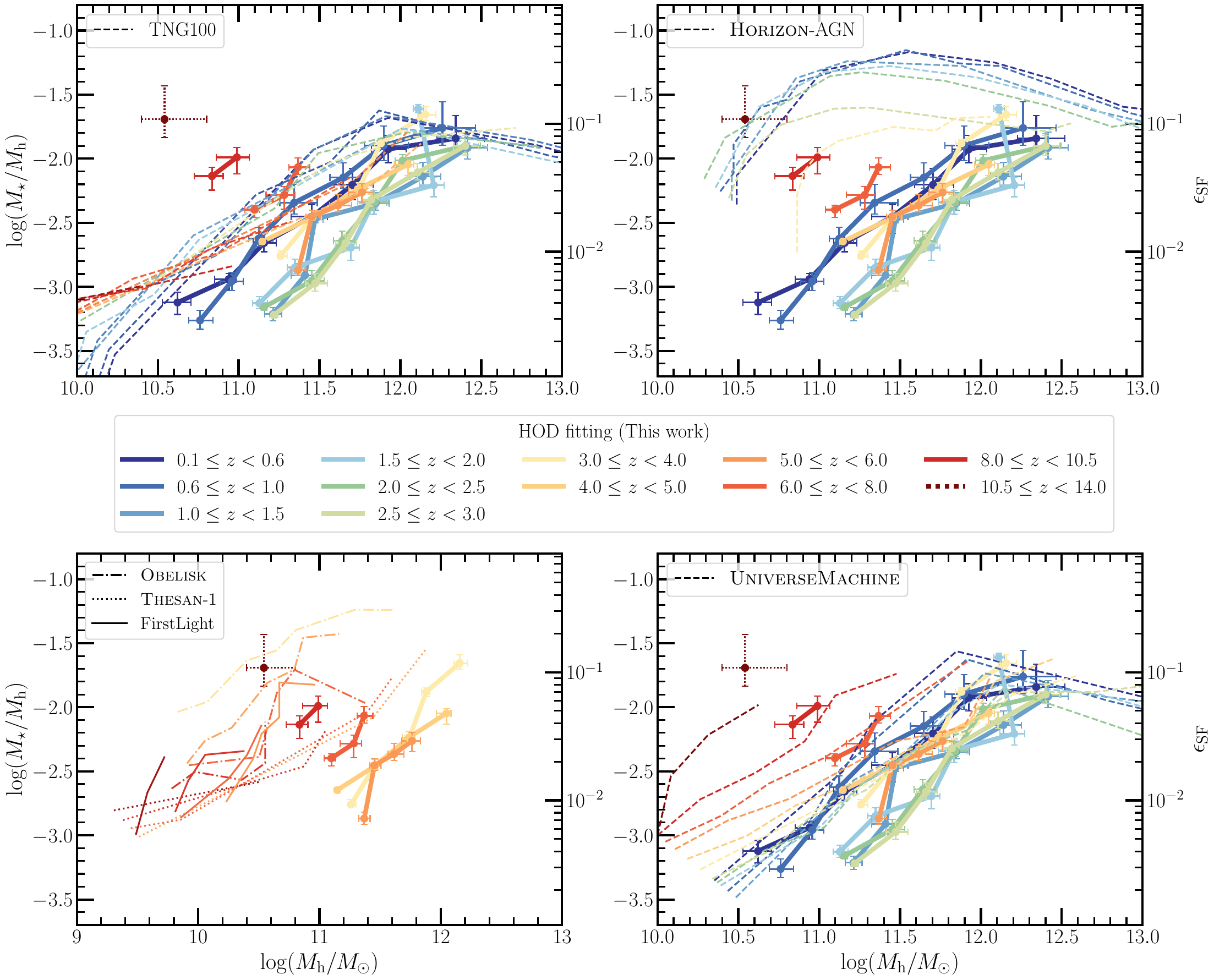}
    \caption{Comparison of the observational SHMR in COSMOS-Web with results from various hydrodynamical simulations and the semi-empirical model \textsc{UniverseMachine}. Solid lines represent the observational SHMR $M_{\star, \rm th}/M_{\rm h,min}$, obtained from HOD fitting of mass-limited galaxy clustering measurements in COSMOS-Web. The point at $10.5 \le z < 14$ is represented in dotted lines because it is considered less certain. The dashed and other lines represent the SHMR in the simulations computed similarly to the observations, where $M_{\rm h,min}$ is the halo mass at which 50\% of halos host a central galaxy with a stellar mass above $M_{\star, \rm th}$. \\ 
    \textit{Top left:}  SHMR from TNG100 simulation snapshots, computed at the mean redshifts of each observational redshift range. \textit{Top right:} SHMR from Horizon-AGN light-cone, calculated over the same redshift ranges as the observations, but limited to $z \leq 6$. \textit{Bottom right:} Same from \textsc{UniverseMachine} light-cone in the COSMOS field, matching the observational redshift ranges. \textit{Bottom left:} Same for high-$z$ simulations (\textsc{Thesan}-1, FirstLight, \textsc{Obelisk}), computed from snapshots at mean redshifts from $z \sim 5$ to $z \sim 12$.}
    \label{fig:shmr_sims}
\end{figure*}

\textbf{Comparison with models.} We compare our results to the semi-empirical model \textsc{UniverseMachine}, shown in the bottom-right panel of Fig.~\ref{fig:shmr_sims}. This model exhibits a similar redshift trend to our observations, with a decreasing SHMR from $z=11$ to $z=2-3$ and a turnover at $z = 2-3$. \textsc{UniverseMachine} has been calibrated using observational constraints before JWST, such as the stellar mass function and clustering measurements up to $z \sim 10$. This indicates that information about this evolution was likely already present in the observational results before JWST. The EMERGE empirical model \citep{Moster2018} also shows a decreasing SHMR from $z = 0.1$ to $z = 4$ and an increase of 30\% from $z = 4$ to $z = 8$. Like \textsc{UniverseMachine}, it is constrained by observed SFRs, SMFs, and cosmic star formation rate densities (CSFRDs) up to $z \sim 10$. EMERGE computes the SFR of individual galaxies as the product of an instantaneous star formation efficiency and their host halo's growth rate, linking stellar history to halo formation. This approach introduces scatter in the SHMR, resulting in higher stellar masses in low-mass halos at high redshift, similar to the effects of bursty SFHs at $z > 4$. Another empirical model, from \cite{Tacchella2018}, follows a similar approach to EMERGE but forces a redshift-independent star formation efficiency, $\epsilon_{\rm SF}(M_{\rm h})$, and is calibrated solely on the UV luminosity function at $z = 4$. Consequently, this model predicts a steadily decreasing SHMR from $z = 4$ to $z = 14$, which contrasts with both EMERGE, \textsc{UniverseMachine} and observational trends.\\

\textbf{Comparison with simulations.} We compare our results with several state-of-the-art hydrodynamical simulations, some of them being specially built to study the high-$z$ universe: TNG100\footnote{\url{https://www.tng-project.org/}} (\citealt{Pillepich2018}; \citealt{Nelson2018}; \mbox{\citealt{Marinacci2018}}; \citealt{Naiman2018}; \citealt{Springel2018}), \textsc{Horizon}-AGN\footnote{\url{https://www.horizon-simulation.org/}} \citep{Dubois2014}, \textsc{Thesan}-1\footnote{\url{https://www.thesan-project.com/}} \citep{Kannan2022}, FirstLight\footnote{\url{https://www.ita.uni-heidelberg.de/~ceverino/FirstLight/}} \citep{Ceverino2017}, \textsc{Obelisk} \citep{Trebitsch2021}. A summary of their main characteristics (resolutions, mass definitions, etc) can be found in App.~\ref{app:sims}.

We did the exercise to measure the SHMR from the simulations the same way as our observational SHMR: in central galaxy samples limited in mass by $M_{\star,\rm th}$, we compute the characteristic halo mass $M_{\rm h,min}$ as the halo mass at which 50\% of halos host a central galaxy of mass above $M_{\star,\rm th}$. We verified that this recovers the true SHMR computed using individual galaxies and their host halo masses, confirming that the SHMR as $M_{\star,\rm th}/M_{\rm h,min}$ is a good proxy for the true SHMR in observations as well: the median difference between the two was less than 7\%, while for $M_{\star,\rm med}/M_{\rm h,min}$ it is of the order of $15 -20$\%. The comparison between our observational SHMR and simulation results is presented in Fig.~\ref{fig:shmr_sims}. Most simulations conducted before JWST do not reproduce the observed abundance of massive galaxies, nor do they predict the high star formation efficiency seen at $z > 6$. So rather than focusing solely on the amplitude of the stellar-to-halo mass ratios, we emphasize their redshift evolution to determine whether they display a similar behavior to the observed SHMR, showing this characteristic turnover around $z = 2-3$. For a more detailed comparison of the SHMR in COSMOS with TNG100 and \textsc{Horizon}-AGN up to $z = 5$ (in terms of amplitudes and slopes), we refer to \cite{Shuntov22}.

The TNG100 simulation is in good agreement with observed SHMR amplitudes and slopes for $z < 1$. However, it does not show a distinct evolution with redshift, except for a gradual change in slope for low-mass halos. This slope becomes shallower at higher redshifts, indicating a higher star formation efficiency at a fixed halo mass as redshift increases, but only for halos with masses below $10^{10.5} M_\odot$. The change in slope is not clearly observed in the data, and the SFE at high redshift is not as high as measured in the observations. For halos above this mass range, the SHMR consistently decreases with increasing redshift.

Similarly, \textsc{Horizon}-AGN simulation shows little evolution in SHMR from $z = 0.1$ to $z = 3$, and there is no indication of a turnover in its evolution. It is then followed by a sharp decline for $3 < z < 6$, but this can be explained by the underestimation of the mass function at $z > 4$ in the simulation due to resolution limits (see App.~\ref{app:sims}). We note that the SHMR amplitude is generally not comparable to observations, as previously noted in studies such as \cite{Hatfield19, Shuntov22}, likely due to the overestimation of galaxy masses in low-mass halos.

Simulations specifically designed to study the high-redshift universe show discrepancies among themselves, and with the data. As JWST has revealed more massive objects than anticipated, these simulations generally contain lower halo and galaxy mass ranges than observed, with amplitudes that are also not directly comparable. Additionally, almost none of them replicate the same redshift evolution observed in our data. For instance, the \textsc{Obelisk} simulation, which is centered on an overdense region within a small volume, only shows a decrease in the SHMR of $\sim 0.5$\,dex from $z = 9$ to $z = 3$. While the \textsc{Thesan}-1 simulations show a more comparable amplitude, the SHMR appears constant with redshift, as expected since \textsc{Thesan} extends the TNG model with radiative transfer and thus naturally looks like TNG's low-$z$ SHMR. Among these simulations, only the FirstLight simulation hints at an increasing SHMR for $z > 4$, but it is a zoom-in simulation of $\sim$ 300 galaxies and could be biased by cosmic variance effects or incompleteness in the computation of the SHMR.

These figures highlight the significant discrepancies between current simulations and observations, as well as inconsistencies among the simulations themselves, particularly those specialized for the epoch of reionization (EoR). This can be the result of wrong calibrations in the simulations at high-$z$ with pre-JWST data but could also depend on changes in the physical processes involved in star formation (e.g., the FFB model). It is noteworthy that even simulations based on the same model, like TNG100 and \textsc{Thesan}, yield significantly different SHMR amplitudes, likely due to their calibration on low versus high-redshift observations. This suggests that the trend of higher star formation efficiency at early times might have already been hinted at in pre-JWST literature. Our results offer valuable constraints on star formation efficiency and the galaxy-halo connection from $z = 0.1$ up to the highest redshifts, which can guide the development of more accurate future simulations and help refine how galaxies populate halos at $z > 10$.

\subsection{Limitations of this work} \label{subsec:limitations}

\subsubsection{High-redshift sample and completeness} \label{subsubsec:limits-highz}

This study relied solely on photometric data for deriving photometric redshifts and estimating stellar masses through SED fitting. At high redshifts, where only a few infrared bands are available in COSMOS-Web, identifying the Lyman break and constraining galaxy SEDs pose challenges \citep[see e.g.,][for stellar mass recovering at $z > 7$]{Narayanan2024}. Potential contamination from lower or higher redshift objects could bias our clustering analyses. To mitigate this, rigorous cleaning procedures were applied to our high-redshift sample (see Sect.~\ref{subsubsec:highzselection}). To assess residual contamination after this cleaning, we computed cross-correlations between our selected redshift bins (App.~\ref{app:crosscorr}). Ideally, a contamination-free sample in a given redshift bin would yield a null signal when cross-correlated to another redshift bin. For low-$z$ bins, the signal remains minimal, except in adjacent bins due to photo-$z$ uncertainties sometimes spanning a large redshift range. For redshift bins at $z \ge 8.0$, the correlation signal is more pronounced, particularly with redshift bins at $2.5 \le z \le 5$, yet still at least $3 \times$ lower than the auto-correlation signal. This suggests that despite inherent photo-$z$ uncertainties, our selection across large redshift bins remains robust. Another argument supporting our sample selection is that we consistently observe a non-null signal in the auto-correlation function at $z > 10$, even with different SED assumptions. If our sample were solely contaminated by low-$z$ galaxies, we would not expect to detect any significant clustering. To test this, we computed the auto-correlation by randomly selecting galaxies within the redshift range $0 < z < 5$, and found a constant clustering signal around $w(\theta) \sim 1$, which contrasts with the behavior observed in the $z > 10$ sample, further indicating the presence of high-redshift galaxies in our sample. Nonetheless, systematic effects may persist in the photometric redshift estimates, influenced, for example, by choices in dust attenuation models used in the SED fitting code, particularly affecting galaxies at $z > 6$.

The issue of completeness arises when considering galaxies at high redshifts and low stellar masses. By implementing a magnitude threshold based on comparisons with the deeper JWST survey PRIMER, we ensured that at least 80\% galaxies below this limit were detected. However, our selection is probably biased towards brighter and more compact galaxies at these high redshifts. Although the fraction of satellite galaxies is expected to be low at $z \ge 8$ and for the masses we observe, faint low-mass satellites could still be missed by our detection, potentially leading to an underestimation of clustering at small scales. Having more satellites would lead to lower estimates of halo masses required to host central galaxies and potentially higher the SHMR. Conversely, for $z < 8$, we have confidence in the completeness of our survey, suggesting that the observed trends in the SHMR evolution with redshift (particularly at $z > 3$) are unlikely to be significantly altered by completeness issues.

We also examined the impact of objects with AGNs dominating their rest UV/optical emission and what has been called Little Red Dots \citep{Matthee24_LRD} at high redshifts on our clustering measurements. These objects were identified as described in Sect.~\ref{subsubsec:highzselection}, and we compared the $z \ge 4$ auto-correlation functions with and without them. We found no significant change in the amplitude or slope of our clustering measurements. Although AGNs can be major contaminants in the stellar mass function due to their SEDs being degenerate with those of high-mass or highly dust-obscured galaxies (see \citealt{Shuntov2025} or e.g., \citealt{Schaerer&deBarros2009, Barro2024}), their presence does not appear to affect the clustering. This insinuates that the clustering of our sample is not only driven by AGN-dominated objects at high-$z$. However, a robust AGN selection and analysis of their clustering properties across all redshifts are outside the scope of this work.

\subsubsection{Stellar mass estimates} \label{subsubsec:limits-stellarmass}

Estimating stellar masses heavily relies on the choice of the initial mass function (IMF), which remains unknown at high redshifts. In this study, we adopt the \cite{Chabrier03} IMF; however, a top-heavy IMF would lead to lower stellar mass estimates and number densities \citep[see e.g.][]{Cameron2024}. While this modification would not change the clustering signal, it would result in a lower SHMR and a reduced star formation efficiency. According to predictions from the ASTREUS simulation \citep{Cueto2024}, their evolving IMF (top-heavy at high-$z$) leads to a decrease in stellar masses of approximately $0.6$\,dex at $z = 8$ and $0.7$\,dex at $z = 12$ for the halo mass range derived in this study. Even accounting for this adjustment, our SHMR $M_{\star, \rm th}/M_{\rm h}$ at $z \geq 8$ remains $0.5$\,dex higher than the $z = 1$ ratio at a same halo mass. Moreover, \cite{Shuntov2025} shows that the ASTRAEUS simulation indicates excessively high number densities at low masses compared to the observational stellar mass function in COSMOS-Web. Further investigation is needed to determine the shape of the primordial IMF, such as studying the metallicity of high-redshift galaxies through a spectroscopic survey, since a higher number of massive stars would contribute more to the metal enrichment of these galaxies \citep[e.g.][]{Pallottini24}. 

Estimating the stellar mass of galaxies at $z \ge 8$ can also be biased due to limited filters for detecting their SED and Balmer breaks. Studies have shown that including MIRI data leads to lower stellar mass estimates compared to NIRCam-only, as the Balmer break for $z \ge 12$ galaxies falls beyond the F444W filter, and MIRI can probe nebular emission lines, helping to rule out older star populations and better constrain dust models \citep[e.g.,][]{Song2023, Papovich2023}. In our sample, around 30\% of galaxies at $z \ge 5$ have MIRI data, but otherwise \cite{Papovich2023} finds that mass over-estimations are typically $< 0.25$ dex at $z < 6$ and $0.37$ dex at $6 \le z < 9$. Dust attenuation models can also introduce mass deviations, up to 0.35 dex at $z = 7 -8 $ according to \cite{Markov2023}. Despite these uncertainties, stellar masses at $z \ge 8$, after correction for these factors, give a SHMR that remains 1 dex higher than the ones at $z < 3$. Lastly, we note that the transition from \cite{Planck18} cosmological model to a standard $\Lambda$CDM model (with $H_0 = 70\,\textrm{km s}^{-1}\textrm{ Mpc}^{-1}, \Omega_{m,0}=0.3, \Omega_{\Lambda,0}=0.7)$, which is used in \LePhare, only changes masses by $-0.030$ dex at $z = 0$ and $-0.0175$ dex at $z = 10$.

\section{Conclusion} \label{sec:conclusion}

This work explores the angular auto-correlation function of mass-limited samples of galaxies within the COSMOS-Web survey, field 0.54\,deg$^2$ observed with JWST. It spans a range of photometric redshifts from $z = 0.1$ to $z \sim 12$, making it the first measurement of clustering at $z > 10$ with a mass-limited sample. It highlights the unique capabilities of COSMOS-Web in examining large-scale structures and the environments of high-redshift galaxies, being the JWST field least affected by cosmic variance compared to other small area current fields. The aim of this research was to examine the galaxy-halo connection consistently across $13.4$\,Gyr of cosmic history, using HOD modeling. Clustering remains the most robust method for estimating halo masses at these high redshifts, which is crucial for establishing the link between galaxies and their host halos and for understanding their role in regulating star formation. The main conclusions of this work are:
\begin{itemize}
    \setlength\itemsep{0.5em}

    \item Clustering measurements from $z = 0.1$ to $z \sim 12$ follow the typical trend, where more massive galaxies exhibit higher clustering amplitudes. The clustering signal at $z \ge 8$ is strong, as expected since the observed massive galaxies are likely hosted in massive, early-forming halos; and is in agreement with other literature results.
    
    \item The clustering at $z \ge 2.5$ is best fitted with a HOD model incorporating a non-linear scale-dependent halo bias, which boosts the correlation signal at quasi-linear scales ($r \sim$ 10-100 kpc). This adjustment accounts for the rarity of halos capable of hosting the massive and bright galaxies that we observe, which are highly biased toward non-linear fluctuations of the dark matter field.
    
    \item From HOD fitting on clustering, we estimate the halo masses to be $\log(M_{\rm h,min} / M_\odot) \simeq 10.99^{+ 0.08}_{- 0.13}$ for galaxies at $8.0 \le z < 10.5$ with $M_\star \ge 10^{9} M_\odot$; $\log(M_{\rm h,min} / M_\odot) \simeq 10.54^{+ 0.26}_{- 0.14}$ for galaxies at $10.5 \le z < 14$ with $M_\star \ge 10^{8.85} M_\odot$. This suggests a higher integrated star formation efficiency at high redshift compared to lower redshift, with values exceeding $\varepsilon_{\rm SF} > 12-36\%$ according to the adopted definition. Such high efficiencies can be explained, for instance, by the feedback-free burst model, which is consistent with the halo mass regime of our estimates.
    
    \item The evolution of the galaxy bias with redshift and with stellar mass thresholds favors the scenario of high star formation efficiency in massive halos instead of high $M_{\rm UV} - M_{\rm h}$ scatter as the primary driver of the abundance of high-$z$ galaxies observed with JWST, though further modeling of stochasticity as a function of stellar mass is needed for conclusive results.
    
    \item The stellar-to-halo mass relationship (SHMR) evolves significantly with redshift in the range $10.5 \le \log(M_{\rm h,min}/M_\odot) \le 12.5$, decreasing by $\sim 1$\,dex from $z \sim 10$ to $z \sim 3$. This decline reflects reduced star formation efficiency over time, potentially driven by more efficient feedback processes (e.g., AGN growth, stellar or radiative feedback) while halos accrete matter exponentially. At $z = 2-3$, dark matter growth reaches a plateau while star formation might continue thanks to the gas stored in halos, causing the SHMR to rise again.
    
    \item Observations, models, and simulations reveal significant discrepancies in the evolution of the SHMR with redshift. Some observations suggest a decline in the SHMR at $z > 4$, while others indicate no evolution or an increase. Hydrodynamical simulations often fail to capture the high SHMR or its redshift trends, emphasizing the need for improved calibrations or treatment of physical processes across cosmic time. Semi-empirical models like EMERGE and \textsc{UniverseMachine}, calibrated on observations up to $z = 10$, successfully reproduce trends similar to ours by linking star formation efficiencies to halo mass, halo accretion rate, and redshift.
\end{itemize}
Our measurements of clustering and SHMR can be found in tabulated form at \url{https://github.com/LouisePaquereau/GalClustering_COSMOS-Web_Paquereau2025}.

Building on this study, a future project will aim to separate the COSMOS-Web sample into star-forming and quiescent galaxies across various redshift and mass ranges. By performing auto-correlation and cross-correlation measurements up to $z = 5$, we will explore how the environment influences star formation and quenching processes. Understanding these mechanisms could provide insights into the evolution of the massive high-redshift galaxies discussed in this work, which may be the progenitors of the quiescent galaxies seen at $z \sim 3 - 5$.

A future spectroscopic survey, e.g. the COSMOS-3D (JWST Cycle 3 Program GO \#5893), will enhance mass and redshift estimates for high-$z$ galaxies in COSMOS-Web. Spectroscopic data will reduce low-$z$ contamination and uncertainties in our clustering and SHMR measurements, leading to more reliable galaxy properties and connection to their host dark matter halos.


\begin{acknowledgements}
We thank the anonymous referee for their careful reading and helpful feedback.
LP acknowledges the thesis funding from the Centre National d’Etudes Spatiales (CNES) and the Ecole Doctorale Astronomie et Astrophysique d'Ile-de-France. The authors acknowledge the funding of the French Agence Nationale de la Recherche for the project iMAGE (grant ANR-22-CE31-0007).
This work was made possible by utilizing the CANDIDE cluster at the Institut d’Astrophysique de Paris. The cluster was funded through grants from the PNCG, CNES, DIM-ACAV, the Euclid Consortium, and the Danish National Research Foundation Cosmic Dawn Center (DNRF140). It is maintained by Stephane Rouberol. 
The authors acknowledge the contributions of the COSMOS collaboration. The French contingent of the COSMOS team is partly supported by the Centre National d’Etudes Spatiales (CNES). Support for this work was provided by NASA through grant JWST-GO-01727 awarded by the Space Telescope Science Institute, which is operated by the Association of Universities for Research in Astronomy, Inc., under NASA contract NAS 5-26555. This work is based in part on observations made with the NASA/ESA/CSA James Webb Space Telescope. The data were obtained from the Mikulski Archive for Space Telescopes at the Space Telescope Science Institute, which is operated by the Association of Universities for Research in Astronomy, Inc., under NASA contract NAS 5-03127 for JWST. The HST-COSMOS program was supported through NASA grant HSTGO-09822. More information on the COSMOS survey is available at \url{https://cosmos.astro.caltech.edu}. 
Based in part on observations collected at the European Southern Observatory under ESO programs 179.A-2005, 198.A-2003, 104.A-0643, 110.25A2 and 284.A-5026 and on data obtained from the ESO Science Archive Facility with DOI https://doi.org/10.18727/archive/52, and on data products produced by CANDIDE and the Cambridge Astronomy Survey Unit on behalf of the UltraVISTA consortium.
\end{acknowledgements}


\bibliographystyle{aa}
\bibliography{references}

\begin{appendix}

\section{Redshift estimates} \label{app:zPDFvszchi2}

\begin{figure}[h!]
    \centering
    \includegraphics[scale=0.55]{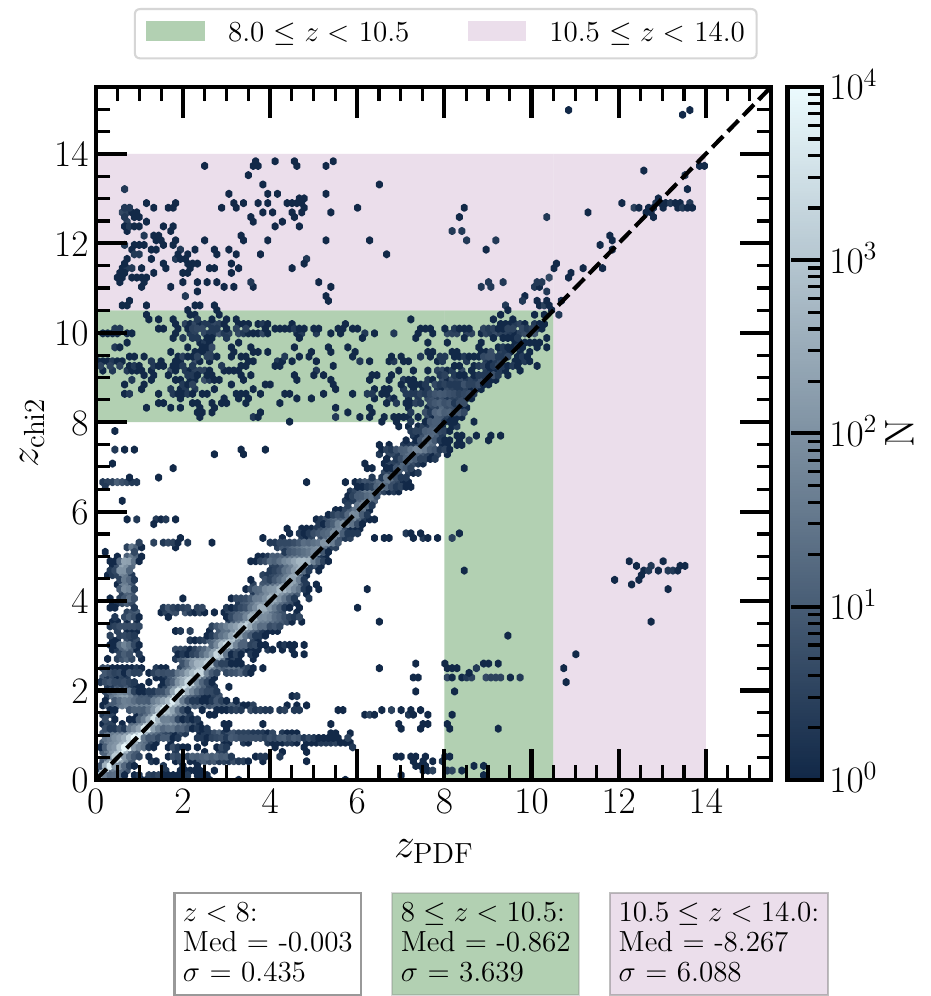}
    \caption{Comparison between \LePhare two photo-$z$ estimates, one being the median of the probability distribution function, $z_{\rm PDF}$, and the other the redshift of minimum $\chi^2$, $z_{\rm chi2}$. All cleaning steps are applied here except for the $z \ge 8$ extensive selection described in Sect.~\ref{subsubsec:highzselection}. The median and standard deviation of the difference $z_{\rm PDF} - z_{\rm chi2}$ is shown.}
    \label{fig:highz_zPDFvszchi2}
\end{figure}

Figure~\ref{fig:highz_zPDFvszchi2} compares redshift estimates for the galaxy sample cleaned of masked regions, hot pixels, and magnitude cuts (Sect.~\ref{subsubsec:completeness}), but before applying mass completeness limits and high-$z$ cleaning (described in Sect.~\ref{subsubsec:highzselection}). Discrepancies between $z_{\rm PDF}$ and $z_{\rm chi2}$ arise due to their definitions: $z_{\rm PDF}$ is the median of the probability density function, summing probabilities from all SED templates at a given redshift, whereas $z_{\rm chi2}$ is the redshift that minimizes the $\chi^2$ from the fitting procedure. At high $\chi^2$ value for a single template can result in a lower $z_{\rm PDF}$ if the summed probabilities favor lower redshifts, and vice versa, explaining these differences. The dispersion is largely reduced after applying the criteria on the width of PDFs at $z < 8$ (removal of objects having at least 70\% of their PDF outside $z \pm \Delta z_{\rm bin}$) and mass completeness limits. A $z \ge 8$, we select objects based on $z_{\rm PDF}$ and $z_{\rm chi2}$, as we consider that the limited number of templates at these redshifts biases $z_{\rm PDF}$ toward lower redshift values. 

\section{Number counts and HOD best-fit parameters for each galaxy sample} \label{app:tableHODparams}

Table~\ref{app-tab:HODparams} shows the number of sources, number densities, best-fit HOD parameters and model-derived quantities, for each redshift and stellar mass-limited galaxy sample. 

\begin{table*}[h!]
    \centering
    \fontsize{6.9pt}{6.9pt}\selectfont
    \renewcommand{\arraystretch}{1.51}
    \begin{tabular}{llllllllll}
    \hline
    z bin               & $\log (M_\star^{\textrm{th}}/M_\odot)$ & $N_{\rm g}$ & $n_{\rm g}$ [Mpc$^{-3}$] & $\log (M_{\rm h,min}/M_\odot)$ & $\log (M_{\rm h,1}/M_\odot)$ &  $\alpha$ & $\log (M_{\rm h,0}/M_\odot)$ & $f_{\rm sat}$ & $b_{\rm g}$ \\ \hline \hline
    $0.1 \le z < 0.6$   & 7.5  & 28892 & $4.71 \times 10^{-2} \pm 6.39 \times 10^{-3}$ & $10.62^{+0.08}_{-0.09}$ & $11.89^{+0.12}_{-0.15}$ & $0.89^{+0.03}_{-0.03}$ & $11.33^{+2.39}_{-2.41}$ & $0.23^{+0.01}_{-0.23}$ & $0.93^{+1.33}_{-0.01}$ \\
                        & 8.0  & 15991 & $2.61 \times 10^{-2} \pm 4.09 \times 10^{-3}$ & $10.94^{+0.04}_{-0.09}$ & $12.25^{+0.08}_{-0.14}$ & $0.88^{+0.04}_{-0.04}$ & $11.61^{+2.36}_{-2.4}$ & $0.19^{+0.02}_{-0.01}$ & $0.95^{+0.01}_{-0.01}$ \\
                        & 8.5  & 9102  & $1.49 \times 10^{-2} \pm 2.68 \times 10^{-3}$ & $11.16^{+0.08}_{-0.07}$ & $12.35^{+0.12}_{-0.12}$ & $0.77^{+0.06}_{-0.07}$ & $11.68^{+2.39}_{-2.39}$ & $0.21^{+0.02}_{-0.02}$ & $0.96^{+0.02}_{-0.02}$ \\
                        & 9.0  & 5057  & $8.25 \times 10^{-3} \pm 1.70 \times 10^{-3}$ & $11.45^{+0.1}_{-0.13}$ & $12.87^{+0.16}_{-0.18}$ & $0.96^{+0.1}_{-0.1}$ & $12.08^{+2.42}_{-2.43}$ & $0.14^{+0.02}_{-0.02}$ & $1.04^{+0.03}_{-0.03}$ \\
                        & 9.5  & 2798  & $4.57 \times 10^{-3} \pm 1.07 \times 10^{-3}$ & $11.7^{+0.13}_{-0.12}$ & $13.15^{+0.18}_{-0.15}$ & $0.97^{+0.09}_{-0.07}$ & $12.3^{+2.44}_{-2.41}$ & $0.12^{+0.02}_{-0.02}$ & $1.08^{+0.03}_{-0.03}$ \\
                        & 10.0 & 1515  & $2.47 \times 10^{-3} \pm 6.58 \times 10^{-4}$ & $11.93^{+0.11}_{-0.1}$ & $13.45^{+0.25}_{-0.2}$ & $0.45^{+0.22}_{-0.25}$ & $12.52^{+2.49}_{-2.45}$ & $0.11^{+0.02}_{-0.02}$ & $1.08^{+0.05}_{-0.04}$ \\
                        & 10.5 & 638   & $1.04 \times 10^{-3} \pm 3.14 \times 10^{-4}$ & $12.34^{+0.18}_{-0.11}$ & $13.9^{+0.43}_{-0.15}$ & $0.6^{+0.14}_{-0.47}$ & $12.87^{+2.63}_{-2.42}$ & $0.08^{+0.02}_{-0.02}$ & $1.25^{+0.08}_{-0.06}$ \\ \hline
    $0.6 \le z < 1.0$   & 7.5  & 55842 & $3.87 \times 10^{-2} \pm 4.93 \times 10^{-3}$ & $10.76^{+0.08}_{-0.07}$ & $12.14^{+0.1}_{-0.11}$ & $0.92^{+0.03}_{-0.04}$ & $11.53^{+2.38}_{-2.38}$ & $0.15^{+0.01}_{-0.01}$ & $1.02^{+0.02}_{-0.01}$ \\
                        & 8.0  & 37701 & $2.62 \times 10^{-2} \pm 3.87 \times 10^{-3}$ & $10.96^{+0.07}_{-0.07}$ & $12.18^{+0.11}_{-0.11}$ & $0.88^{+0.03}_{-0.03}$ & $11.56^{+2.38}_{-2.38}$ & $0.19^{+0.01}_{-0.02}$ & $1.09^{+0.02}_{-0.01}$ \\
                        & 8.5  & 24301 & $1.69 \times 10^{-2} \pm 2.88 \times 10^{-3}$ & $11.12^{+0.09}_{-0.06}$ & $12.28^{+0.15}_{-0.09}$ & $0.83^{+0.02}_{-0.02}$ & $11.63^{+2.41}_{-2.37}$ & $0.2^{+0.02}_{-0.02}$ & $1.12^{+0.02}_{-0.01}$ \\
                        & 9.0  & 14667 & $1.02 \times 10^{-2} \pm 1.99 \times 10^{-3}$ & $11.34^{+0.14}_{-0.09}$ & $12.53^{+0.22}_{-0.16}$ & $0.79^{+0.1}_{-0.1}$ & $11.82^{+2.47}_{-2.42}$ & $0.17^{+0.02}_{-0.04}$ & $1.16^{+0.06}_{-0.03}$ \\
                        & 9.5  & 8316  & $5.77 \times 10^{-3} \pm 1.29 \times 10^{-3}$ & $11.65^{+0.12}_{-0.08}$ & $12.79^{+0.21}_{-0.12}$ & $0.77^{+0.07}_{-0.09}$ & $12.02^{+2.46}_{-2.39}$ & $0.17^{+0.02}_{-0.03}$ & $1.27^{+0.04}_{-0.03}$ \\
                        & 10.0 & 4408  & $3.06 \times 10^{-3} \pm 7.76 \times 10^{-4}$ & $11.9^{+0.15}_{-0.12}$ & $13.04^{+0.19}_{-0.19}$ & $1.09^{+0.08}_{-0.06}$ & $12.21^{+2.45}_{-2.44}$ & $0.17^{+0.03}_{-0.04}$ & $1.46^{+0.07}_{-0.04}$ \\
                        & 10.5 & 1777  & $1.23 \times 10^{-3} \pm 3.55 \times 10^{-4}$ & $12.26^{+0.2}_{-0.14}$ & $13.46^{+0.24}_{-0.17}$ & $1.35^{+0.15}_{-0.11}$ & $12.53^{+2.48}_{-2.43}$ & $0.12^{+0.03}_{-0.06}$ & $1.66^{+0.36}_{-0.06}$ \\ \hline
    $1.0 \le z < 1.5$   & 8.0  & 35621 & $1.34 \times 10^{-2} \pm 1.89 \times 10^{-3}$ & $11.21^{+0.05}_{-0.05}$ & $12.34^{+0.08}_{-0.07}$ & $0.93^{+0.03}_{-0.02}$ & $11.68^{+2.36}_{-2.35}$ & $0.17^{+0.01}_{-0.01}$ & $1.37^{+0.02}_{-0.02}$ \\
                        & 8.5  & 24256 & $9.12 \times 10^{-3} \pm 1.49 \times 10^{-3}$ & $11.41^{+0.09}_{-0.08}$ & $12.62^{+0.14}_{-0.12}$ & $0.85^{+0.06}_{-0.06}$ & $11.89^{+2.4}_{-2.39}$ & $0.13^{+0.01}_{-0.03}$ & $1.41^{+0.03}_{-0.02}$ \\
                        & 9.0  & 14206 & $5.34 \times 10^{-3} \pm 1.00 \times 10^{-3}$ & $11.47^{+0.05}_{-0.03}$ & $12.57^{+0.13}_{-0.03}$ & $0.4^{+0.06}_{-0.12}$ & $11.85^{+2.4}_{-2.32}$ & $0.18^{+0.01}_{-0.01}$ & $1.4^{+0.02}_{-0.02}$ \\
                        & 9.5  & 7584  & $2.85 \times 10^{-3} \pm 6.13 \times 10^{-4}$ & $11.85^{+0.08}_{-0.05}$ & $12.87^{+0.22}_{-0.11}$ & $0.64^{+0.16}_{-0.17}$ & $12.08^{+2.47}_{-2.38}$ & $0.18^{+0.02}_{-0.02}$ & $1.65^{+0.04}_{-0.04}$ \\
                        & 10.0 & 3801  & $1.43 \times 10^{-3} \pm 3.50 \times 10^{-4}$ & $12.14^{+0.11}_{-0.08}$ & $13.25^{+0.31}_{-0.14}$ & $0.9^{+0.15}_{-0.26}$ & $12.37^{+2.54}_{-2.41}$ & $0.12^{+0.03}_{-0.02}$ & $1.88^{+0.07}_{-0.06}$ \\
                        & 10.5 & 1624  & $6.10 \times 10^{-4} \pm 1.70 \times 10^{-4}$ & $12.42^{+0.12}_{-0.09}$ & $13.64^{+0.38}_{-0.12}$ & $1.07^{+0.3}_{-0.49}$ & $12.66^{+2.59}_{-2.39}$ & $0.07^{+0.02}_{-0.02}$ & $2.1^{+0.13}_{-0.09}$ \\ \hline
    $1.5 \le z < 2.0$   & 8.0  & 41786 & $1.33 \times 10^{-2} \pm 1.80 \times 10^{-3}$ & $11.13^{+0.05}_{-0.04}$ & $12.18^{+0.08}_{-0.06}$ & $0.9^{+0.03}_{-0.03}$ & $11.55^{+2.36}_{-2.35}$ & $0.16^{+0.01}_{-0.01}$ & $1.6^{+0.02}_{-0.02}$ \\
                        & 8.5  & 26523 & $8.43 \times 10^{-3} \pm 1.33 \times 10^{-3}$ & $11.35^{+0.06}_{-0.05}$ & $12.42^{+0.09}_{-0.07}$ & $0.93^{+0.05}_{-0.05}$ & $11.74^{+2.37}_{-2.36}$ & $0.14^{+0.01}_{-0.01}$ & $1.72^{+0.03}_{-0.03}$ \\
                        & 9.0  & 14715 & $4.68 \times 10^{-3} \pm 8.50 \times 10^{-4}$ & $11.69^{+0.05}_{-0.1}$ & $12.87^{+0.22}_{-0.16}$ & $0.63^{+0.28}_{-0.27}$ & $12.08^{+2.46}_{-2.42}$ & $0.11^{+0.02}_{-0.02}$ & $1.91^{+0.06}_{-0.05}$ \\
                        & 9.5  & 7511  & $2.39 \times 10^{-3} \pm 4.98 \times 10^{-4}$ & $11.84^{+0.09}_{-0.1}$ & $12.92^{+0.14}_{-0.11}$ & $1.2^{+0.07}_{-0.06}$ & $12.12^{+2.4}_{-2.38}$ & $0.11^{+0.02}_{-0.03}$ & $2.11^{+0.12}_{-0.06}$ \\
                        & 10.0 & 3547  & $1.13 \times 10^{-3} \pm 2.69 \times 10^{-4}$ & $12.21^{+0.06}_{-0.09}$ & $13.59^{+0.12}_{-0.11}$ & $0.67^{+0.12}_{-0.16}$ & $12.63^{+2.39}_{-2.38}$ & $0.05^{+0.01}_{-0.01}$ & $2.4^{+0.09}_{-0.08}$ \\
                        & 10.5 & 1502  & $4.78 \times 10^{-4} \pm 1.30 \times 10^{-4}$ & $12.11^{+0.03}_{-0.03}$ & $13.47^{+0.04}_{-0.07}$ & $0.56^{+0.09}_{-0.08}$ & $12.54^{+2.33}_{-2.35}$ & $0.07^{+0.01}_{-0.01}$ & $2.3^{+0.04}_{-0.03}$ \\ \hline
    $2.0 \le z < 2.5$   & 8.0  & 28397 & $8.59 \times 10^{-3} \pm 1.14 \times 10^{-3}$ & $11.16^{+0.02}_{-0.02}$ & $12.15^{+0.04}_{-0.04}$ & $0.67^{+0.06}_{-0.05}$ & $11.53^{+2.33}_{-2.33}$ & $0.15^{+0.01}_{-0.01}$ & $1.9^{+0.02}_{-0.02}$ \\
                        & 8.5  & 19093 & $5.77 \times 10^{-3} \pm 8.93 \times 10^{-4}$ & $11.45^{+0.08}_{-0.04}$ & $12.41^{+0.18}_{-0.06}$ & $0.69^{+0.05}_{-0.08}$ & $11.73^{+2.43}_{-2.34}$ & $0.15^{+0.02}_{-0.02}$ & $2.16^{+0.05}_{-0.04}$ \\
                        & 9.0  & 10540 & $3.19 \times 10^{-3} \pm 5.70 \times 10^{-4}$ & $11.64^{+0.07}_{-0.05}$ & $12.57^{+0.13}_{-0.08}$ & $0.96^{+0.05}_{-0.05}$ & $11.85^{+2.4}_{-2.36}$ & $0.14^{+0.02}_{-0.02}$ & $2.39^{+0.06}_{-0.05}$ \\
                        & 9.5  & 5385  & $1.63 \times 10^{-3} \pm 3.35 \times 10^{-4}$ & $11.84^{+0.07}_{-0.04}$ & $12.76^{+0.16}_{-0.06}$ & $0.94^{+0.11}_{-0.12}$ & $12.0^{+2.42}_{-2.35}$ & $0.14^{+0.02}_{-0.02}$ & $2.62^{+0.07}_{-0.05}$ \\
                        & 10.0 & 2460  & $7.44 \times 10^{-4} \pm 1.75 \times 10^{-4}$ & $12.01^{+0.05}_{-0.05}$ & $12.99^{+0.07}_{-0.06}$ & $1.33^{+0.09}_{-0.07}$ & $12.17^{+2.35}_{-2.35}$ & $0.09^{+0.01}_{-0.01}$ & $2.85^{+0.06}_{-0.05}$ \\
                        & 10.5 & 910   & $2.75 \times 10^{-4} \pm 7.42 \times 10^{-5}$ & $12.4^{+0.09}_{-0.05}$ & $13.48^{+0.25}_{-0.14}$ & $1.28^{+0.26}_{-0.47}$ & $12.54^{+2.49}_{-2.41}$ & $0.04^{+0.01}_{-0.01}$ & $3.41^{+0.14}_{-0.12}$ \\ \hline
    $2.5 \le z < 3.0$   & 8.0  & 17537 & $5.32 \times 10^{-3} \pm 7.07 \times 10^{-4}$ & $11.22^{+0.02}_{-0.03}$ & $12.28^{+0.04}_{-0.05}$ & $0.52^{+0.11}_{-0.06}$ & $11.63^{+2.33}_{-2.34}$ & $0.11^{+0.01}_{-0.01}$ & $2.32^{+0.02}_{-0.02}$ \\
                        & 8.5  & 12416 & $3.77 \times 10^{-3} \pm 5.84 \times 10^{-4}$ & $11.47^{+0.08}_{-0.06}$ & $12.65^{+0.17}_{-0.06}$ & $1.1^{+0.17}_{-0.26}$ & $11.91^{+2.43}_{-2.35}$ & $0.05^{+0.01}_{-0.01}$ & $2.6^{+0.09}_{-0.07}$ \\
                        & 9.0  & 6705  & $2.03 \times 10^{-3} \pm 3.65 \times 10^{-4}$ & $11.66^{+0.07}_{-0.06}$ & $12.64^{+0.11}_{-0.07}$ & $1.26^{+0.03}_{-0.04}$ & $11.9^{+2.38}_{-2.35}$ & $0.08^{+0.01}_{-0.01}$ & $2.92^{+0.09}_{-0.07}$ \\
                        & 9.5  & 3321  & $1.01 \times 10^{-3} \pm 2.09 \times 10^{-4}$ & $11.82^{+0.05}_{-0.05}$ & $13.1^{+0.13}_{-0.16}$ & $0.68^{+0.3}_{-0.29}$ & $12.25^{+2.4}_{-2.42}$ & $0.04^{+0.01}_{-0.01}$ & $3.08^{+0.08}_{-0.07}$ \\
                        & 10.0 & 1378  & $4.18 \times 10^{-4} \pm 9.95 \times 10^{-5}$ & $12.11^{+0.07}_{-0.06}$ & $13.17^{+0.21}_{-0.12}$ & $0.96^{+0.34}_{-0.44}$ & $12.31^{+2.46}_{-2.39}$ & $0.05^{+0.01}_{-0.01}$ & $3.62^{+0.14}_{-0.11}$ \\
                        & 10.5 & 450   & $1.37 \times 10^{-4} \pm 3.74 \times 10^{-5}$ & $12.4^{+0.08}_{-0.07}$ & $13.64^{+0.26}_{-0.2}$ & $1.15^{+0.63}_{-0.49}$ & $12.67^{+2.5}_{-2.45}$ & $0.02^{+0.01}_{-0.01}$ & $4.23^{+0.24}_{-0.17}$ \\ \hline
    $3.0 \le z < 4.0$   & 8.5  & 16346 & $2.61 \times 10^{-3} \pm 4.21 \times 10^{-4}$ & $11.26^{+0.04}_{-0.02}$ & $12.16^{+0.11}_{-0.02}$ & $0.51^{+0.07}_{-0.1}$ & $11.54^{+2.38}_{-2.31}$ & $0.13^{+0.01}_{-0.01}$ & $3.13^{+0.05}_{-0.04}$ \\
                        & 9.0  & 8393  & $1.34 \times 10^{-3} \pm 2.51 \times 10^{-4}$ & $11.46^{+0.03}_{-0.03}$ & $12.49^{+0.08}_{-0.06}$ & $0.53^{+0.1}_{-0.12}$ & $11.79^{+2.36}_{-2.35}$ & $0.09^{+0.01}_{-0.01}$ & $3.43^{+0.05}_{-0.05}$ \\
                        & 9.5  & 4134  & $6.60 \times 10^{-4} \pm 1.42 \times 10^{-4}$ & $11.74^{+0.05}_{-0.04}$ & $12.69^{+0.14}_{-0.1}$ & $0.95^{+0.27}_{-0.29}$ & $11.95^{+2.41}_{-2.38}$ & $0.07^{+0.01}_{-0.01}$ & $3.98^{+0.11}_{-0.09}$ \\
                        & 10.0 & 1920  & $3.07 \times 10^{-4} \pm 7.60 \times 10^{-5}$ & $11.88^{+0.03}_{-0.04}$ & $13.0^{+0.09}_{-0.18}$ & $0.67^{+0.32}_{-0.25}$ & $12.18^{+2.36}_{-2.43}$ & $0.05^{+0.01}_{-0.01}$ & $4.28^{+0.08}_{-0.07}$ \\
                        & 10.5 & 702   & $1.12 \times 10^{-4} \pm 3.19 \times 10^{-5}$ & $12.16^{+0.06}_{-0.05}$ & $13.35^{+0.26}_{-0.19}$ & $1.06^{+0.42}_{-0.52}$ & $12.45^{+2.5}_{-2.45}$ & $0.02^{+0.01}_{-0.01}$ & $4.97^{+0.16}_{-0.15}$ \\ \hline  
    $4.0 \le z < 5.0$   & 8.5  & 10677 & $1.89 \times 10^{-3} \pm 3.44 \times 10^{-4}$ & $11.15^{+0.02}_{-0.02}$ & $12.03^{+0.04}_{-0.04}$ & $0.57^{+0.09}_{-0.11}$ & $11.44^{+2.33}_{-2.33}$ & $0.1^{+0.01}_{-0.01}$ & $4.06^{+0.04}_{-0.04}$ \\
                        & 9.0  & 4360  & $7.70 \times 10^{-4} \pm 1.62 \times 10^{-4}$ & $11.46^{+0.05}_{-0.03}$ & $12.3^{+0.23}_{-0.09}$ & $1.05^{+0.24}_{-0.51}$ & $11.65^{+2.48}_{-2.37}$ & $0.06^{+0.01}_{-0.01}$ & $4.79^{+0.1}_{-0.1}$ \\
                        & 9.5  & 1397  & $2.47 \times 10^{-4} \pm 6.01 \times 10^{-5}$ & $11.74^{+0.04}_{-0.06}$ & $12.82^{+0.16}_{-0.22}$ & $0.93^{+0.58}_{-0.62}$ & $12.05^{+2.42}_{-2.47}$ & $0.02^{+0.01}_{-0.01}$ & $5.52^{+0.18}_{-0.14}$ \\
                        & 10.0 & 360   & $6.36 \times 10^{-5} \pm 1.80 \times 10^{-5}$ & $12.05^{+0.04}_{-0.08}$ & $13.08^{+0.47}_{-0.1}$ & $1.59^{+0.21}_{-1.0}$ & $12.24^{+2.65}_{-2.38}$ & $0.01^{+0.01}_{-0.0}$ & $6.61^{+0.26}_{-0.21}$ \\ \hline
    $5.0 \le z < 6.0$   & 8.5  & 2228  & $4.40 \times 10^{-4} \pm 9.65 \times 10^{-5}$ & $11.37^{+0.05}_{-0.05}$ & $12.27^{+0.2}_{-0.11}$ & $1.18^{+0.27}_{-0.74}$ & $11.63^{+2.45}_{-2.38}$ & $0.03^{+0.01}_{-0.01}$ & $6.04^{+0.16}_{-0.13}$ \\
                        & 9.0  & 1154  & $2.28 \times 10^{-4} \pm 5.78 \times 10^{-5}$ & $11.45^{+0.05}_{-0.04}$ & $12.29^{+0.08}_{-0.06}$ & $1.94^{+0.06}_{-0.06}$ & $11.64^{+2.36}_{-2.35}$ & $0.02^{+0.0}_{-0.0}$ & $6.34^{+0.17}_{-0.14}$ \\
                        & 9.25 & 681   & $1.35 \times 10^{-4} \pm 3.66 \times 10^{-5}$ & $11.61^{+0.08}_{-0.05}$ & $12.41^{+0.42}_{-0.1}$ & $1.82^{+0.18}_{-1.06}$ & $11.73^{+2.62}_{-2.38}$ & $0.02^{+0.01}_{-0.01}$ & $6.96^{+0.34}_{-0.23}$ \\
                        & 9.5  & 362   & $7.15 \times 10^{-5} \pm 2.10 \times 10^{-5}$ & $11.76^{+0.07}_{-0.08}$ & $12.91^{+0.26}_{-0.31}$ & $0.83^{+0.54}_{-0.73}$ & $12.11^{+2.5}_{-2.53}$ & $0.01^{+0.01}_{-0.01}$ & $7.55^{+0.41}_{-0.29}$ \\ \hline
    $6.0 \le z < 8.0$   & 8.70 & 1926  & $2.24 \times 10^{-4} \pm 7.26 \times 10^{-5}$ & $11.1^{+0.03}_{-0.06}$ & $11.83^{+0.16}_{-0.06}$ & $1.75^{+0.25}_{-0.64}$ & $11.29^{+2.42}_{-2.35}$ & $0.02^{+0.0}_{-0.0}$ & $7.56^{+0.2}_{-0.18}$ \\
                        & 9.0  & 1041  & $1.21 \times 10^{-4} \pm 4.27 \times 10^{-5}$ & $11.28^{+0.07}_{-0.11}$ & $12.67^{+0.06}_{-0.52}$ & $0.35^{+1.09}_{-0.25}$ & $11.93^{+2.34}_{-2.69}$ & $0.0^{+0.0}_{-0.0}$ & $8.34^{+0.57}_{-0.34}$ \\
                        & 9.30 & 487   & $5.66 \times 10^{-5} \pm 2.17 \times 10^{-5}$ & $11.37^{+0.07}_{-0.06}$ & $12.35^{+0.33}_{-0.12}$ & $1.05^{+0.36}_{-0.79}$ & $11.69^{+2.55}_{-2.39}$ & $0.01^{+0.01}_{-0.0}$ & $8.78^{+0.42}_{-0.3}$ \\ \hline
    $8.0 \le z < 10.5$  & 8.70 & 494   & $5.77 \times 10^{-5} \pm 1.00 \times 10^{-5}$ & $10.94^{+0.27}_{-0.13}$ & $11.91^{+0.59}_{-0.21}$ & $1.51^{+0.49}_{-0.77}$ & $11.35^{+2.75}_{-2.46}$ & $0.0^{+0.0}_{-0.0}$ & $10.93^{+2.15}_{-0.93}$ \\
    "conservative"        & 9.0  & 226   & $2.64 \times 10^{-5} \pm 1.38 \times 10^{-5}$ & $11.03^{+0.13}_{-0.13}$ & $12.23^{+0.23}_{-0.53}$ & $0.39^{+1.08}_{-0.16}$ & $11.59^{+2.48}_{-2.71}$ & $0.0^{+0.01}_{-0.0}$ & $11.55^{+1.28}_{-0.65}$ \\ \hline
    $8.0 \le z < 10.5$  & 8.70 & 697   & $8.14 \times 10^{-5} \pm 3.72 \times 10^{-5}$ & $10.84^{+0.07}_{-0.11}$ & $11.91^{+0.33}_{-0.19}$ & $1.18^{+0.42}_{-0.82}$ & $11.35^{+2.55}_{-2.44}$ & $0.0^{+0.0}_{-0.0}$ & $10.29^{+0.71}_{-0.44}$ \\
    "extended"          & 9.0  & 203   & $3.42 \times 10^{-5} \pm 1.79 \times 10^{-5}$ & $10.99^{+0.08}_{-0.13}$ & $11.9^{+0.41}_{-0.28}$ & $1.39^{+0.3}_{-0.91}$ & $11.34^{+2.62}_{-2.51}$ & $0.0^{+0.01}_{-0.0}$ & $11.25^{+1.04}_{-0.51}$ \\ \hline
    $10.5 \le z < 14.0$ & 8.85 & 35    & $3.79 \times 10^{-6} \pm 2.64 \times 10^{-6}$ & $10.66^{+0.15}_{-0.1}$ & $11.59^{+1.52}_{-0.06}$ & $0.57^{+0.9}_{-0.3}$ & $11.1^{+3.45}_{-2.35}$ & $0.0^{+0.0}_{-0.0}$ & $15.27^{+1.37}_{-0.88}$ \\
    "conservative"        & \\ \hline
    $10.5 \le z < 14.0$ & 8.85 & 140   & $1.47 \times 10^{-5} \pm 1.00 \times 10^{-5}$ & $10.54^{+0.26}_{-0.14}$ & $11.49^{+0.6}_{-0.33}$ & $0.75^{+0.58}_{-0.64}$ & $11.03^{+2.76}_{-2.55}$ & $0.0^{+0.03}_{-0.0}$ & $14.29^{+2.28}_{-1.68}$ \\ 
    "extended"          & \\ \hline
    \end{tabular}
    \vspace{3mm}
    \caption{Number of galaxies for each redshift and mass threshold bins (after cleaning); HOD best-fit parameters and model-derived properties.}
    \label{app-tab:HODparams}
\end{table*}

\section{Cross-correlations between redshift bins} \label{app:crosscorr}

\begin{figure*}
    \centering
    \includegraphics[scale=0.52]{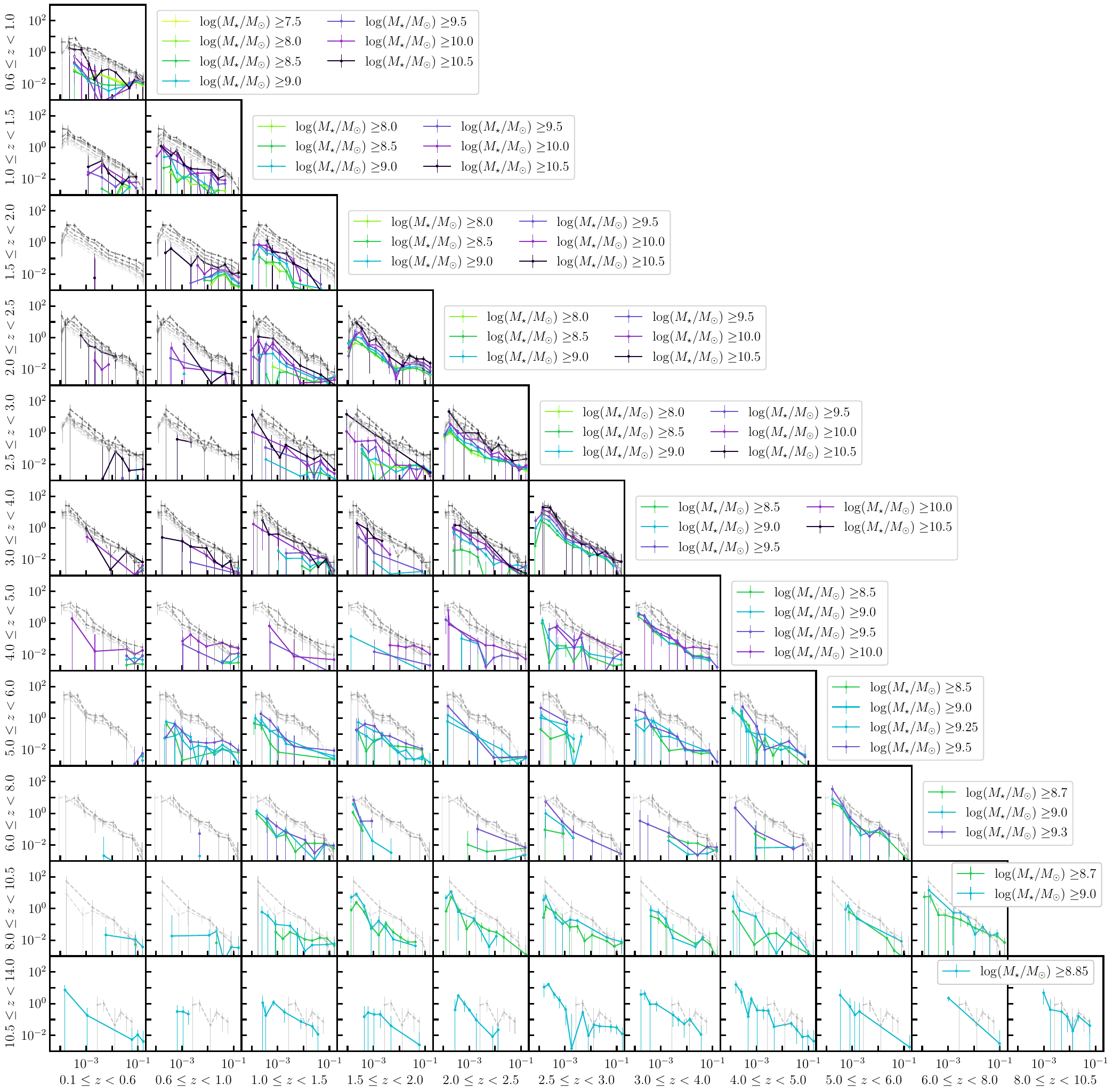}
    \caption{Cross-correlations between all redshift bins in our analysis, for each mass-limited sample. To facilitate comparisons, the auto-correlation signal for the redshift bin in each row is also shown in grey.}
    \label{app-fig:crosscorr}
\end{figure*}

To assess potential contamination in our analysis arising from photo-$z$ estimates or source detection, we computed cross-correlations between galaxy samples at various redshift bins but with the same mass thresholds. This approach has been generalized by \cite{Szapudi&Szalay1998}:
\begin{equation} \label{eq:wtheta_crosscorr}
\omega_{1,2}(\theta) = \frac{D_1D_2(\theta) -D_1R(\theta) -D_2R(\theta) +RR(\theta)}{RR(\theta)}\, .
\end{equation}
Cross-correlation analyses have been utilized in previous studies such as \cite{Coupon15}, who employed them to validate and improve photo-$z$ estimates, while others like \cite{Hatfield&Jarvis17} used them to trace different galaxy populations. In our case, in the light of recent JWST findings that have initially identified sources as very high-$z$ that turned out to be dusty sources around $z = 3$, we utilized cross-correlations to investigate how different selections for high-redshift galaxies impacted their clustering and the potential contamination from lower redshift bins. Results are presented in Fig.~\ref{app-fig:crosscorr}, where all cross-correlations with the various redshift bins are shown, and for different galaxy masses. 

Notably, the large probability distributions of photo-$z$ estimates become apparent when comparing adjacent bins, indicating potential contamination. Additionally, distinct features emerge in cross-correlations between high-redshift bins ($10.5 \le z < 14$) and lower bins ($2 \le z < 3$ or $3 \le z < 4.5$), as well as for rare massive galaxies at $z \ge 6$ or $z \ge 8$. These observations suggest that even after our cleaning efforts and visual inspections, certain sources may still contaminate our high-$z$ sample, such as highly dust-obscured galaxies or AGNs at lower $z$. Despite these challenges, we maintain confidence in our auto-correlation measurements, as the observed signals are about one order of magnitude lower.

\section{Theory and HOD fits with non-linear scale-dependent halo bias.} \label{app:clust_SDNL}

The fitting function for the non-linear scale-dependent correction term $\zeta(r, M_{\rm h}, z)$, according to \cite{Jose16}, can be expressed as:
\begin{align}
    \zeta(r, M_{\rm h}, z)  & = \zeta(\xi_{\rm mm}, \nu(M_{\rm h}), \alpha_{\rm m}, M_{\rm h})  \\
    & = [ 1 + K_0 \log[1 + \xi_{\rm mm}^{k_1}(r) ] \; \nu^{k_2} \; (1 + k_3/\alpha_{\rm m}) ]  \notag \\
    & \times [ 1 + L_0 \log[1 + \xi_{\rm mm}^{l_1}(r) ] \; \nu^{l_2} \; (1 + l_3/\alpha_{\rm m})]\, . \notag
\end{align}
Here, $\xi_{\rm mm}$ is the matter auto-correlation, $\nu(M_{\rm h}, z) = \delta_{\rm c} / \sigma(M_{\rm h},z)$ is the peak height of halos (measurement of the "rarity" of halos), $\sigma(M_{\rm h},z)$ is the variance of the linear matter overdensity field on a halo mass scale $M$ and $\delta_{\rm c} = 1.686$ is the critical overdensity for halo spherical collapse. Parameters are fitted using the MS-W7 simulation, in the range $3 \le z < 5$, as $K_0 = -0.0697$, $k_1 = 1.1682$, $k_2 = 4.7577$, $k_3 = - 0.1561$, $L_0 = 5.1447$, $l_1 = 1.4023$, $l_2 = 0.5823$, and $l_3 = - 0.1030$.
The effective power law index $\alpha_{\rm m}$ is given by :
\begin{equation}
    \alpha_{\rm m}(z) = \frac{\log(\delta_{\rm c})}{\log[M_{\rm h, nl}(z)/M_{\rm h,col}(z)]}\, ,
\end{equation}
where $M_{\rm h,nl}$ is the mass scale when the peak height equals the critical density, and $M_{\rm h,col}$ is the collapse mass scale when the peak height equals one. This non-linear term in the bias goes to unity at large scales.

\begin{figure*}
    \centering
    \includegraphics[scale=0.5]{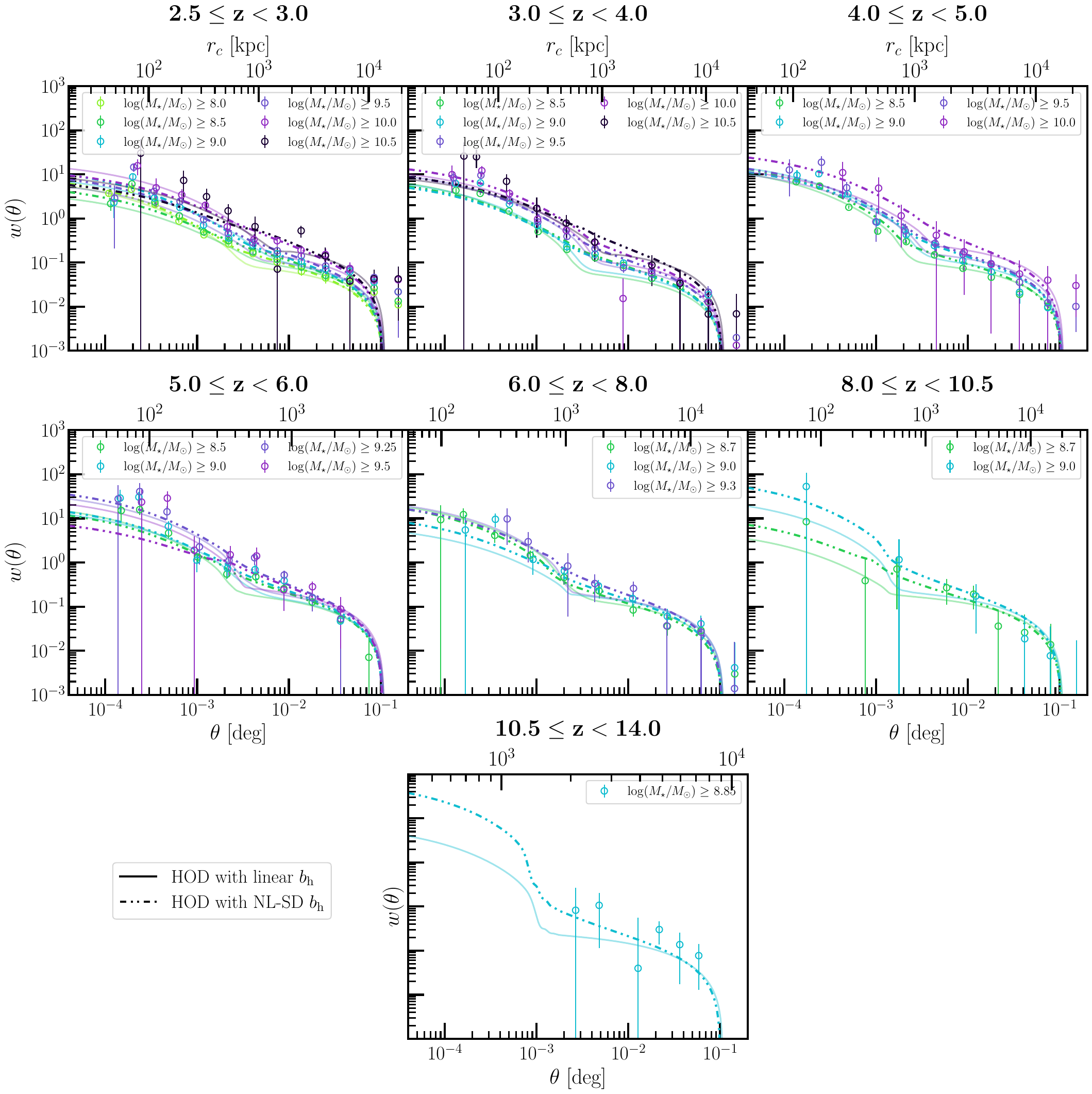}
    \caption{Angular auto-correlation function of galaxies in the COSMOS-Web survey, in redshift and mass limited bins. Dash-dotted lines show the HOD best-fit models with \cite{Jose16} non-linear scale-dependent halo bias and solid lines are HOD best-fit models without it (assuming a large-scale bias of \citealt{Tinker10}).}
    \label{app-fig:clust_SDNL}
\end{figure*}

Figure~\ref{app-fig:clust_SDNL} shows the clustering measurements at $z \ge 2.5$ with fitted HOD models that assume a non-linear scale-independent halo bias from \cite{Jose16}, noted as NL-SD. 

\clearpage
\section{Description of the simulations and semi-empirical models} \label{app:sims}

Table~\ref{app-tab:simulations} provides a non-exhaustive overview of the simulations used in Sect.~\ref{subsubsec:SHMR_comparison} to compare with our measured SHMR for central galaxies. In each simulation, we selected central galaxies as those hosted by the primary subhalo in each Friends-of-Friends (FoF) group, defined as the subhalo with the largest number of bound particles, typically the most massive. To match the halo masses returned by our HOD model in \texttt{halomod}, which uses virial halo masses (total mass enclosed within a spherical overdensity as defined by \citealt{Bryan&Norman1998}), we adopted the same mass definition for \textsc{Thesan} and Firstlight. For \textsc{Horizon}-AGN, \textsc{Obelisk}, and TNG100, the halo mass represents the total dark matter mass in the group. Stellar mass for central galaxies is defined as the sum of all stellar particles within twice the stellar half-mass radius in TNG100, or the total mass of all bound stellar particles in the substructure for the other simulations. Where necessary, we converted these masses to be consistent with the \cite{Chabrier03} IMF.

\begin{table*}[!hb]
    \centering
    \fontsize{8pt}{8pt}\selectfont
    \renewcommand{\arraystretch}{1.5}
    
    \begin{tabular}{|>{\centering\arraybackslash}m{2cm}|>{\centering\arraybackslash}m{1.1cm}|>{\centering\arraybackslash}m{1.2cm}|>{\centering\arraybackslash}m{1.6cm}|>{\centering\arraybackslash}m{1.3cm}|>{\raggedright\arraybackslash}m{2.75cm}|>{\raggedright\arraybackslash}m{1.5cm}|>{\raggedright\arraybackslash}m{3cm}|}
    \hline
    \textbf{Simulation} & \textbf{Box Size} & \textbf{Resolution} & \textbf{Mass Resolution} & \textbf{Cosmology} & \textbf{Dataset used in this work} & \textbf{Halo mass definition} & \textbf{Key Comments}  \\ 
    \hline
    \hline
    \textsc{Horizon}-AGN \citep{Dubois2014} & 100 $\rm h^{-1}$cMpc & $1024^3$ & DM:\hspace{1cm}$8 \times 10^7 M_\odot$ Stars:\hspace{1cm}$2 \times 10^6 M_\odot$ & WMAP-7 & Galaxy and halo catalogs from a light-cone of 1 deg$^2$ area from redshift $z = 0$ to $z = 6$ \citep{Laigle2019, Gouin2019}, using \citep{Chabrier03} IMF. & Sum of all DM particles in the main subhalo & AGN feedback; Overestimates SMF for $\log(M_\star/M_\odot) \ge 9.5$ \citep{Kaviraj2017} and underestimates at $z > 4$ due to spatial resolution limitations \citep[see][]{Hatfield19, Shuntov22}. \\ 
    \hline
    TNG100 \citep{Pillepich2018, Nelson2018, Marinacci2018, Naiman2018, Springel2018} & 75 $\rm h^{-1}$cMpc & $1820^3$ & DM:\hspace{1cm}$7.5 \times 10^{6} M_\odot$ Stars:\hspace{1cm}$1.6 \times 10^6 M_\odot$ & Planck16 & Public galaxy and halo catalogs from snapshots. & Sum of all DM particles in the main subhalo & Calibrated on low-$z$ data; \citep{Chabrier03} IMF. \\ 
    \hline
    \textsc{Thesan} \citep{Kannan2022} & 64.7 $\rm h^{-1}$cMpc & $2100^3$ & DM:\hspace{1cm}$3.12 \times 10^6 M_\odot$ Stars:\hspace{1cm}$5.8 \times 10^5 M_\odot$ & Planck16 & Galaxy catalogs from the public database. & Virial mass & Epoch of Reionization; IllustrisTNG model + radiative transfer; \citep{Chabrier03} IMF. \\ 
    \hline
    FirstLight \citep{Ceverino2017} & 10–80 $\rm h^{-1}$Mpc & $8192^3$ & DM: $10^4 M_\odot$\hspace{1cm} Stars: $10^6 M_\odot$ & Planck16 & Public database that contains information on 377 galaxies. & Virial mass & Epoch of Reionization; 300 zoom-in simulations, mass-complete sample; \citep{Salpeter1955} IMF. \\ 
    \hline
    \textsc{Obelisk} \citep{Trebitsch2021} & 15 $\rm h^{−1}$cMpc & $4096^3$ & DM:\hspace{1cm}$1.2 \times 10^6 M_\odot$ Stars:\hspace{1cm}$1 \times 10^4 M_\odot$ & WMAP-7 & Galaxy and halo catalogs from \cite{Trebitsch2021}. & Sum of all DM particles in the main subhalo & Epoch of Reionization; High resolution, focused on proto-cluster assembly; Re-simulates \textsc{Horizon}-AGN subvolume; \citep{Kroupa2001} IMF. \\
    \hline
    \end{tabular}
    
    \vspace{3mm}
    \caption{Non-exhaustive summary of hydrodynamical simulations used in this work.}
    \label{app-tab:simulations}
\end{table*}

\end{appendix}

\end{document}